\documentclass{jfm}

\usepackage{graphicx,psfrag,color}

\usepackage{natbib}

\ifCUPmtlplainloaded \else
  \checkfont{eurm10}
  \iffontfound
    \IfFileExists{upmath.sty}
      {\typeout{^^JFound AMS Euler Roman fonts on the system,
                   using the 'upmath' package.^^J}%
       \usepackage{upmath}}
      {\typeout{^^JFound AMS Euler Roman fonts on the system, but you
                   dont seem to have the}%
       \typeout{'upmath' package installed. JFM.cls can take advantage
                 of these fonts,^^Jif you use 'upmath' package.^^J}%
       \providecommand\upi{\pi}%
      }
  \else
    \providecommand\upi{\pi}%
  \fi
\fi

\ifCUPmtlplainloaded \else
  \checkfont{msam10}
  \iffontfound
    \IfFileExists{amssymb.sty}
      {\typeout{^^JFound AMS Symbol fonts on the system, using the
                'amssymb' package.^^J}%
       \usepackage{amssymb}%
         
         \let\geq=\geqslant
      }{}
  \fi
\fi

\ifCUPmtlplainloaded \else
  \IfFileExists{amsbsy.sty}
    {\typeout{^^JFound the 'amsbsy' package on the system, using it.^^J}%
     \usepackage{amsbsy}}
    {\providecommand\boldsymbol[1]{\mbox{\boldmath $##1$}}}
\fi

\newcommand\Rey{\mbox{\textit{Re}}}  % Reynolds number
\newsavebox{\astrutbox}
\sbox{\astrutbox}{\rule[-5pt]{0pt}{20pt}}

\newcommand\half{\small {1 \over 2}}

\title[Minimal seeds in plane Couette flow]{Variational
identification of minimal seeds to trigger transition in plane Couette flow}

\author[S. M. E. Rabin, C. P. Caulfield, and R. R. Kerswell]%
{S.\ns M.\ns E.\ns R\ls A\ls B\ls I\ls N$^1$%
  \thanks{Email address for correspondence: s.rabin@damtp.cam.ac.uk},\ns
C.\ns P.\ns C\ls A\ls U\ls L\ls F\ls I\ls E\ls L\ls D$^{2,1}$\break
\and R.\ns R.\ns K\ls E\ls R\ls S\ls W\ls E\ls L\ls L$^{3}$}

\affiliation{$^1$Department of Applied Mathematics \& Theoretical Physics, Centre for Mathematical Sciences, University of Cambridge, Wilberforce Road, Cambridge CB3 0WA, UK\\[\affilskip]
$^2$BP Institute, University of Cambridge, Madingley Rise, Madingley Road, Cambridge CB3 0EZ, UK\\[\affilskip]
$^3$School of Mathematics, University of Bristol, BS8 1TW Bristol, UK}

\pubyear{2011}
\volume{1}
\pagerange{1--2}
\date{\today}
\begin{document}

\maketitle

\begin{abstract}

A variational formulation incorporating the full Navier-Stokes
equations is used to identify initial perturbations with finite
kinetic energy $E_0$ which generate the largest gain in perturbation
kinetic energy (across all possible time intervals) for plane Couette
flow.  Two different representative flow geometries are chosen
corresponding to those used previously by \cite{butler1992} and
\cite{monokrousos2011}.  In the former (smaller geometry) case as
$E_0$ increases from 0, we find an optimal which is a smooth nonlinear
continuation of the well-known linear result at $E_0=0$. At $E_0=E_c$,
however, completely unrelated states are uncovered which trigger
turbulence and our algorithm consequently fails to converge. As $E_0
\rightarrow E_c^+$, we find good evidence that the
turbulence-triggering initial conditions approach a `minimal seed'
which corresponds to the state of lowest energy on the
laminar-turbulent basin boundary or `edge'. This situation is repeated
in the \cite{monokrousos2011} (larger) geometry albeit with one
notable new feature - the appearance of a nonlinear optimal (as found
recently in pipe flow by \cite{pringle2010a} and boundary layer flow
by \citet{cherubini2010}) at finite $E_0<E_c$ which has a very
different structure to the linear optimal. Again the minimal seed at
$E_0=E_c$ does not resemble the linear or now the nonlinear optimal.
Our results support the first of two conjectures recently posed by
\citet{pringle2011} but contradict the second. Importantly, their
prediction that the form of the functional optimised is not important
for identifying $E_c$ providing heightened values are produced by
turbulent flows is confirmed: we find the what looks to be the same
$E_c$ and minimal seed using energy gain as opposed to total
dissipation in the \cite{monokrousos2011} geometry.

\end{abstract}

%\begin{keywords}
%Nonlinear instability, Transition to turbulence, Navier-Stokes
%equations, Turbulent transition
%\end{keywords}

\section{Introduction}

The investigation of hydrodynamic stability is one of the canonical
problems of fluid dynamics. A particularly interesting archetypal flow
is plane Couette flow (PCF), where the flow is between two parallel
plates moving at a relative velocity $2U$, separated by a distance
$2h$, with thus a characteristic Reynolds number $\Rey = Uh/ \nu$
where $\nu$ is the kinematic viscosity. PCF is linearly stable for
even large $Re$ (\citet{romanov1973}) yet turbulence has been observed
experimentally as low as $Re=325$ (\citet{bottin1998}).  It has been
hypothesized that transient perturbation growth, due to the
non-normality of the underlying linear operator of the Navier-Stokes
equations, may explain this disconnect. Several authors (for example
\citet{gustavsson1991,butler1992,reddy1993}) demonstrated that
substantial transient kinetic energy gain $G(T)=E(T)/E(0)$ (where
$E(T)$ is the infinitesimally small kinetic energy at the final time
$T$) could be achieved by a linear (infinitesimal) optimal
perturbation (LOP) (see \cite{schmid2007} for a review).  Proponents of such
an essentially linear mechanism for energy growth point to the
Reynolds-Orr equation (\citet{schmid2007}) as an indication that
energy growth is a linear effect, as this equation shows that $dE/dt$
is independent of the nonlinear advective terms in the Navier--Stokes
equations.  The argument is that if a real flow is seeded with a
`small', yet finite amplitude perturbation with the same structure as
a LOP, the transient energy gain of the LOP could be sufficiently
large to `push' the perturbation into the `nonlinear regime' and hence
trigger transition.

However, this line of thinking is based on a couple of implicit
assumptions: that the `entrance' into the nonlinear regime of this
perturbation will lead to turbulence; and that the linear optimal
perturbation (LOP) is still the `best' choice for the growth of
nonlinear perturbations. The latter assumption can be explicitly
probed by posing the optimal growth problem for initial perturbations
of finite amplitude which can affect the base flow as they grow. This
has been done recently for pipe flow (\cite{pringle2010a}) and boundary
layer flow (\cite{cherubini2010}) with both studies discovering
the existence of a nonlinear optimal perturbation (NLOP) which has a
very different structure to the LOP and outgrows it beyond a small but
finite energy threshold.

With more of an eye on reaching turbulence, \citet{monokrousos2011}
posed a different problem for PCF by maximising the total energy
dissipation over a long but fixed time period. They purposely looked
for a turbulent end state at the end of their optimisation window and
then worked downwards in initial energy to identify the threshold for
transition. Earlier, \citet{pringle2010a} had failed to identify this
threshold by working upwards in initial energy because of convergence
issues. However, a follow-up study (\citet{pringle2011}, henceforth
referred to as PWK11) with a more efficient code run at higher
resolution succeeded in identifying a (converged) nonlinear optimal
for larger initial energies.  At $E_0=E_{fail}$, they again found a
failure to converge but noticed that this corresponded to their
optimisation algorithm encountering turbulent (end) flows. The
conclusion was that this failure energy $E_{fail}$ is sufficiently
large to enable an initial perturbation to undergo the transition to
turbulence: i.e. $E_{fail} \geq E_c$, the energy threshold of
transition (PWK11 actually conjectured that $E_{fail}=E_c$: see
Conjecture 1 below).  In dynamical systems parlance, this initial
perturbation is then in the basin of attraction of the turbulence or
more generally (if the turbulence is actually not an attractor but a
chaotic saddle), has crossed the `edge', a hypersurface which
separates initial conditions which become turbulent from those with
relaminarise
(\citet{itano2001,skufca2006,schneider2007,duguet2008}). The picture
then put forward by PWK11 is that as $E_0$ increases from 0, the
$E=E_0$ hypersurface in phase space intersects the edge for the first
time at $E_0=E_c$ and that the initial perturbation which corresponds
to their (generically unique) intersection at $E_0=E_c$ is the
`minimal seed' for triggering turbulence. This seed is `minimal' in
the sense that it is the lowest energy state on the edge and therefore
represents the most energy efficient way of triggering turbulence by
adding an infinitesimal perturbation to it.  PWK11 also find evidence
to suggest that the NLOP tends to the minimal seed associated with
this loss of convergence (and transition to turbulence) as $E_0
\rightarrow E_c^-$. They summarise their thinking as two conjectures.

`` \textbf{Conjecture 1}: For T sufficiently large, the initial energy
value $E_{fail}$ at which the energy growth problem first
fails (as $E_{0}$ is increased) to have a smooth optimal solution will
correspond exactly to $E_{c}$.
  
\textbf{Conjecture 2}: For $T$ sufficiently large, the optimal initial
condition for maximal energy growth at $E_{0} = E_{c} - \epsilon^{2}$
converges to the minimal seed at $E_{c}$ as $\epsilon \rightarrow
0$. ''\\

In this paper, we wish to investigate the validity of these
conjectures in the context of PCF.  Two sets of geometry and Reynolds
numbers are considered for PCF, one discussed in each of
\citet{butler1992} (henceforth referred to as BF92) - a relatively
narrow spanwise domain with $Re=1000$ - and \citet{monokrousos2011}
(henceforth referred to as M11) who used a domain with double the
spanwise extent at $Re=1500$.  By considering these different
situations, we are able to investigate whether there is anything
generic that can be said about the progression of optimal
perturbations starting with the (infinitesimal) LOPs at $E_0=0$,
through NLOPs as $E_0$ increases to the minimal seed at $E_0=E_c$. Of
principal interest will be whether this approach can identify $E_c$
and the form of the minimal seed either directly (by smooth evolution
of the optimal as $E_0 \rightarrow E_c^-$) or indirectly (by failing
to converge).  By considering the M11 geometry but choosing to
maximise the energy gain rather than total energy dissipation, we
also assess the sensitivity of the procedure to the exact choice of
optimising functional.

From the technical perspective, we also take this opportunity to
further develop the variational formulation to include optimisation
over $T$, the duration of the observation window or `target time'.
This means we are then able to identify the initial perturbation which
achieves the highest gain possible over {\em all} $T$ with the
corresponding optimal final time now an interesting output.  All
previous studies (\cite{pringle2010a, pringle2011, cherubini2010,
  monokrousos2011}) chose to work with a pre-defined $T$ over which to
perform their optimization.  Conceptually, letting $T$ be an output of
the optimisation seems a significant advance yet operationally, it is
requires only a small adjustment in the algorithm.

The plan of the paper is as follows. In section \ref{frame} we briefly
present the variational framework and discuss how the new target time
optimisation is carried out. The following two sections, \ref{results}
and \ref{comp}, discuss the results obtained for the BF92 and M11
situations respectively. A final section \ref{conc} then discusses
the results in light of the above-quoted conjectures of PWK11 and
draws  a number of conclusions.

\section{Lagrangian framework} \label{frame}

We seek the initial disturbance of kinetic energy $E_0$ to the laminar
flow which attains the largest energy growth $G(T):=E(T)/E_0$ a time
$T$ later while evolving under the Navier-Stokes equations, remaining
incompressible and respecting the applied boundary conditions. Here
$E(T):= \half\langle \mathbf{u}(T), \mathbf{u}(T) \rangle$, with the
angle brackets denoting,
\begin{equation}
\langle \mathbf{v},\mathbf{u} \rangle := 
\frac{1}{V}\int_{\mathcal{D}}\mathbf{v}^{\dagger}\mathbf{u}\,\,\, dV\,, \\
\end{equation}
where $\dagger$ denotes the Hermitian conjugate, and $V$ is the volume
of the domain ${\mathcal D}$. The flow configuration considered is PCF
with coordinate system such that the streamwise direction is $x$, the
wall normal direction is $y$ and the spanwise direction $z$. The $x$
and $z$ directions are assumed to be periodic and the separation
between the walls ($2h$) is used to scale length so that their
positions are given by $y=\pm 1$. The speed difference between the
walls ($2U$) scales the velocity so that the non-dimensionalised
background Couette flow is $\mathbf{U}(y)=y\mathbf{e}_{x}$ and the
Reynolds number $\Rey:=Uh / \nu$.

The functional to be extremised is the energy gain which, when
constrained by the Navier Stokes equations, the initial energy value
$E(0)=E_{0}$ and incompressibility,\refstepcounter{equation} produces
the Lagrangian
$$
   \mathcal{L} := \frac{E(T)}{E_{0}} -  \left[\partial_{t}\mathbf{u} + N( \mathbf{u}) +  \nabla p,\mathbf{v} \right]   -\left[ \nabla.\mathbf{u},q \right] \label{nonlinL} \\
\nonumber -  \left( \half\langle \mathbf{u}_{0},\mathbf{u}_{0} \rangle - E_{0} \right) c + \langle \mathbf{u}_{0}-\mathbf{u}(0),\mathbf{v}_{0} \rangle \,. \quad
  \eqno{(\theequation{\mathit{a}})}
$$ 
$N$ is the nonlinear operator
$$
  N(u_{i}):=U_{j}\partial_{j}u_{i} + u_{i}\partial_{i}U_{j}+u_{j}\partial_{j}u_{i} -\frac{1}{Re}\partial_{j}\partial_{j} u_{i}\,,  \quad
  \eqno{(\theequation{\mathit{b}})}
$$
and square brackets denote a time average of the inner product, 
\begin{equation}
\left[ \mathbf{v},\mathbf{u} \right] := \frac{1}{T}
\int_{0}^{T} \langle \mathbf{v},\mathbf{u} \rangle dt\,.
\end{equation}
In the Lagrangian, $\mathbf{v}$, $q$, $\mathbf{v}_{0}$ and $c$ are
Lagrange multipliers, $\mathbf{u}_{0}$ is the initial value of the
perturbation velocity $\mathbf{u}$ and $\mathbf{U}$ the background
Couette flow. While not strictly necessary to divide our cost
functional by $E_{0}$ we found it easier to tune our algorithm by
doing so.

Taking first variations of the Lagrangian with respect to
$\mathbf{v}$, $q$, $\mathbf{v}_{0}$ and $c$ and setting them to zero
recovers (respectively) the constraints of the Navier Stokes
equations, incompressibility, the initial kinetic energy of $E_{0}$
and initial state $\mathbf{u}_{0} =\mathbf{u}(0)$,
\begin{eqnarray}
\frac{\delta \mathcal{L}}{\delta \mathbf{v}} &=& 
\partial_{t}\mathbf{\mathbf{u}} + N(\mathbf{u}) + \nabla p :=0 \, , \label{for1} \\
\frac{\delta \mathcal{L}}{\delta q} &=& \nabla. \mathbf{u} :=0 \, , \\ 
\frac{\delta \mathcal{L}}{\delta \mathbf{v}_{0}} &=& 
\mathbf{u}_{0} - \mathbf{u}(0) :=0 .  \label{for3} \\
\frac{\delta \mathcal{L}}{\delta c} &=& 
\frac{1}{2}\langle \mathbf{u}_{0},\mathbf{u}_{0} \rangle - E_{0} := 0 .  \label{for4}
\end{eqnarray}
First variations with respect to the physical
 variables yields a complementary set of adjoint equations,
\begin{eqnarray}
\frac{\delta \mathcal{L}}{\delta \mathbf{u}} &=& \partial_{t}\mathbf{v} + N^{\dagger}(\mathbf{v},\mathbf{u}) + \nabla q + \left(\frac{\mathbf{u}}{E_{0}}-\mathbf{v}\right)|_{t=T} + \left(\mathbf{v}-\mathbf{v}_{0} \right) |_{t=0} :=0 \, , \label{adj1} \\
 \frac{\delta \mathcal{L}}{\delta p} &=& \nabla. \mathbf{v} :=0 \, , \\ 
 \frac{\delta \mathcal{L}}{\delta \mathbf{u}_{0}} &=& \mathbf{v}_{0} - c\mathbf{u}_{0} :=0 .  \label{adj3}
\end{eqnarray}
Here,  
\begin{equation}
N^{\dagger}({v}_{i},\mathbf{u}):=\partial_{j}\left(u_{j}{v}_{i}\right) 
- {v}_{j}\partial_{i}u_{j}+\partial_{j}\left(U_{j}{v}_{i}\right) 
- {v}_{j}\partial_{i}U_{j} +\frac{1}{Re}\partial_{j}\partial_{j} {v}_{i}
\end{equation}
can be identified as the adjoint of $N$ and $\mathbf{v}$,
$\mathbf{v}_{0}$ and $q$ are the adjoint variables of $\mathbf{u}$,
$\mathbf{u}_{0}$ and $p$.  Equation (\ref{adj1}) is in reality three
equations. The first part, $\partial_{t}\mathbf{v} +
N^{\dagger}(\mathbf{v},\mathbf{u}) + \nabla q={\bf 0}$, must be satisfied at
all times and is the adjoint Navier Stokes equation. Since the full
Navier-Stokes equations have been imposed, the adjoint operator
depends on the velocity field $\mathbf{u}$.  The sign of the diffusion
term is also reversed and therefore the adjoint equation can only be
solved backwards in time. The second part of (\ref{adj1}),
$\left(\mathbf{u}/E_{0}-\mathbf{v}\right)|_{t=T}=0$, is a terminal
condition, linking our physical and adjoint variables and needs only
to be satisfied at time $T$. The third part,
$\left(\mathbf{v}-\mathbf{v}_{0} \right) |_{t=0}={\bf 0}$, is a condition
linking $\mathbf{v}_{0}$ and $\mathbf{v}(0)$, which must be satisfied
at $t=0$.

Further to previous recent formulations (\citet{pringle2010a,
  pringle2011, cherubini2010, monokrousos2011}), we also optimize over
the target time $T$. The first variation with respect to $T$ yields
the simple relation
\begin{equation}
\frac{\partial \mathcal{L}}{\partial T} := \frac{1}{E_0}\frac{d}{d T}
E(T) =0 \,
\label{L_T}
\end{equation}
provided ${\bf u}$ is incompressible and satisfies the Navier-Stokes
equations at $t=T$.  Our algorithm then proceeds as follows. We first
start with a suitable guess for the optimal initial condition,
$\mathbf{u}_{0}$ and a target time $T$. We then time march our initial
condition to time $T$ using the Navier-Stokes equations and use
$\left(\mathbf{u}/E_{0}-\mathbf{v}\right)|_{t=T}=0$ to `initialise'
the adjoint equations which are then solved backwards in time to
calculate $\mathbf{v}_{0}$. This procedure ensures that all the
variational equations are satisfied apart from (\ref{adj3}) and
(\ref{L_T}). If the current value for $\mathbf{u}_{0}$ is optimal then
(\ref{adj3}) will be satisfied: on the other hand if (\ref{adj3}) is
not satisfied, it provides an estimate for the gradient $\delta
\mathcal{L} / \delta \mathbf{u}_{0}$. Using this gradient we then use
a method of steepest ascent to update our guess for $\mathbf{u}_{0}$,
while $c$ is simultaneously calculated by ensuring that our new
initial condition has an energy of $E_{0}$. Once a new value of
$\mathbf{u}_{0}$ is obtained, $T$ is updated by integrating the
Navier-Stokes equations forward in time using the updated
$\mathbf{u}_{0}$ as an initial condition until a maximum value of
$E(t)$ is reached. The time of this maximum is taken as the new value
of $T$ and (\ref{L_T}) is then satisfied.

\section{BF92 geometry} \label{results}

The underlying objective of this paper is to investigate how optimal
initial conditions for energy growth some $T$ later change as a
function of $E_{0}$. One specific issue is whether there is always an
energy range below $E_c$ where a NLOP is the optimal (a NLOP being
an initial condition qualitatively different in structure and gain
from the LOP). A second is whether $E_{fail}=E_c$ and a third is
examining the form of the optimal as $E_0 \rightarrow E_c$ from above
or below.  

\begin{figure}
  \centerline{\includegraphics[width=0.49\textwidth]{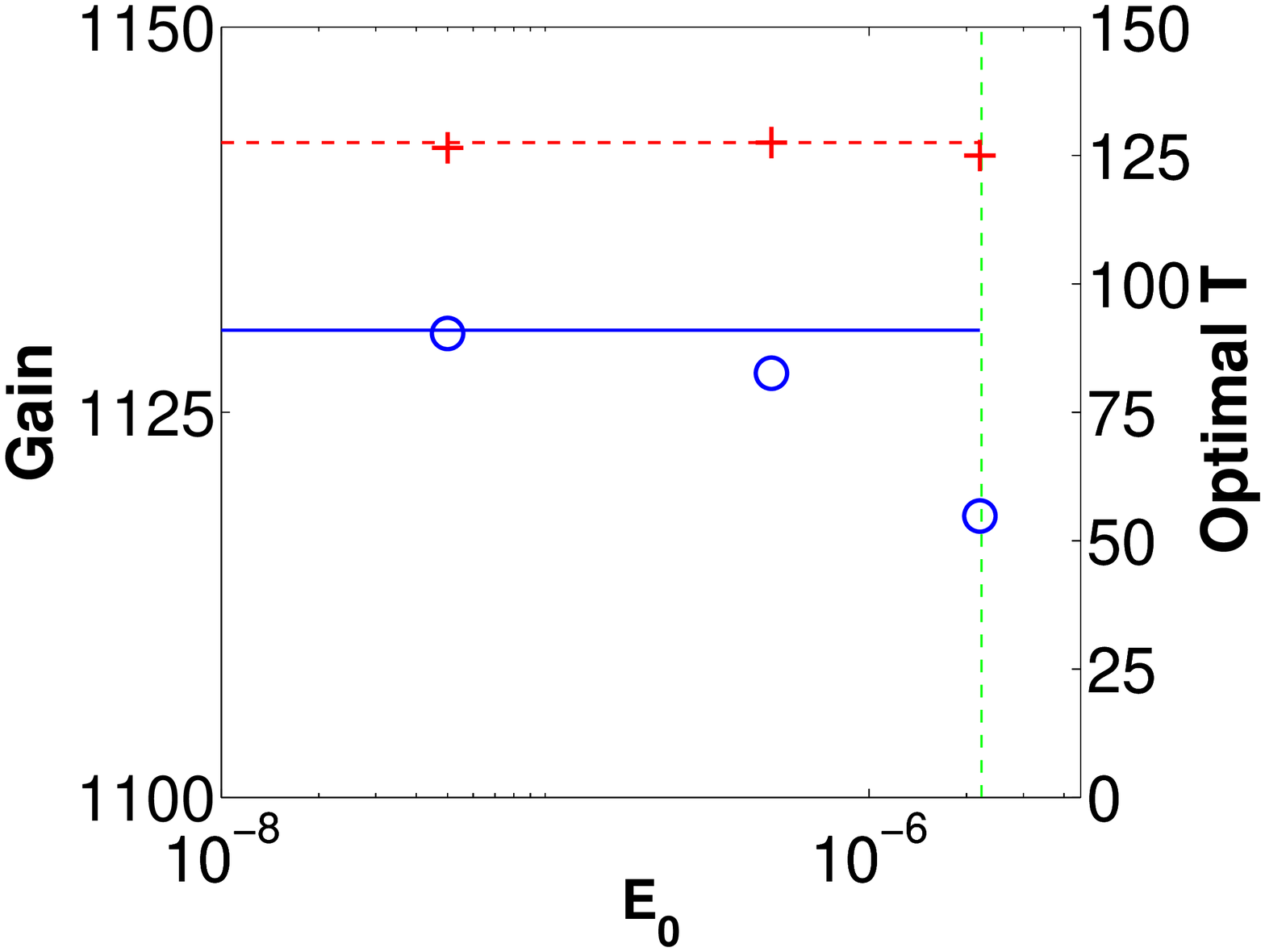}
  \includegraphics[width=0.49\textwidth]{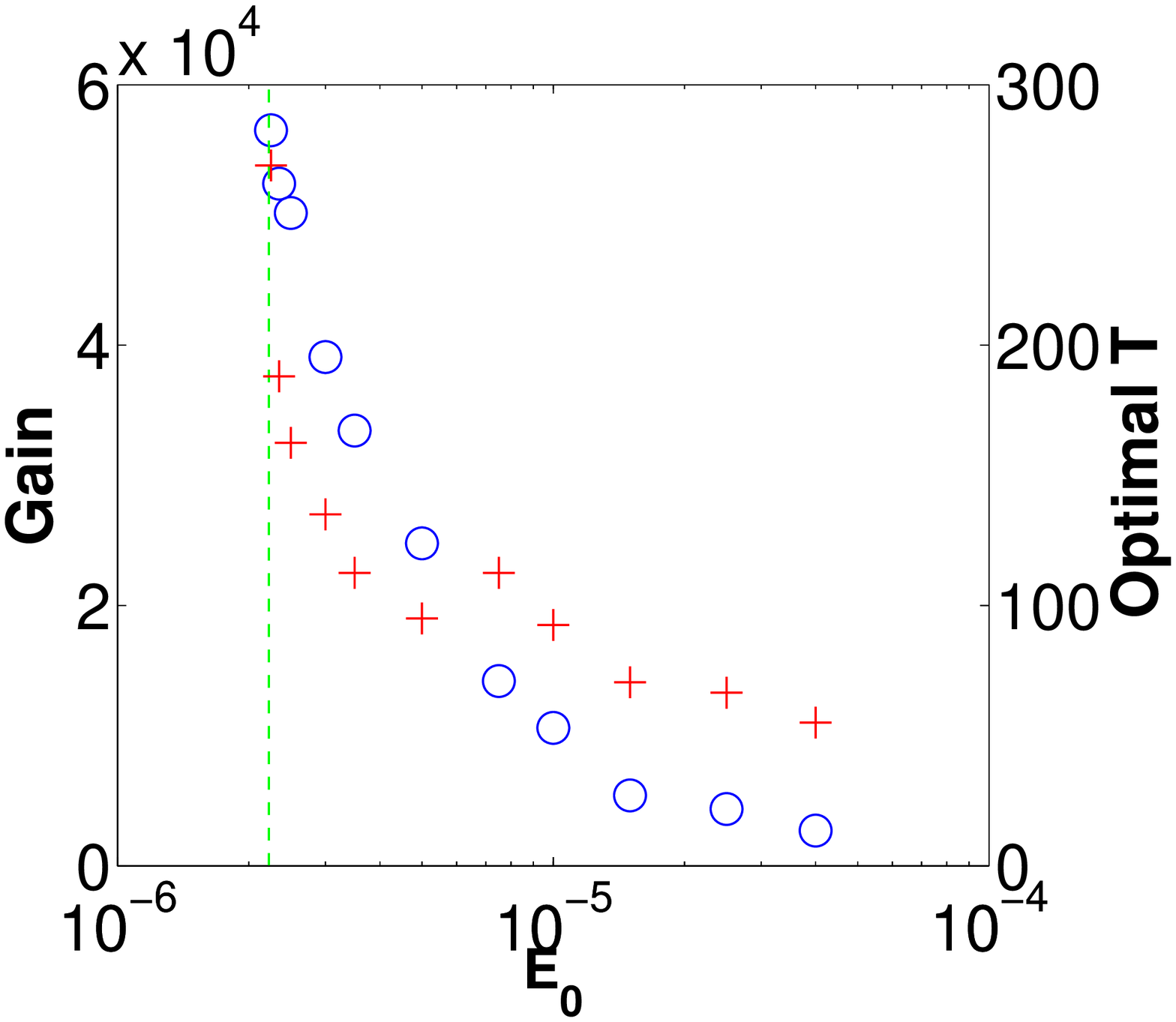}}
%  Images in 100% size
  \caption{(\textit{a}) Gain (blue circles) and associated optimal
    time (red crosses) against $E_{0}$. $E_{c} \approx 2.3 \times
    10^{-6}$ is marked by vertical green dashed line. LOP gain is the
    horizontal blue solid line, LOP optimal time is the horizontal red
    dashed line. (\textit{b}) Gain (blue circles) and associated
    optimal time (red crosses) against $E_{0}$ for $E_0 > 2.3 \times
    10^{-6}$.}
\label{GT_BF}
\end{figure}

\begin{figure}
\centerline{\includegraphics[width=0.49\textwidth]{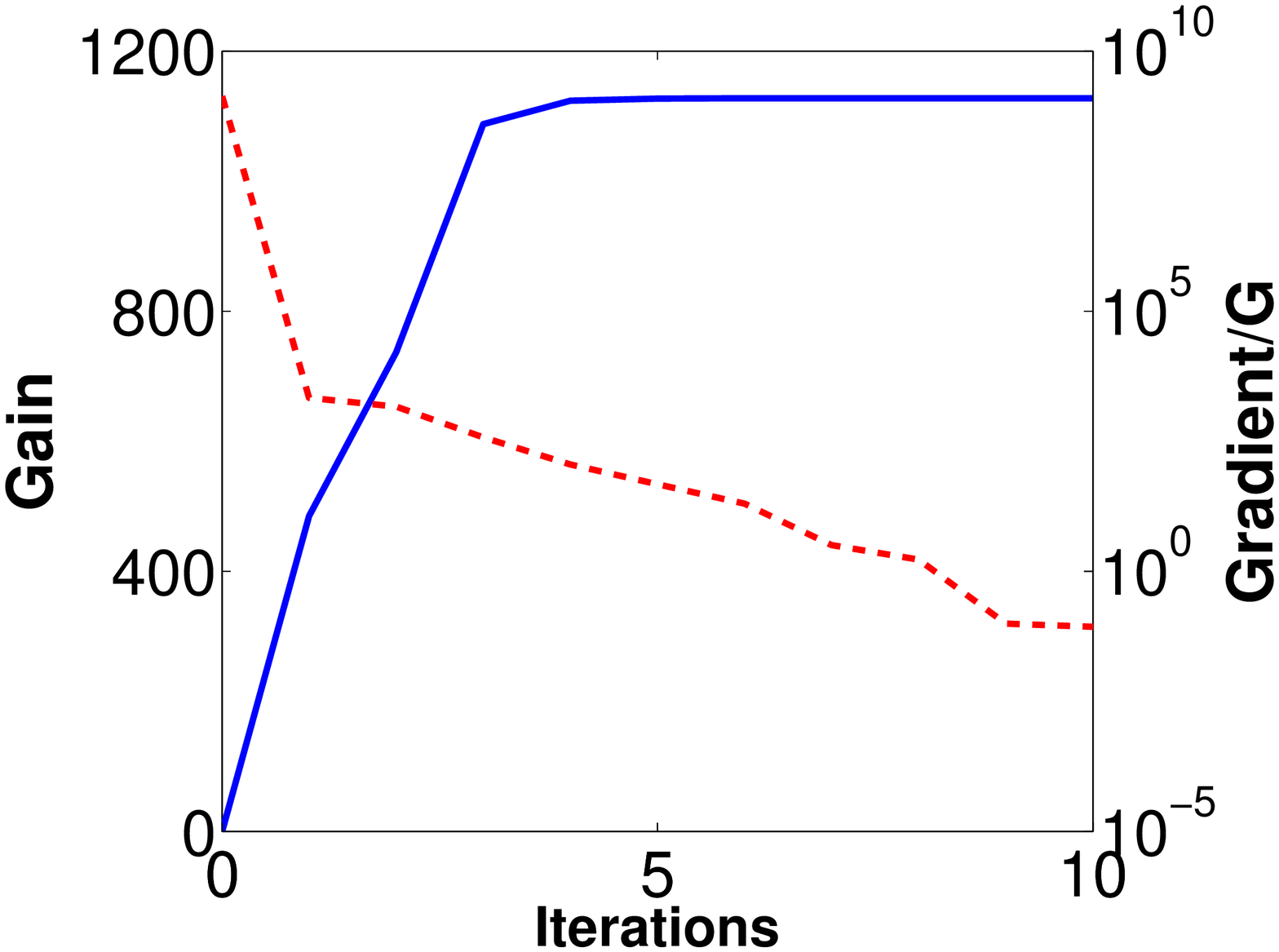}
            \includegraphics[width=0.49\textwidth]{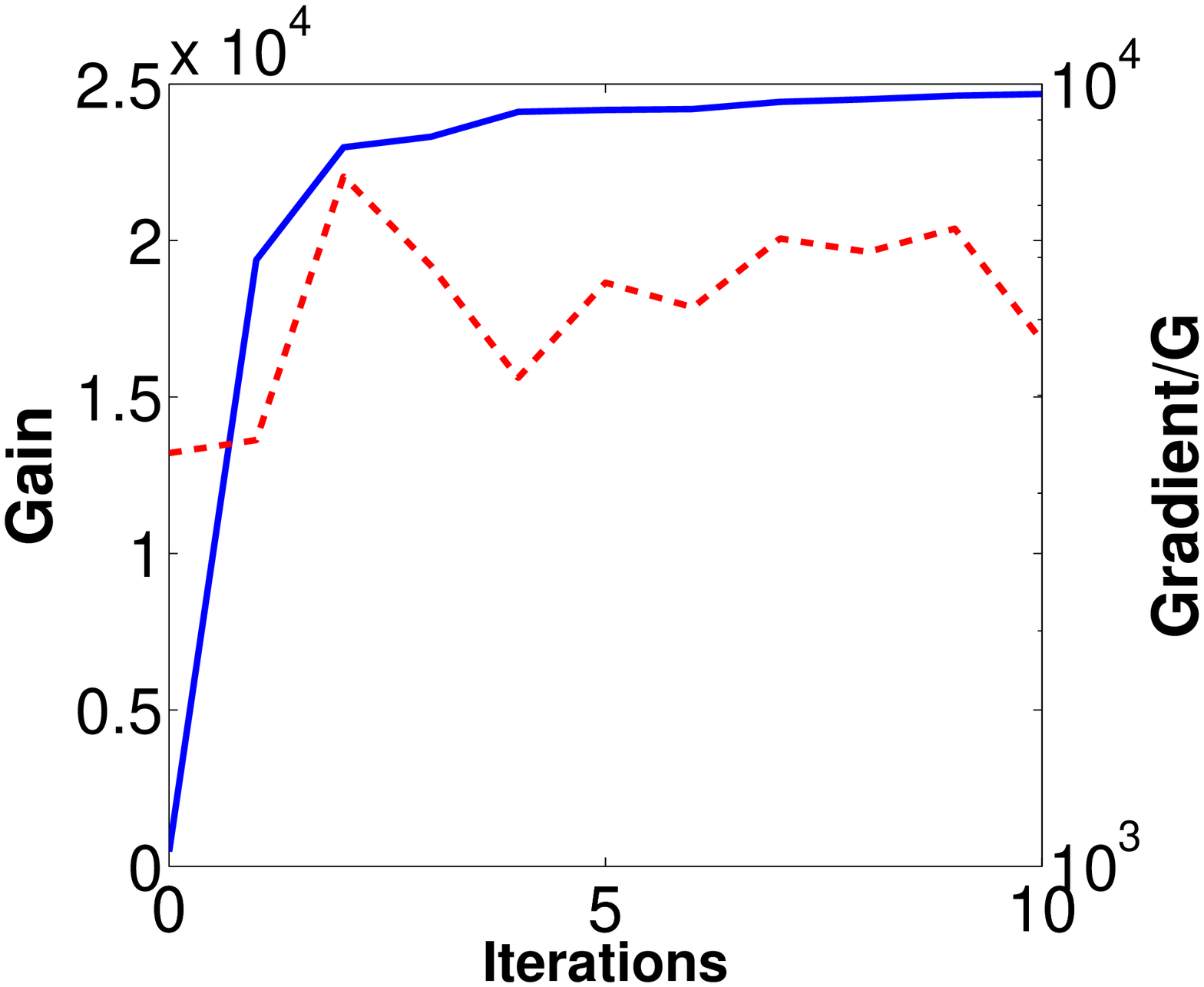}}
  \caption{G (blue solid line) and $\langle \delta \mathcal{L} /
    \delta \mathbf{u}_{0}, \delta \mathcal{L} / \delta \mathbf{u}_{0}
    \rangle^{\frac{1}{2}}/G$, (red dashed line) plotted against iteration for (\textit{a})
    $E_{0}=5.0 \times 10^{-7}$ (\textit{b}) $E_{0}=5.0 \times
    10^{-6}$. The QLOP presented in (\textit{a}) appears to be
    converging well, whereas there is no convergence in (\textit{b}).}
\label{Iter_BF}
\end{figure}

Results are first presented from a geometry studied previously in the
linear regime ($E_0 \rightarrow 0$) in BF92: a periodic box with
dimensions $L_{x}=2\upi/0.49=13.66$, $L_y=2$ and
$L_{z}=2\upi/1.9=3.31$ (or $4.08\pi \times 2 \times 1.05\pi$) with
$\Rey=1000$. A modified version of the Diablo CFD solver,
(\cite{taylor2008}) which is spectral in $x$ and $z$ and finite
difference in $y$, was used to solve the forward and adjoint equations
using a resolution of $128 \times 256 \times 32$ in $x$, $y$ and $z$
respectively. For sufficiently low energies we found that the optimal
perturbation was extremely similar to the LOP in both gain and
structure and as result was named a `quasi linear optimal
perturbation' (QLOP). The QLOP achieved a maximum gain of
approximately 1100 at $T=125$ (in units of $h/U$). With increasing but
still small $E_{0}$, the gain and optimal time of the QLOP remains
fairly constant as shown in figure \ref{GT_BF}(a). However, beyond a
certain energy threshold, (approximately $2.3 \times 10^{-6}$) there
is a sudden and large jump in the gain achievable, as shown in figure
\ref{GT_BF}(b) (note the change of ordinate scale).  In addition to
the much higher gain at $E_{0} = 2.3 \times 10^{-6}$, it is noticeable
that the optimal time tends to very large values as $E_{0}$ approaches
this transition energy from above.  The initial conditions found by
our algorithm there clearly evolve into a turbulent state given the
much higher target-time kinetic energy values and the highly
disordered endstate. This implies that $E_c \lesssim 2.3 \times
10^{-6}$ is the threshold energy for transition.  To examine
convergence, $\langle \delta \mathcal{L} / \delta \mathbf{u}_{0},
\delta \mathcal{L} / \delta \mathbf{u}_{0} \rangle^{\frac{1}
{2}}/G$ is plotted in figure \ref{Iter_BF} against
iteration for $E_{0}=5.0 \times 10^{-7} < E_c$ and $E_{0}=5.0 \times
10^{-6} > E_c$.

%
% Fig 1
%

For the smaller initial energy case (figure \ref{Iter_BF}(a)) we see
that after 10 iterations the gain has plateaued and that the value of
the normalised gradient has dropped by 10 orders of magnitude, which suggests that
the QLOP is converging well.

Conversely, for the higher energy case (figure \ref{Iter_BF}(b)) while
the gain appears to plateau, the gradient is failing to decrease in size
indicating the algorithm is not converging.  In reality, because of
the turbulent nature of the flow at time $T$ we would not expect
convergence to be possible, as a very small change in the initial
condition is likely to produce a significant change in the final
state. Despite this lack of convergence, the algorithm is successful
in finding initial states which trigger turbulence when $E_{0}>E_{c}$.

%
% Fig 2
% 

\begin{figure}
  \centerline{\includegraphics[width=0.49\textwidth]{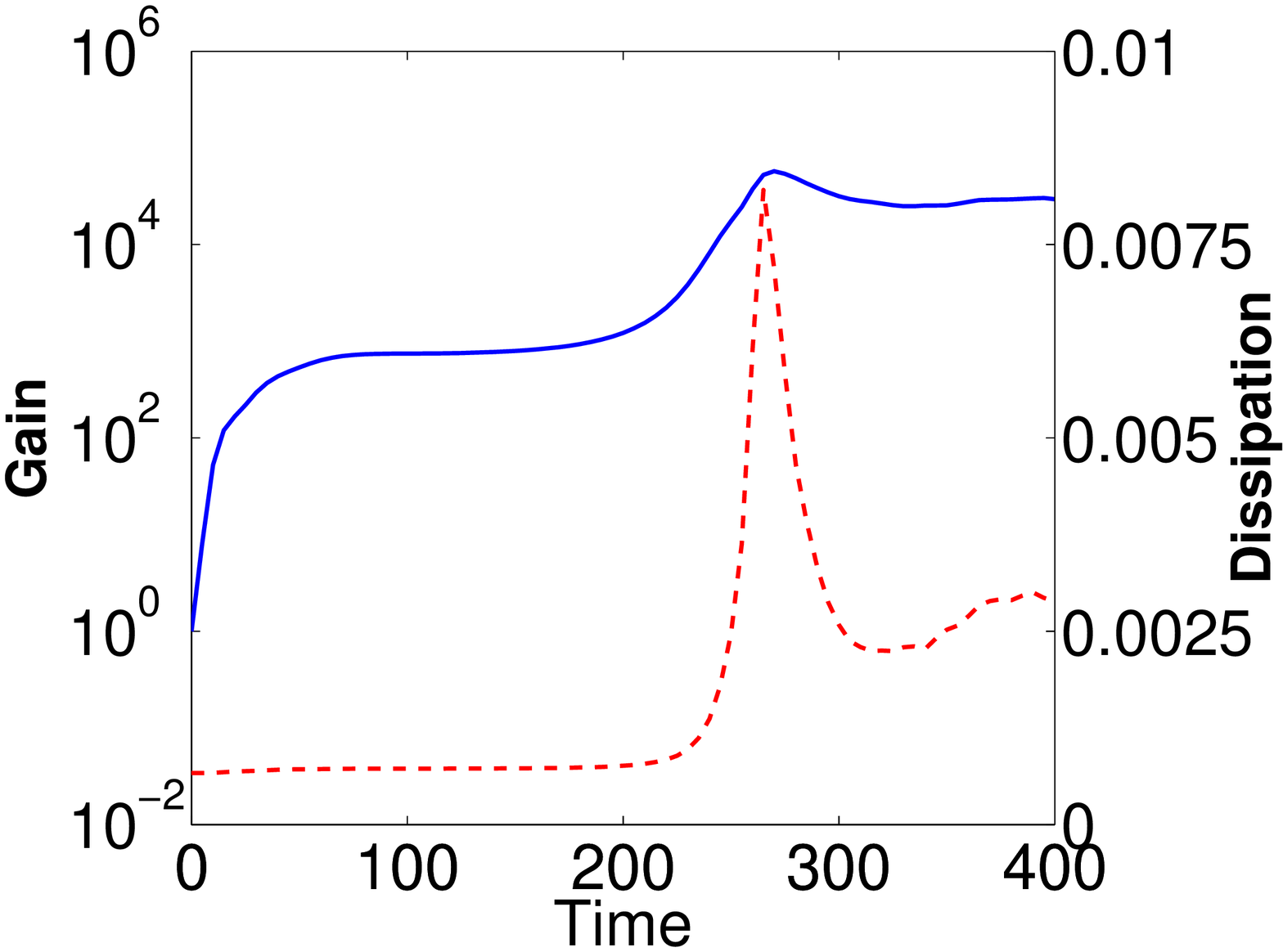}
              \includegraphics[width=0.49\textwidth]{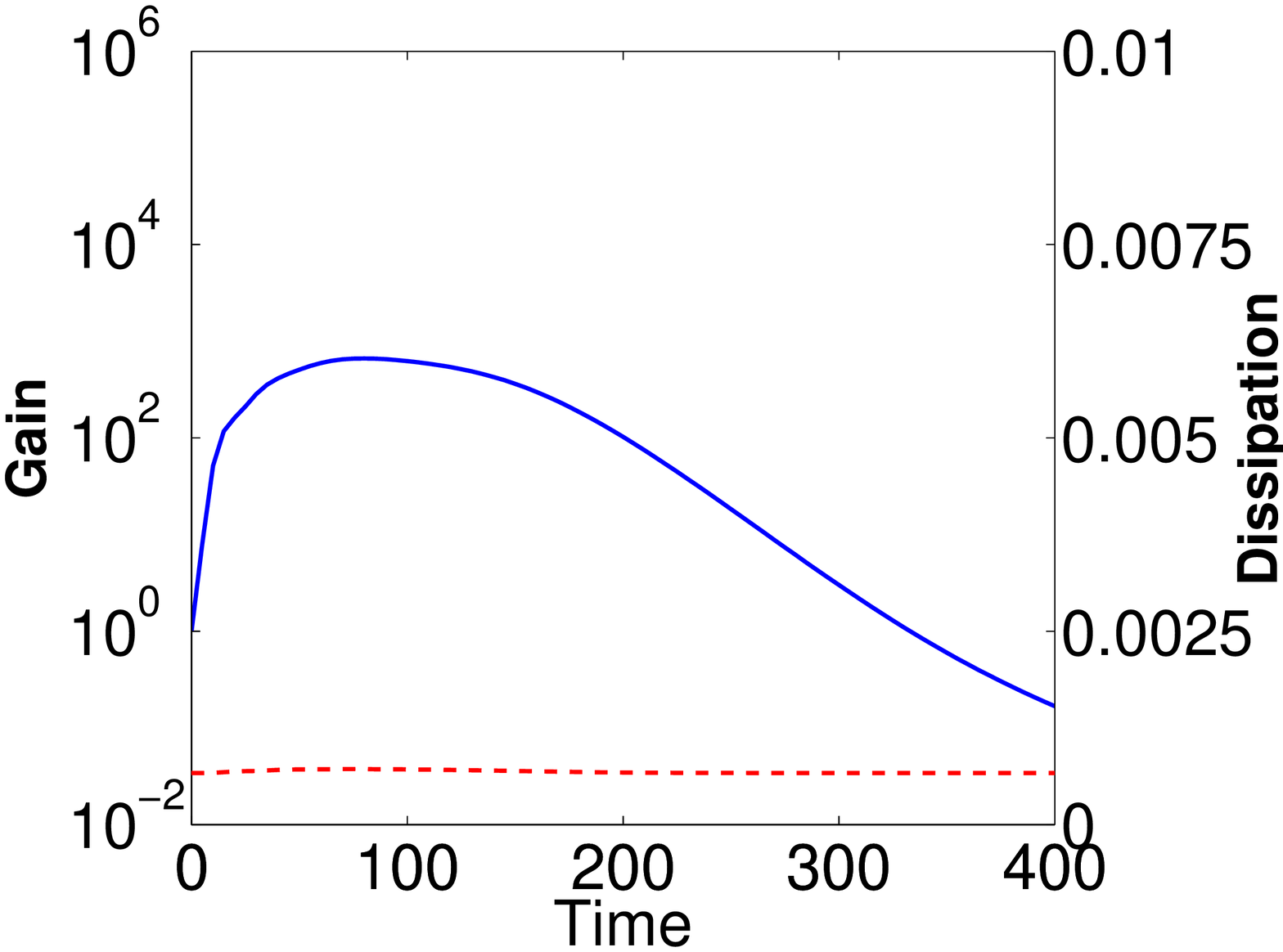}}% Images in 100% size
  \caption{Gain (blue solid line) and dissipation (red dashed line)
    against time for (\textit{a}) the turbulent seed at $E_0=2.3
    \times 10^{-6}$ and (\textit{b}) rescaled turbulent seed at
    $E_0=2.2 \times 10^{-6}$. Both initially behave similarly
    achieving a gain of approximately 1000. They maintain this energy
    level for an extended period of time, until $t \sim 150$. After
    this the turbulent seed has a noticeable spike in dissipation,
    indicating a transition to turbulence, whereas the rescaled
    turbulence seed decays away.}
\label{Gdisp_BF}
\end{figure}

In the picture of PWK11, one unique initial condition - the minimal
seed - should emerge as the limiting state for turbulence-triggering
initial conditions as $E_0 \rightarrow E_c^+$.  To examine the
dynamical route states close to the minimal seed take to turbulent
disorder, we have considered in detail a turbulent seed found at
$E_0=2.3 \times 10^{-6}$ as it evolves in time. Figure
\ref{Gdisp_BF}(a) plots the kinetic energy and dissipation rate
against time and in \ref{Gdisp_BF}(b) the same quantities are plotted
for the `rescaled' turbulent seed whose initial energy is $E_{0} =
2.2 \times 10^{-6}$. The behaviours of the turbulent seed and rescaled
turbulent seed are very similar up to $t \sim 75\,h/U$. Beyond this
time, however, the kinetic energies of the two initial perturbations
begin to differ significantly, with the rescaled turbulent seed
eventually decaying so the flow relaminarises whereas the turbulent
seed triggers transition at $T \sim 270\, h/U$.  (Reassuringly, if the
rescaled turbulent seed is used to initiate the optimizing procedure
at $E_{0}=2.2 \times 10^{-6}$, the algorithm converges to the expected
QLOP result.)

%
% Fig 3
%
 
\begin{figure}
  \centerline{\includegraphics[width=1.0\textwidth]{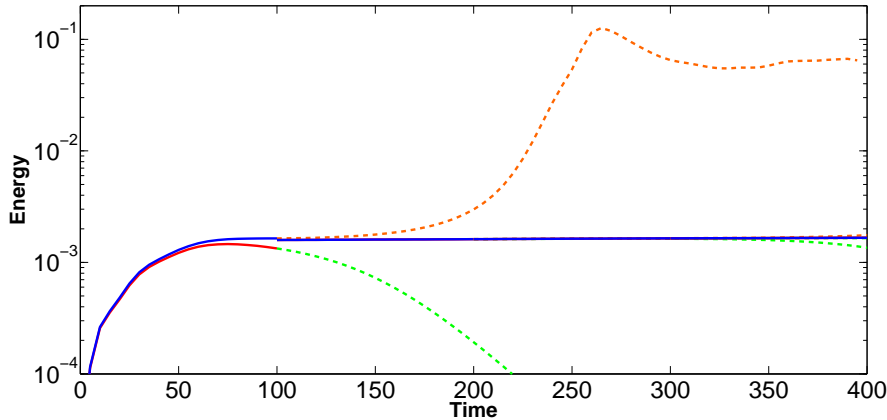}}% Images in 100% size
  \caption{Perturbation energy against time for various initial states
    close to the edge. Upper bound of edge state blue, lower bound
    red. Every 100 time units, the edge state is rescaled to produce
    new upper and lower bounds. The minimal seed at $E_c$ (blue line)
    stays on the edge (over this time period) and gets attracted to
    the edge state which emerges as having constant energy. The
    turbulent seed at $E_0=2.3 \times 10^{-6}$ is shown as the first blue/orange
    dashed line and the rescaled turbulent seed is the first red/green
    dashed line.}
\label{3plots_BF}
\end{figure}

We observe in figure \ref{Gdisp_BF} that both flows spend an extended
period of time at an intermediate (perturbation) energy level before
going their separate ways. This is because both initial states are
close to the edge (but on `opposite sides') and spend some time
tracking it while being gently repelled (in opposite directions). To
confirm this, the turbulent seed and its rescaling can be used to
refine the initial condition so that it stays nearer to the edge for
longer (\citet{itano2001,skufca2006,schneider2007,duguet2008}).  In
figure \ref{3plots_BF}, this refinement is carried out to track the
edge for $t=400\, h/U$ showing that the edge state (attracting state
for edge-confined dynamics) has constant energy (consistent with the
work of \cite{schneider2008} who treat a PCF system $4\pi \times
2 \times 2\pi$ albeit at $Re=400$ and find a steady edge state).

%
% Fig 4
%

%
% Fig 5
%
\begin{figure}
  \centerline{\includegraphics[width=0.32\textwidth]{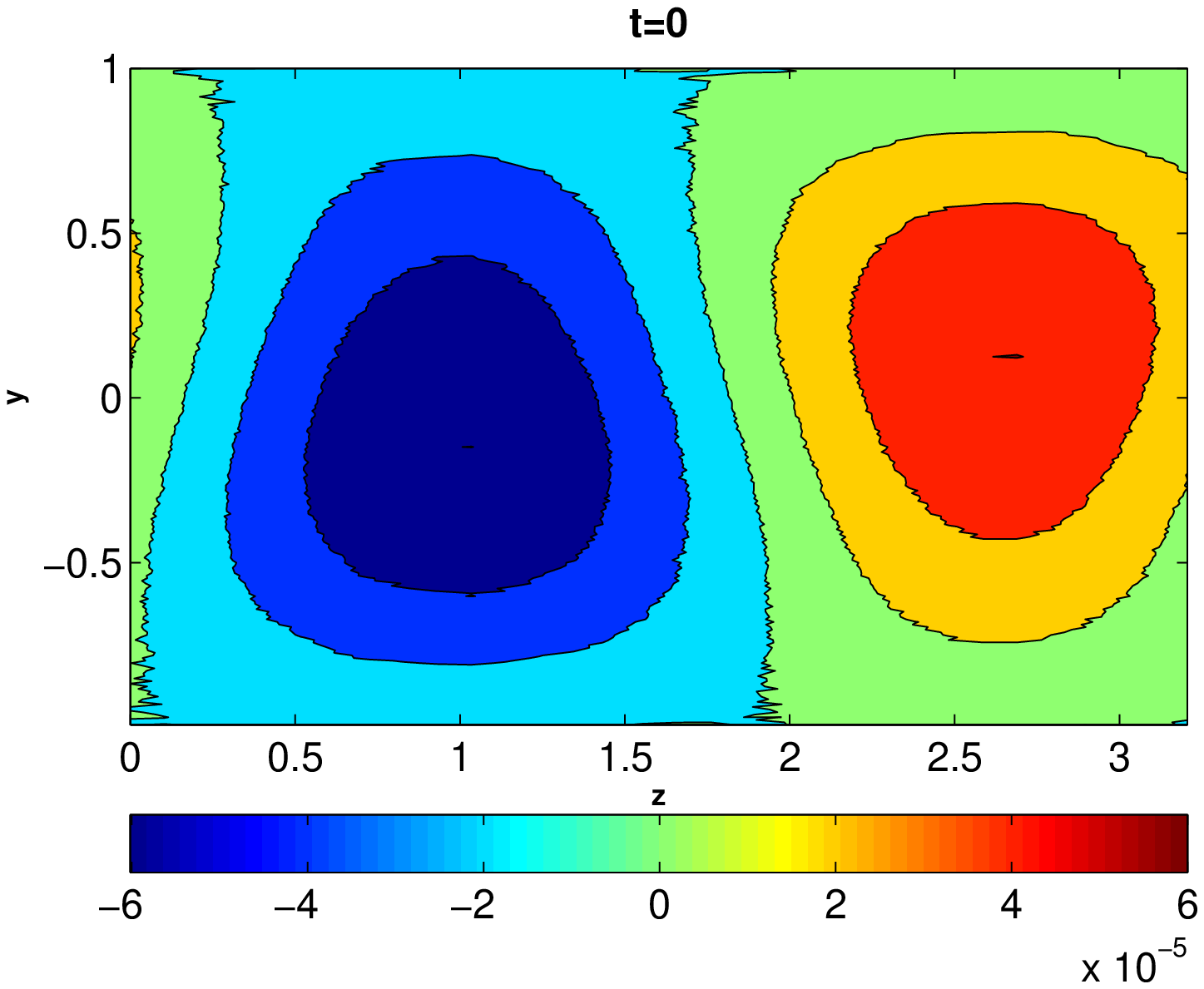}
  \includegraphics[width=0.32\textwidth]{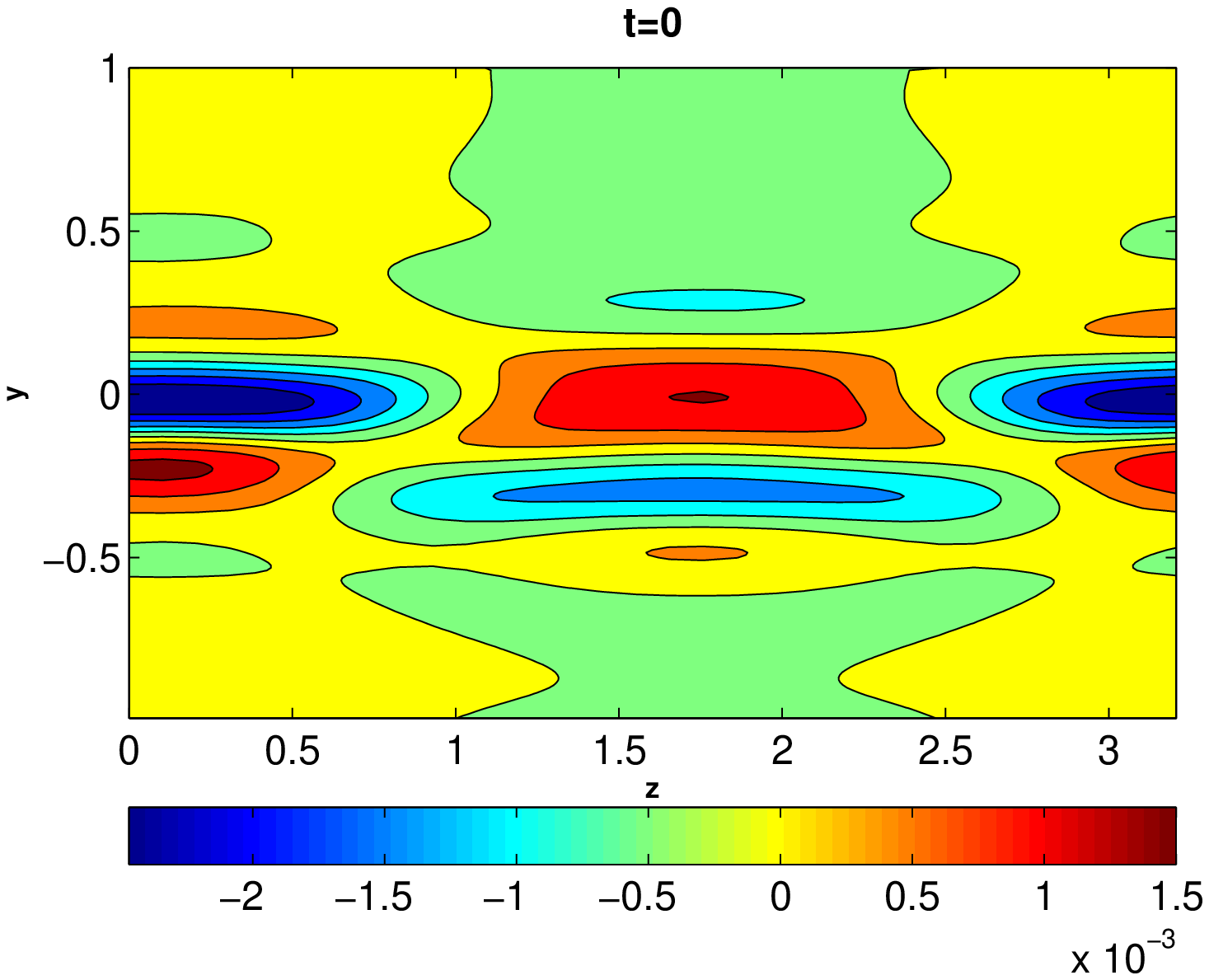}
  \includegraphics[width=0.32\textwidth]{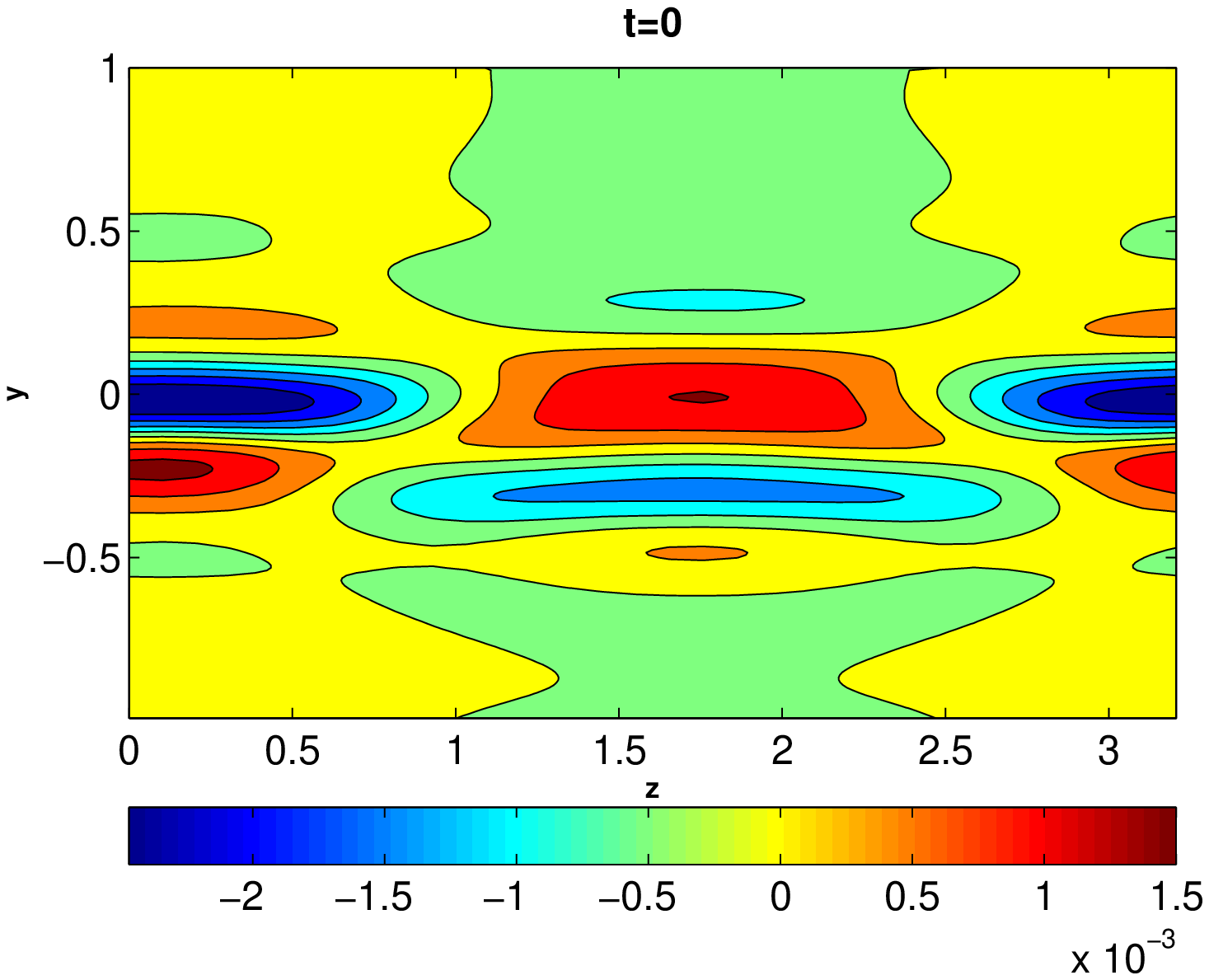}} 
  \centerline{\includegraphics[width=0.32\textwidth]{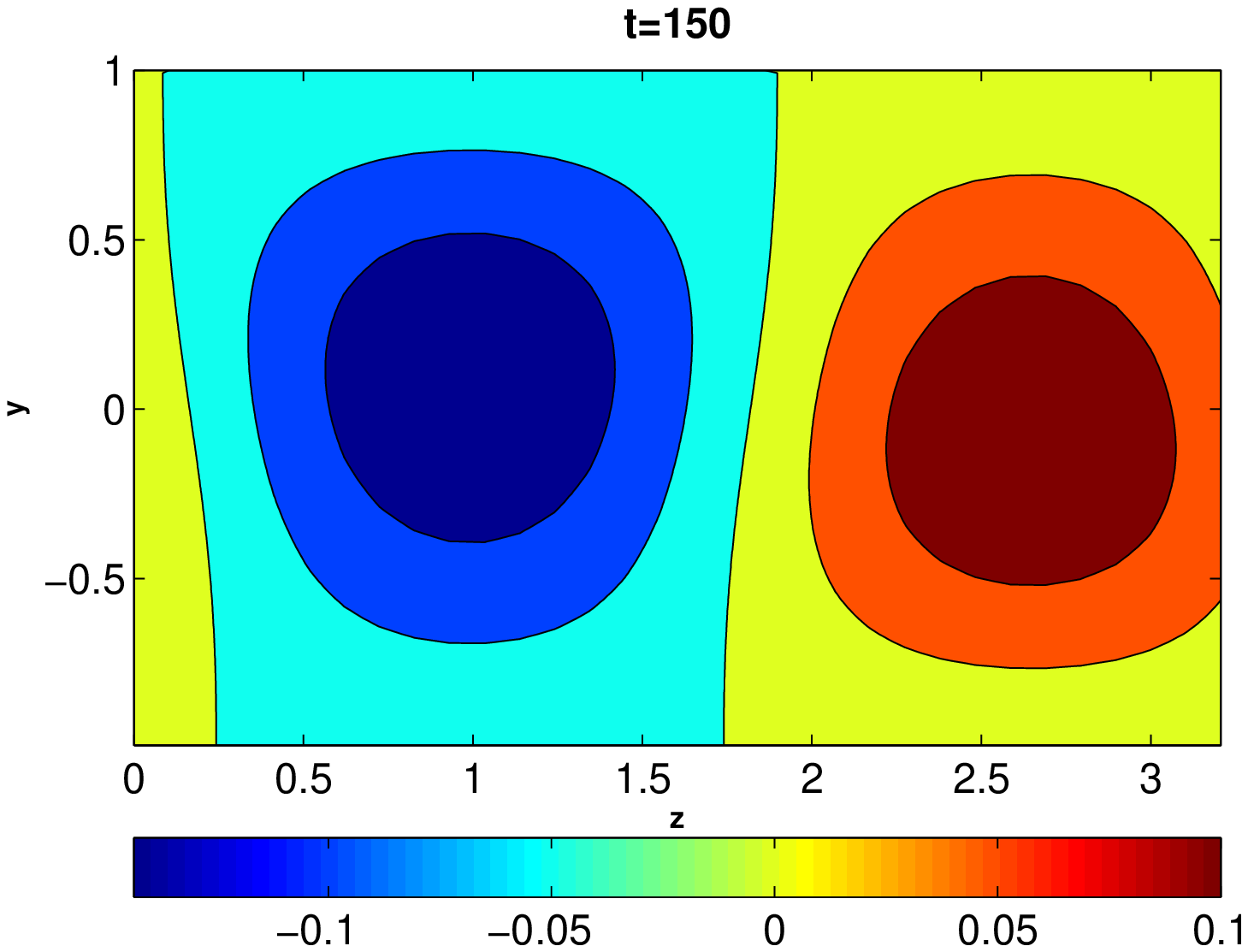}
  \includegraphics[width=0.32\textwidth]{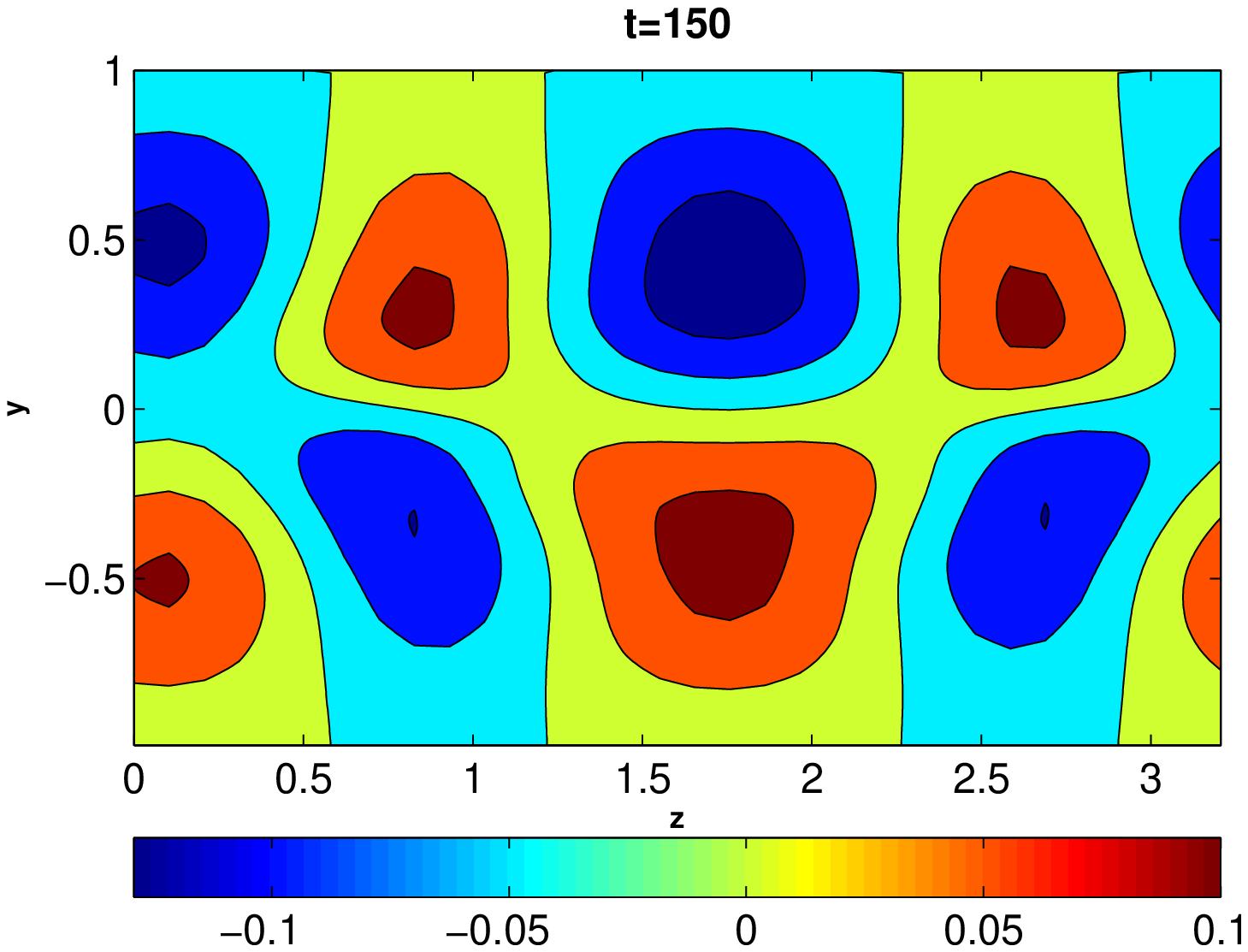}
  \includegraphics[width=0.32\textwidth]{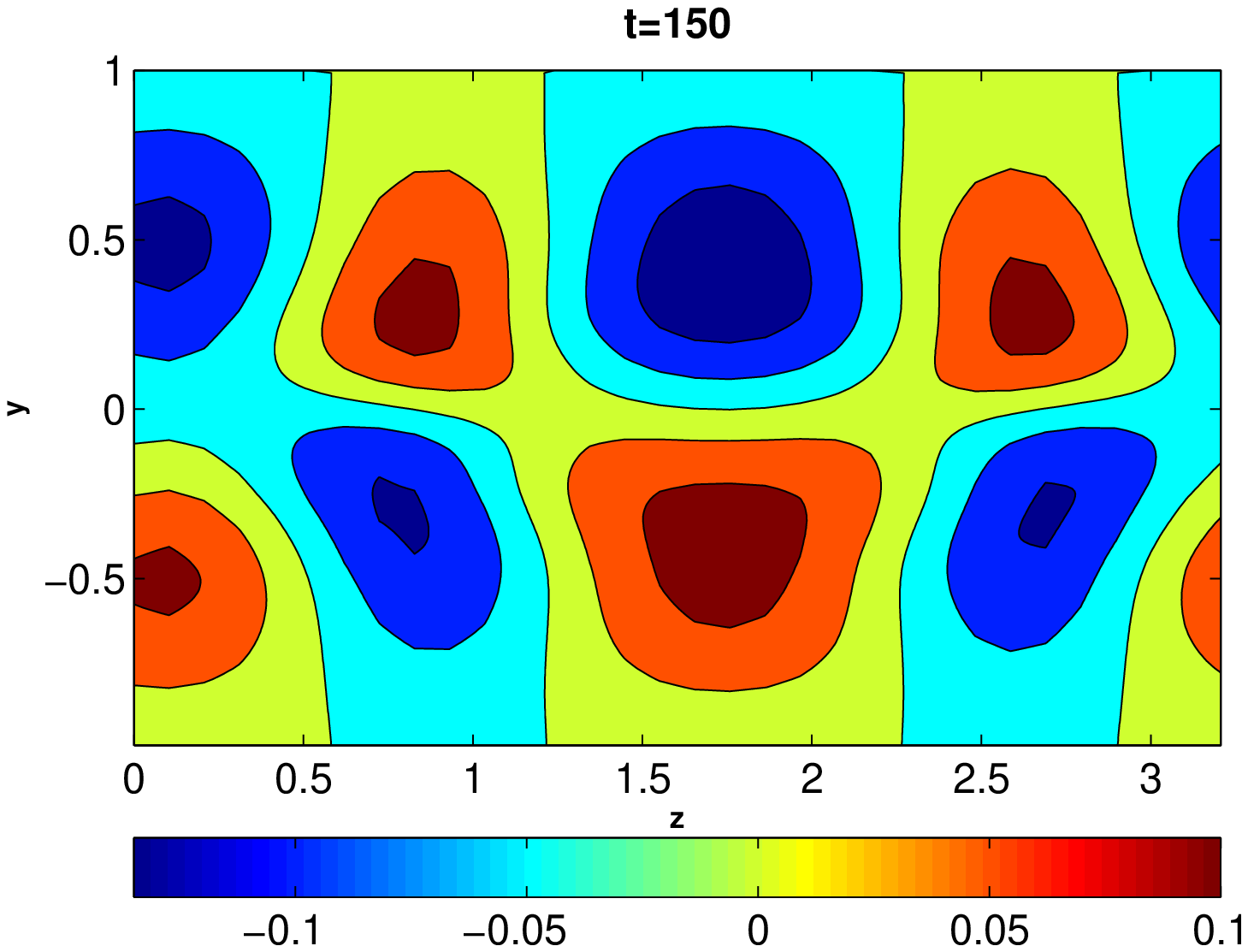}} 
  \centerline{\includegraphics[width=0.32\textwidth]{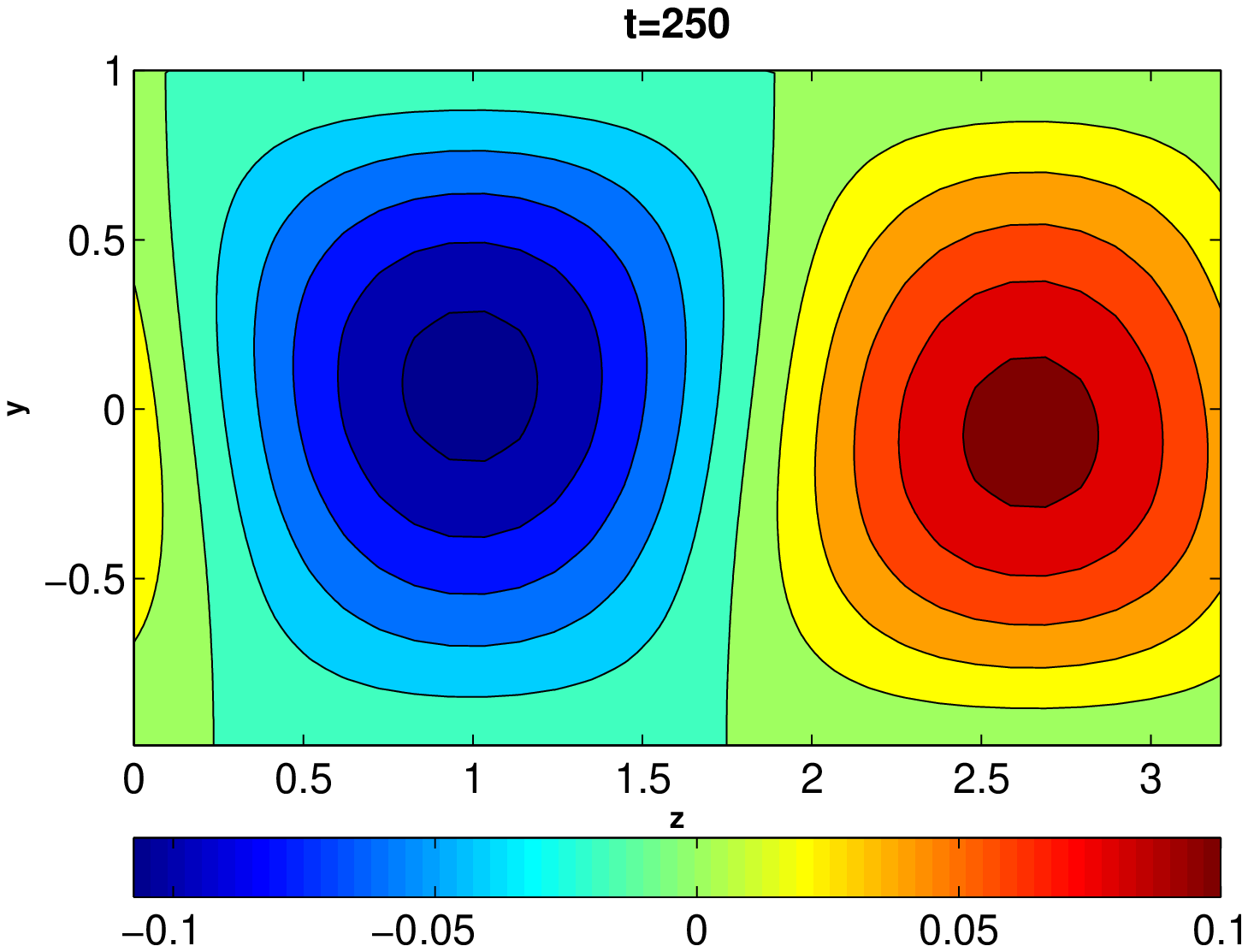}
  \includegraphics[width=0.32\textwidth]{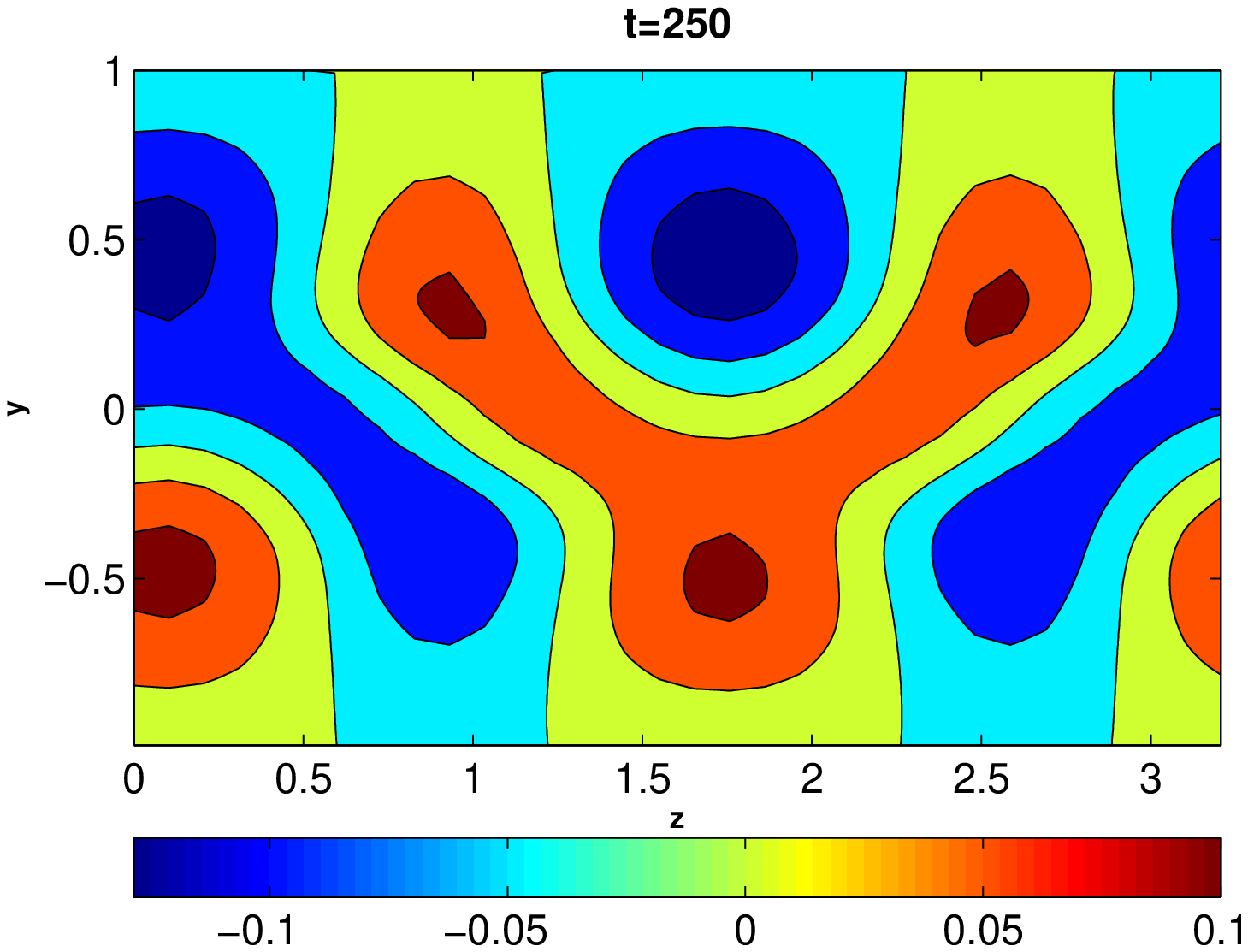}
  \includegraphics[width=0.32\textwidth]{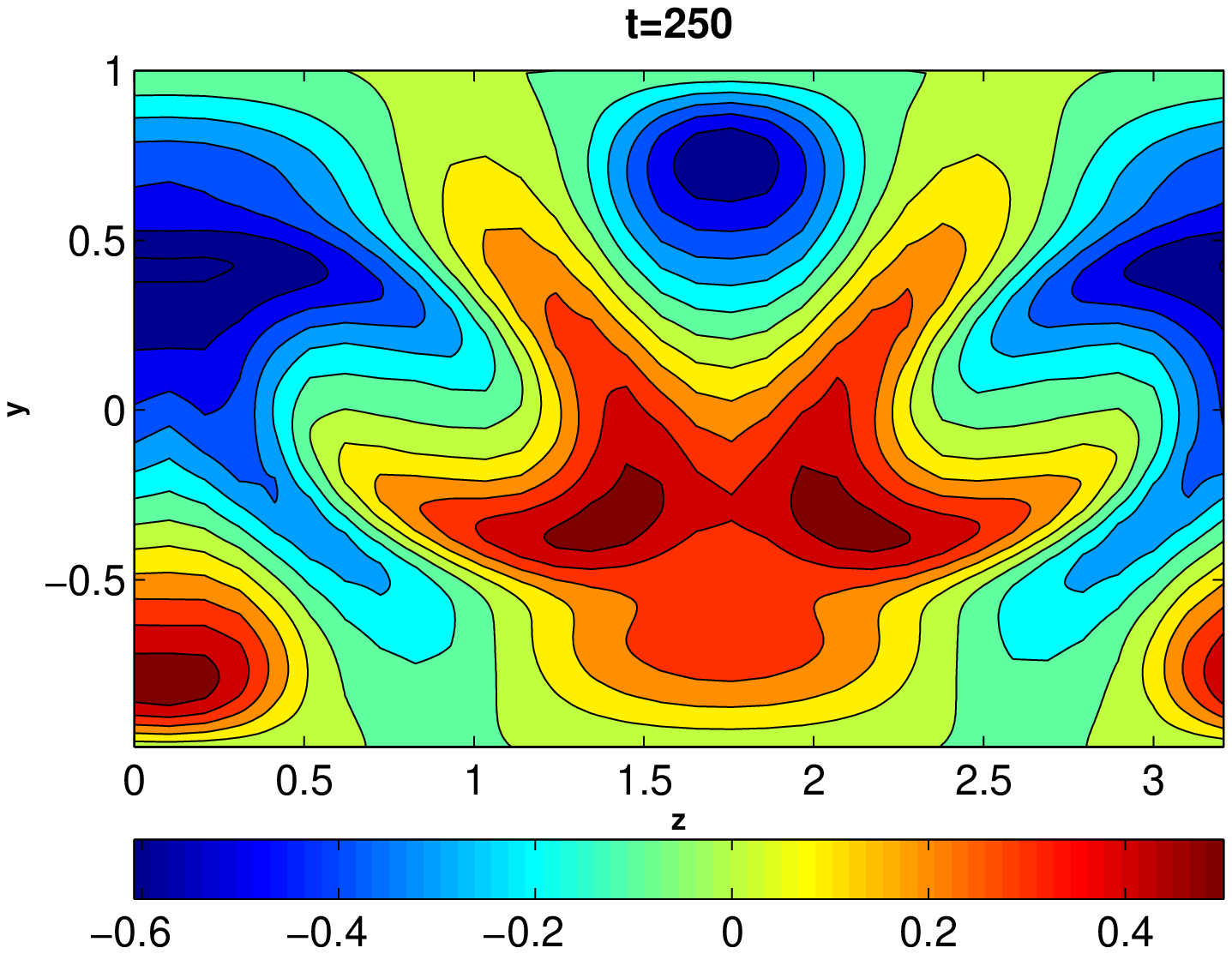}} 
  \centerline{\includegraphics[width=0.32\textwidth]{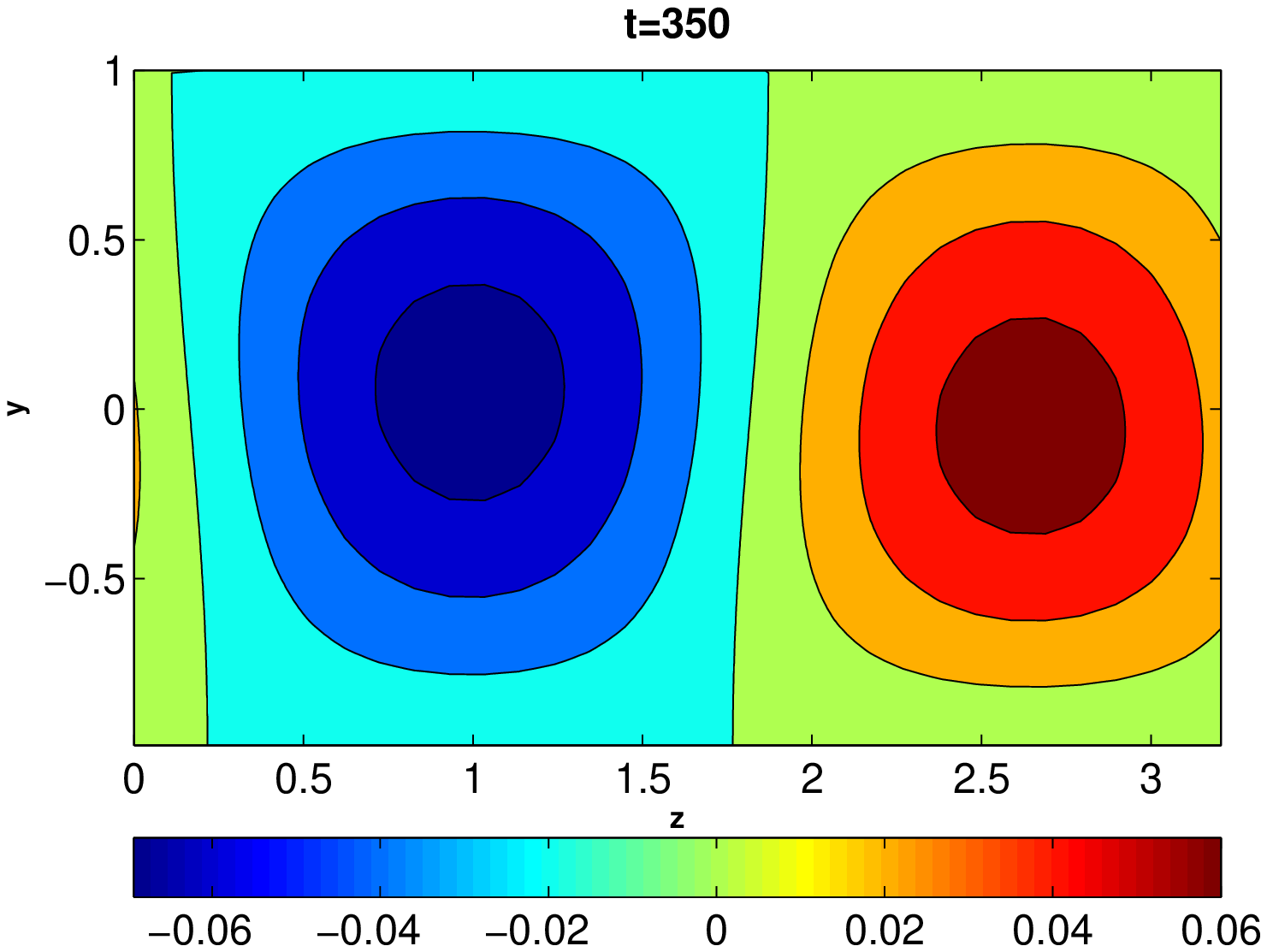}
  \includegraphics[width=0.32\textwidth]{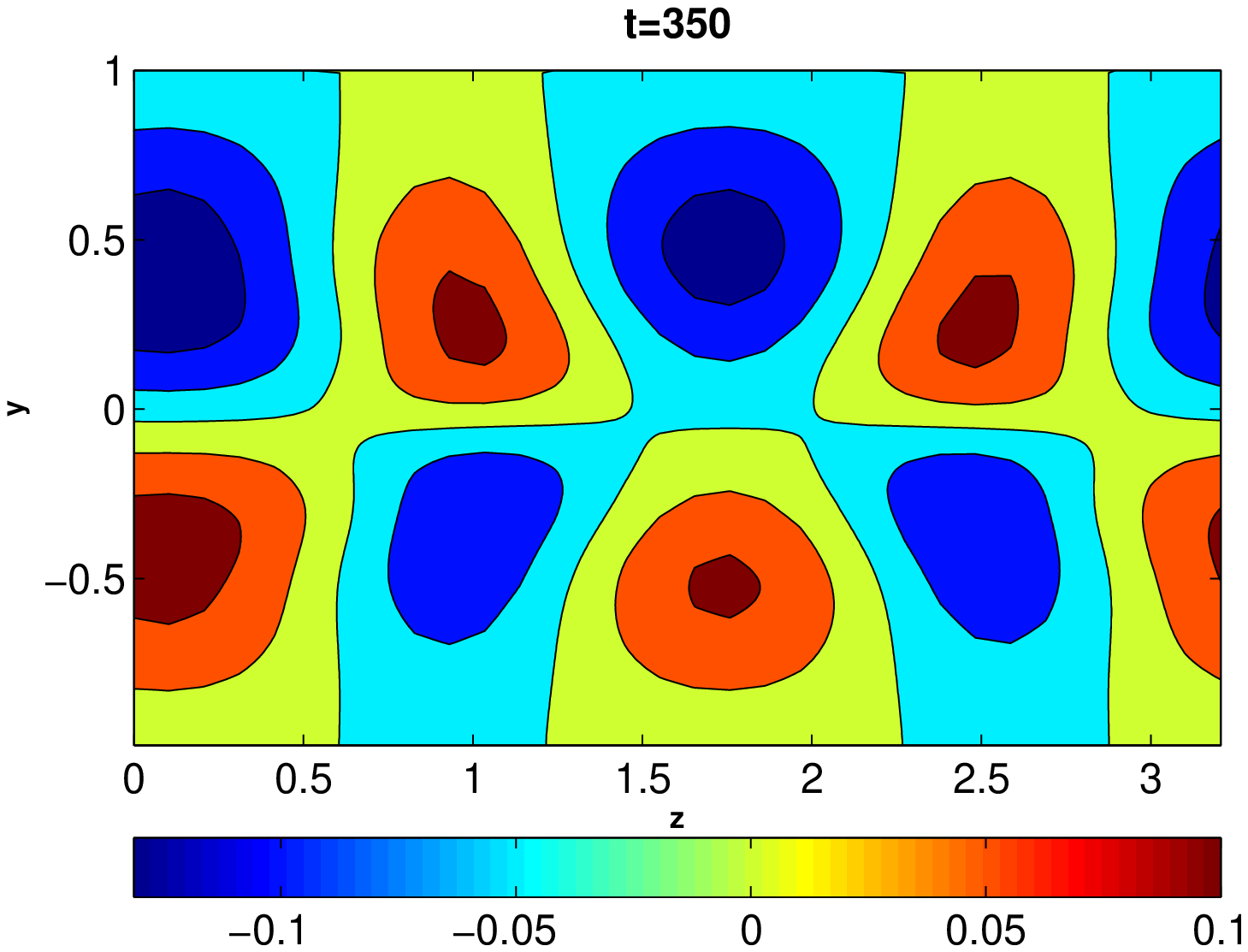}
  \includegraphics[width=0.32\textwidth]{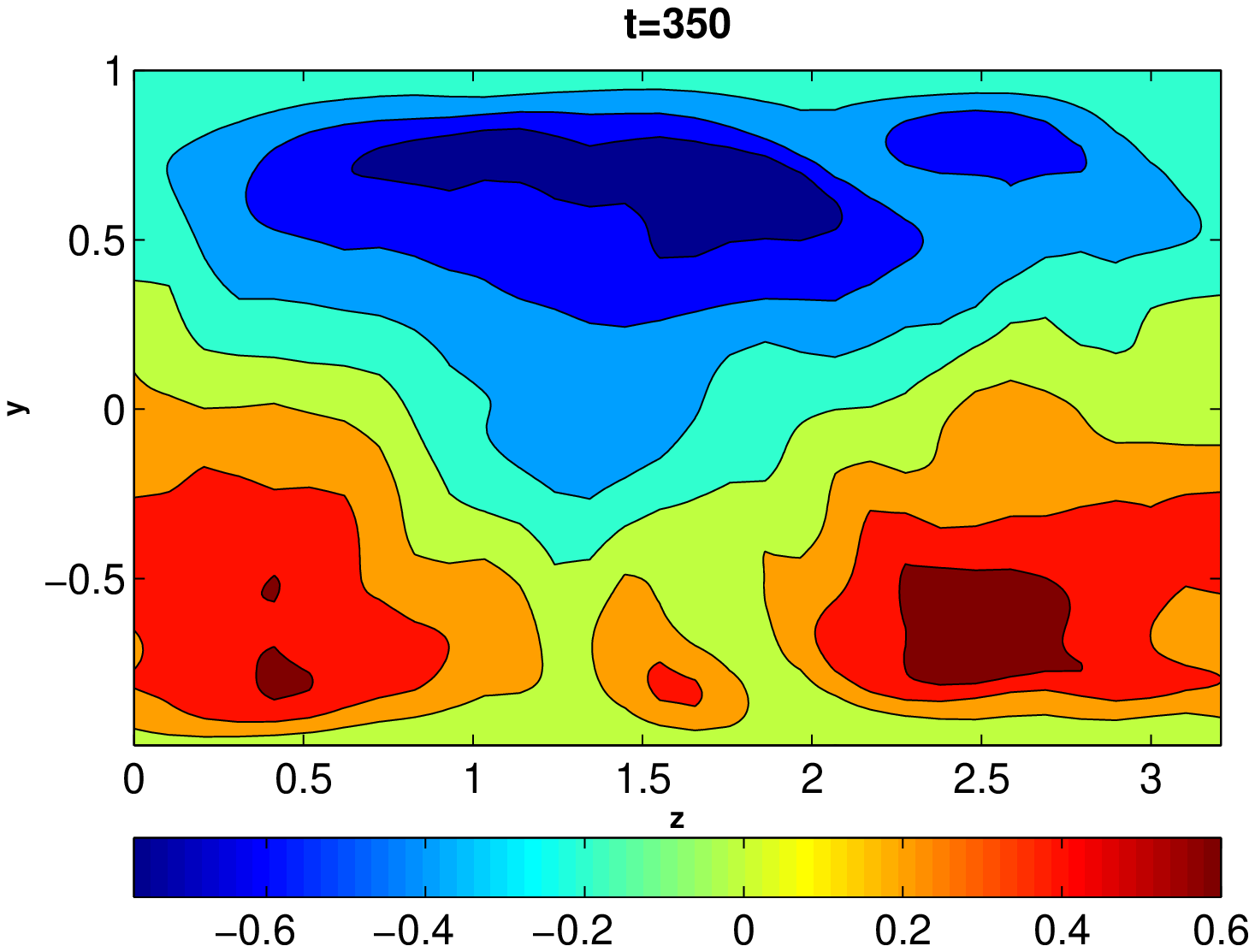}} 
\caption{Contours of streamwise velocity $u$ at times 0,
  150, 250, 350 for QLOP (left), minimal seed (centre) with $E_0=E_c$
  and turbulent seed (right) for $E_0=2.3 \times 10^{-6} \gtrsim
  E_c$. Contour levels are: going down the left column
  (min,spacing,max)=$(-6,2,6)\times 10^{-5}$, $(-0.1,0.05,0.1)$,
  $(-0.1,0.02,0.1)$ and $(-0.06,0.02,0.06)$; going down the centre
  (min,spacing,max)=$(-2,0.5,1.5)\times 10^{-3}$ and $(-0.1,0.05,0.1)$
  subsequently; going down the right column
  (min,spacing,max)=$(-2,0.5,1.5)\times 10^{-3}$, $(-0.1,0.05,0.1)$,
  $(-0.6,0.1,0.5)$ and $(-0.6,0.2,0.6)$.}
\label{contours_BF}
\end{figure}

Figure \ref{contours_BF} shows a side by side comparison of the
contours of the perturbation streamwise velocity at $x=0$ at four
different times for the QLOP at $E_{0} = 2.2 \times 10^{-6}$, the
minimal seed at $E_0=E_c$ (at least to the accuracy of figure
\ref{3plots_BF}) and the turbulent seed (which clearly is close to the
minimal seed) at $E_c \lesssim E_{0} = 2.3 \times 10^{-6}$. These
plots demonstrate that while the minimal seed evolves towards the edge
state, a small increase in its initial energy will lead to
transition. As $E_0 \rightarrow E_{c}$ from above (below), the time to
transition (relaminarisation) tends to $\infty$ due to the extra time
needed to evolve upwards (downwards) in energy away from the edge.  It
is also clear that the QLOP is completely different from the minimal
seed.

\begin{figure}
  \centerline{\includegraphics[width=0.32\textwidth]{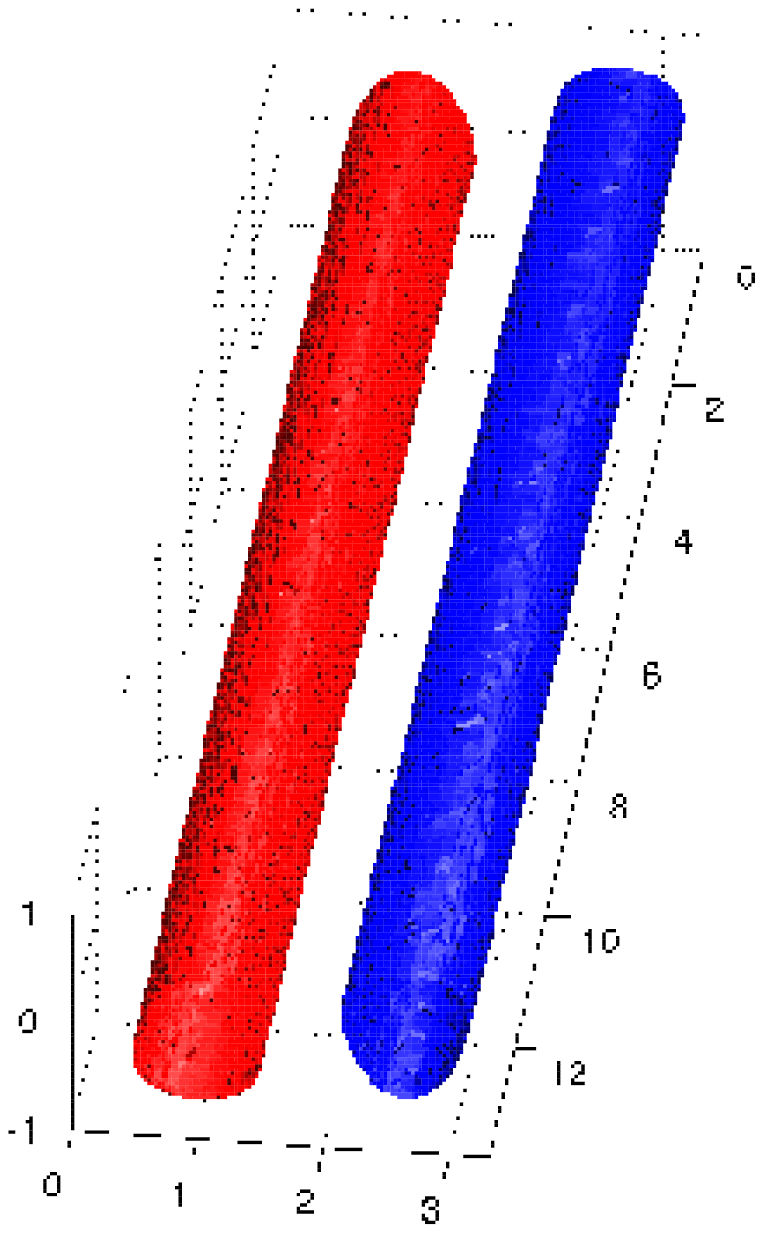}
  \includegraphics[width=0.32\textwidth]{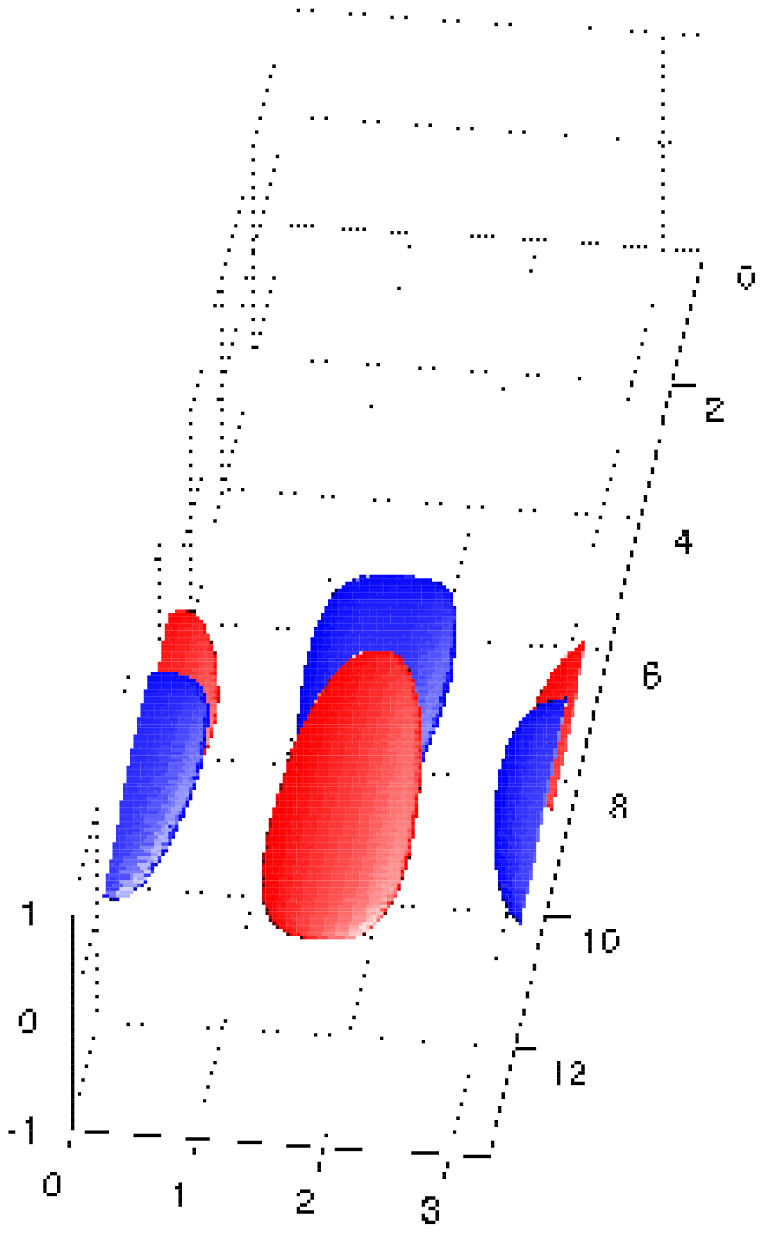}
  \includegraphics[width=0.32\textwidth]{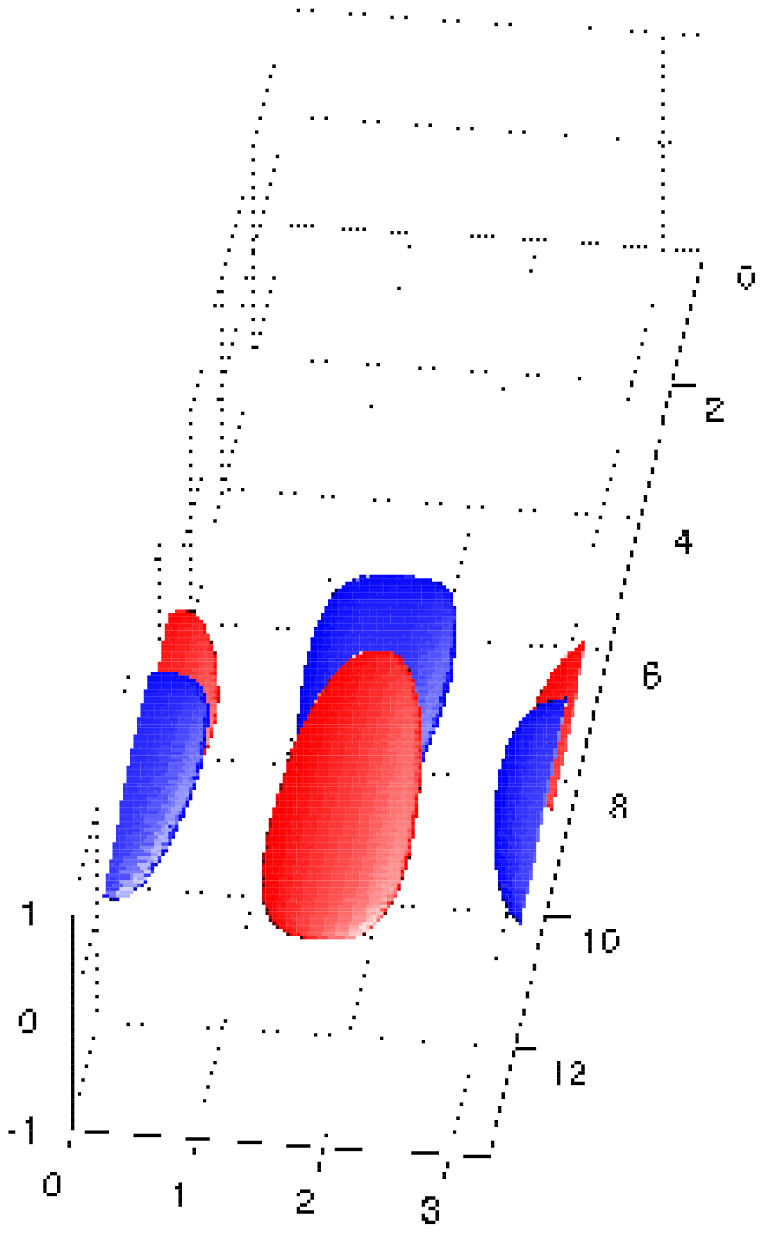}} 
  \centerline{\includegraphics[width=0.32\textwidth]{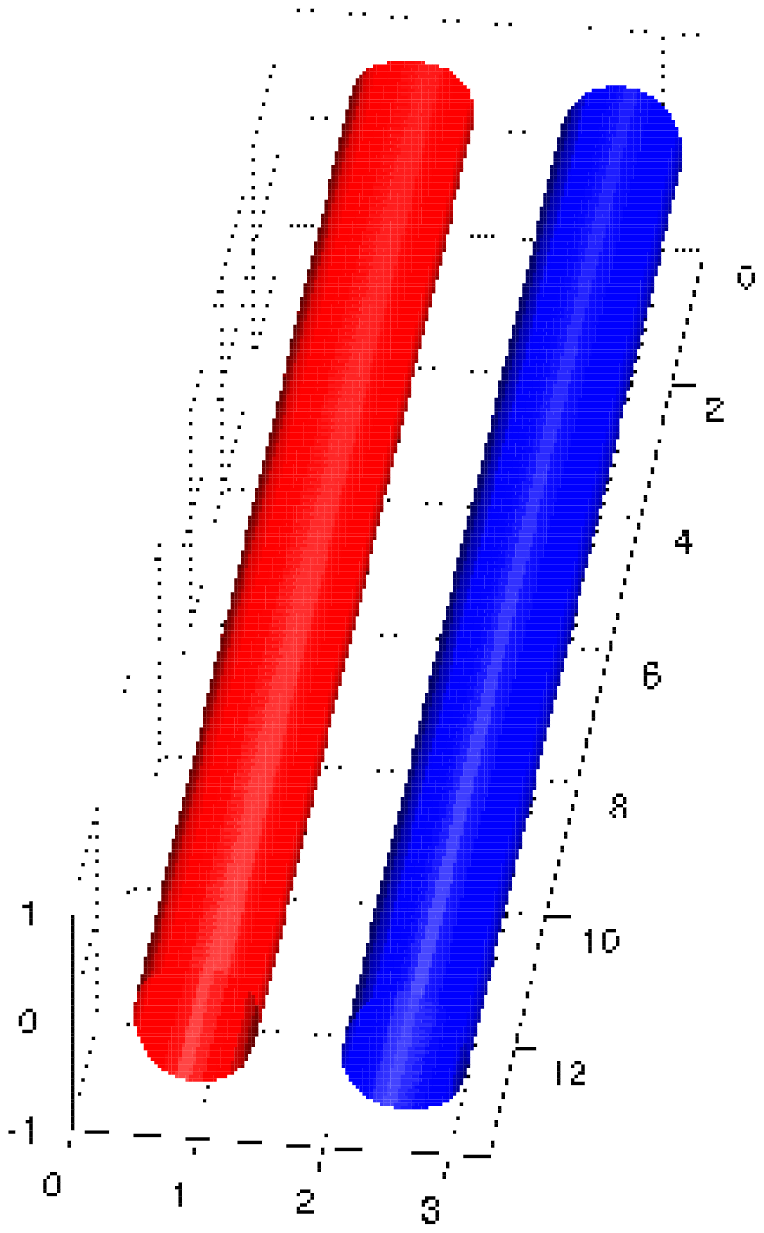}
  \includegraphics[width=0.32\textwidth]{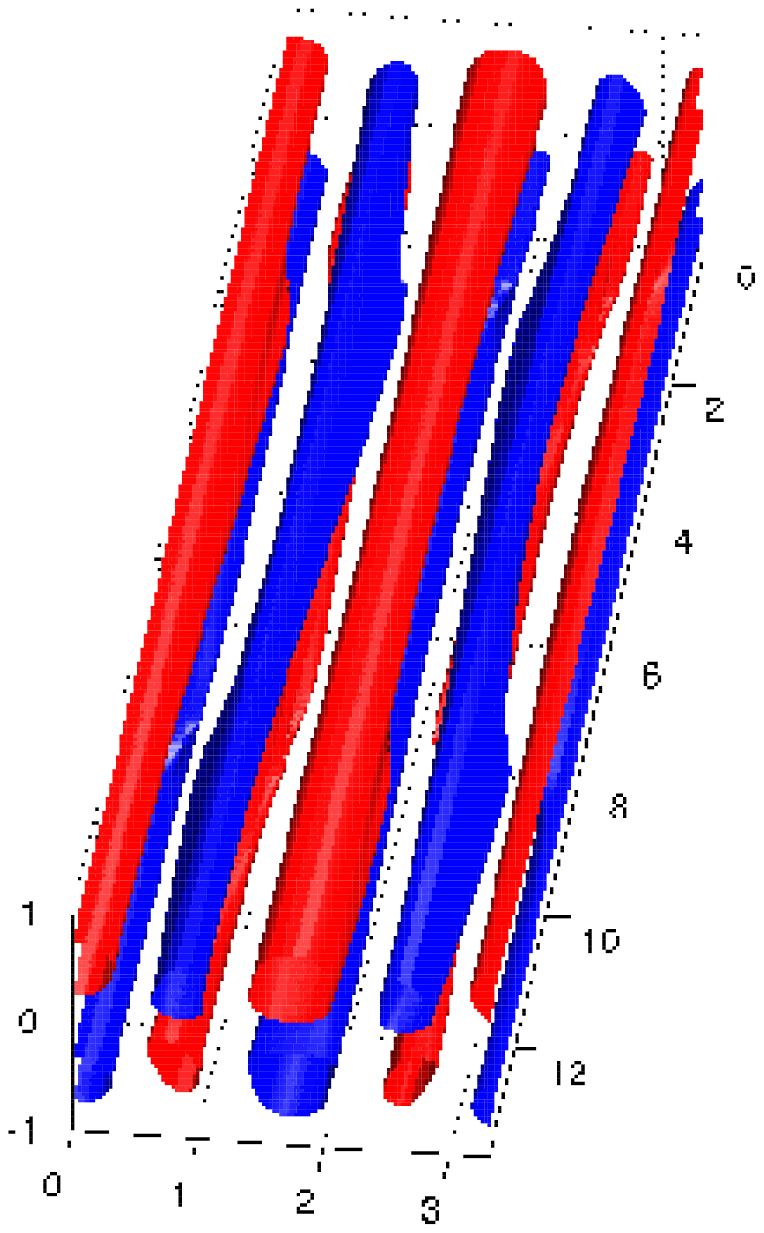}
  \includegraphics[width=0.32\textwidth]{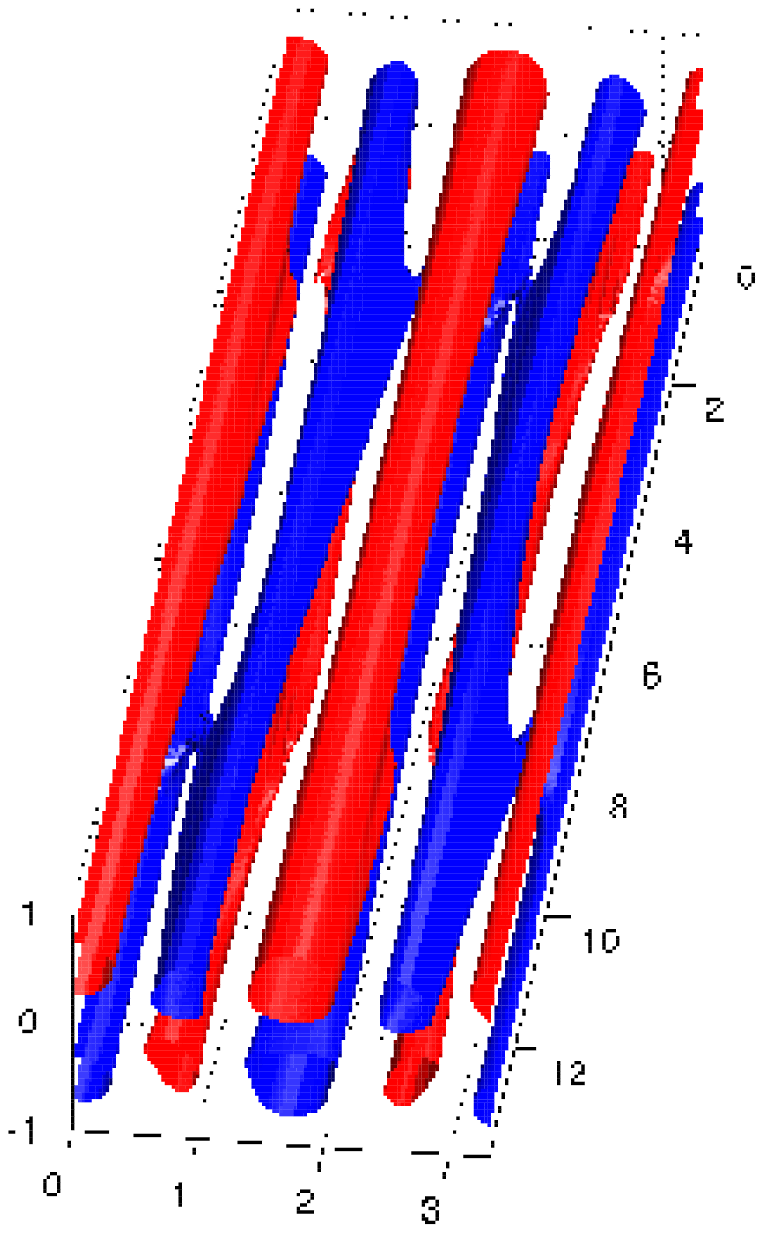}} 
  \centerline{\includegraphics[width=0.32\textwidth]{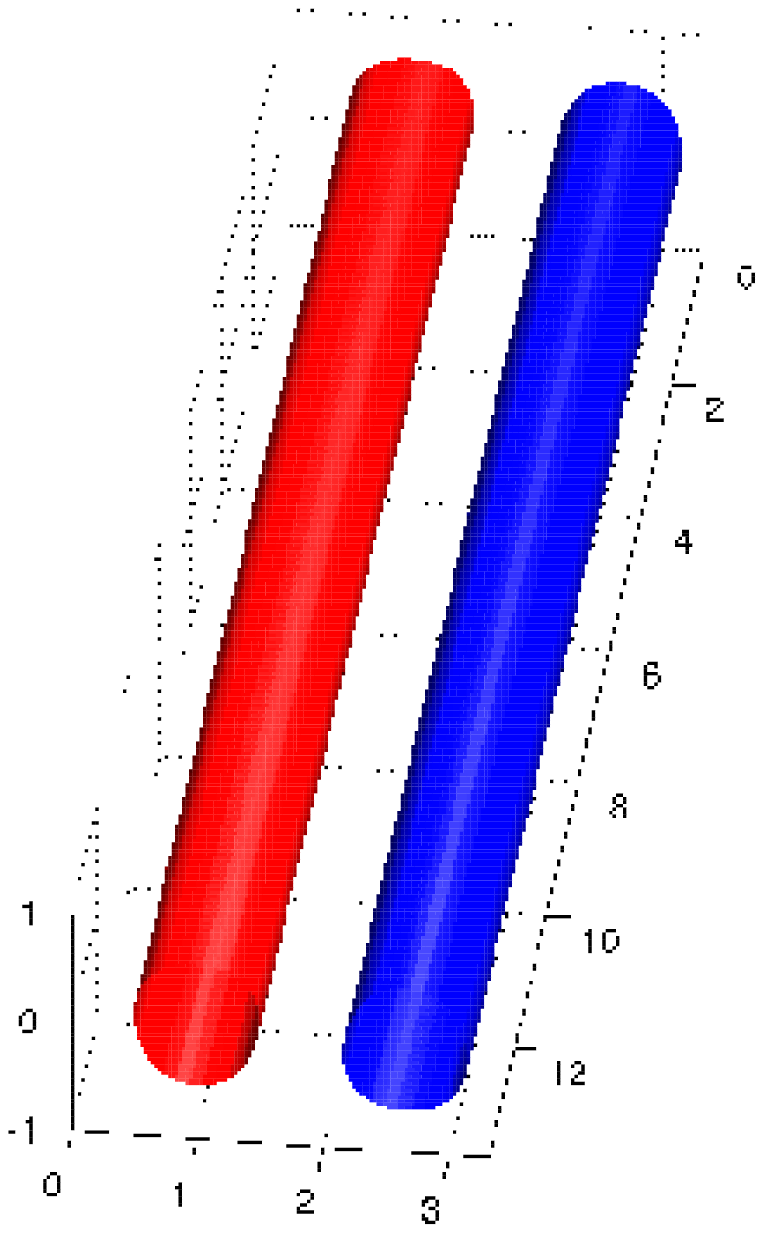}
  \includegraphics[width=0.32\textwidth]{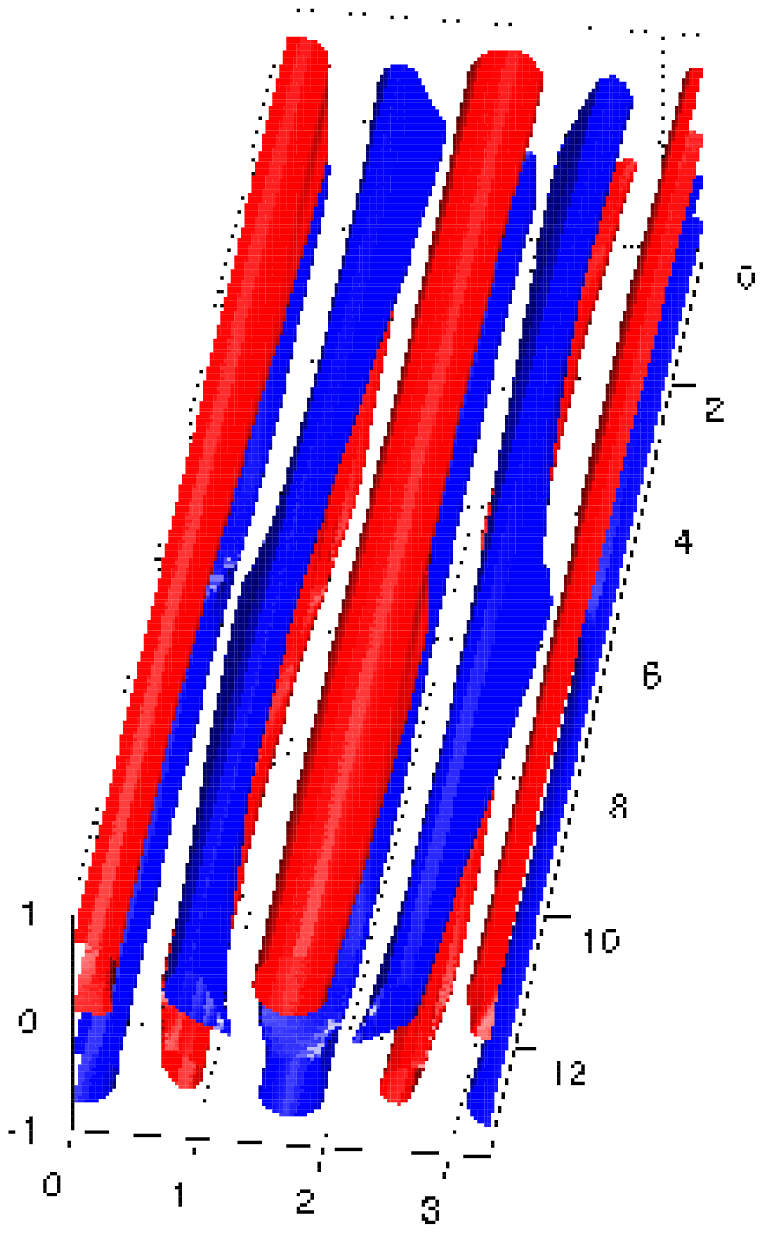}
  \includegraphics[width=0.32\textwidth]{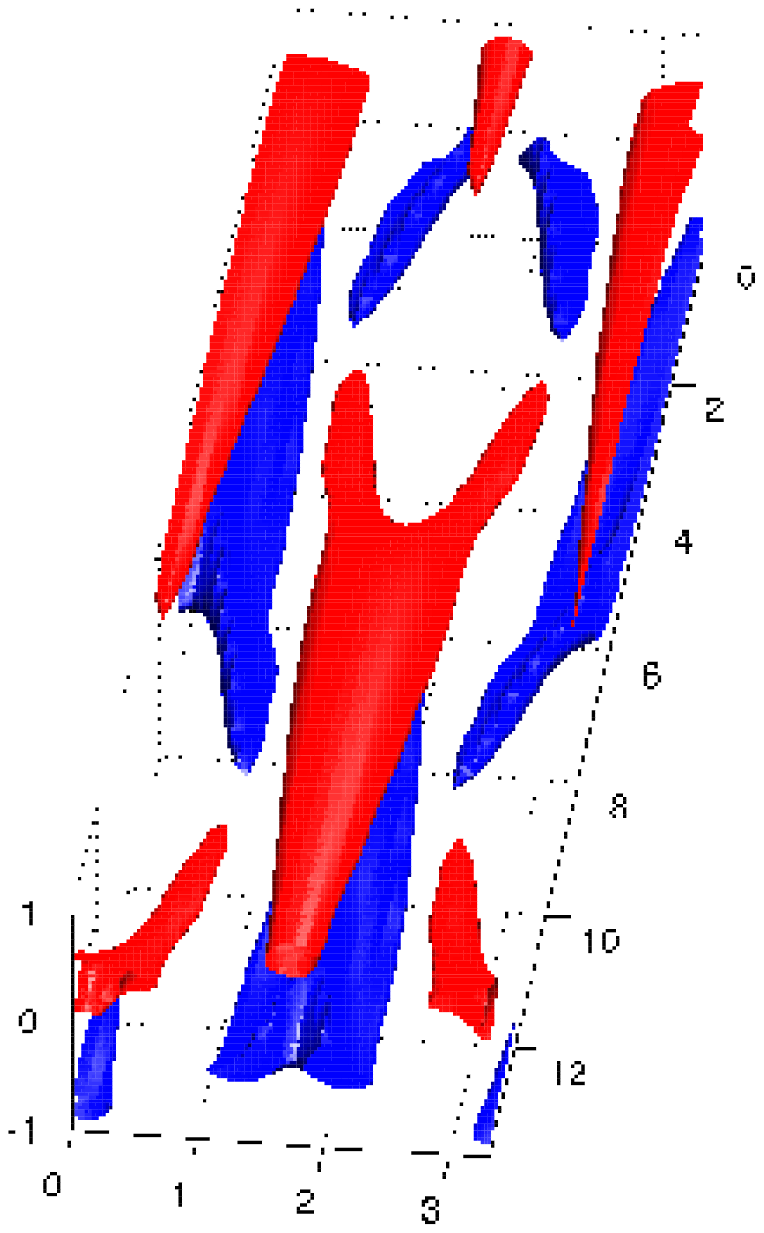}} 
  \centerline{\includegraphics[width=0.32\textwidth]{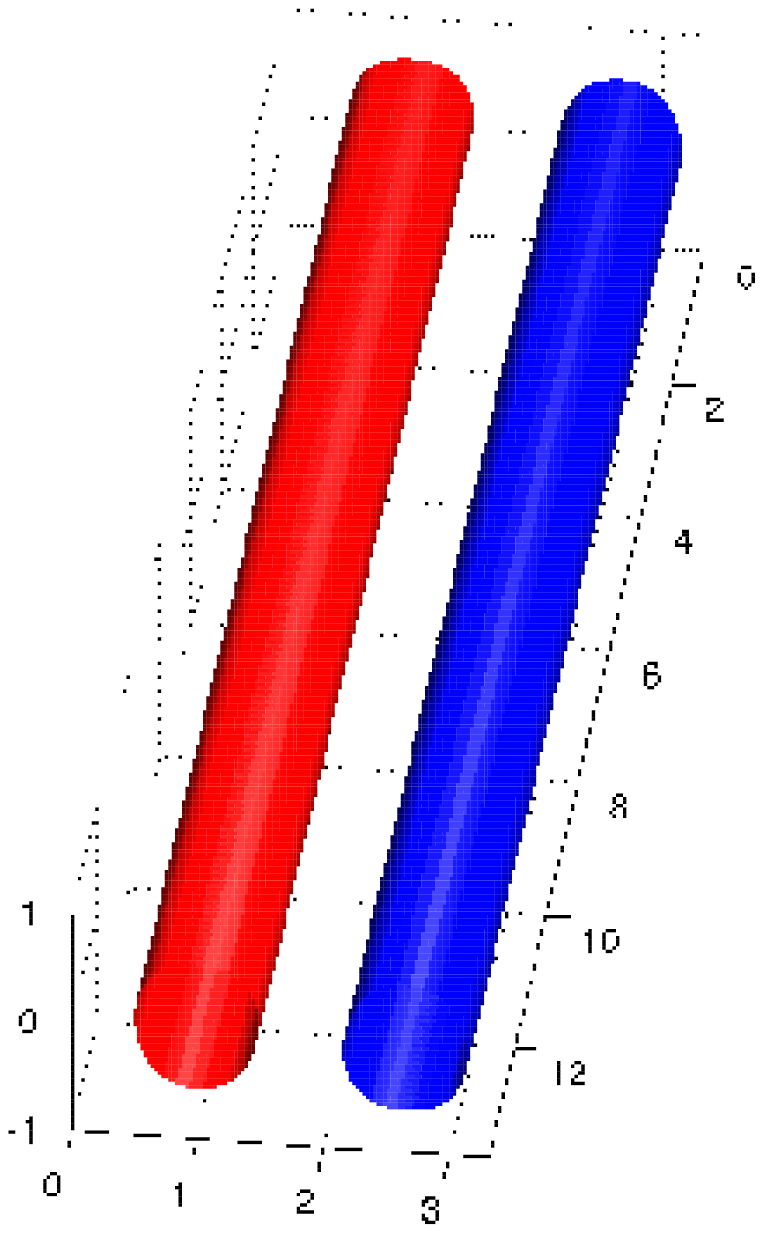}
  \includegraphics[width=0.32\textwidth]{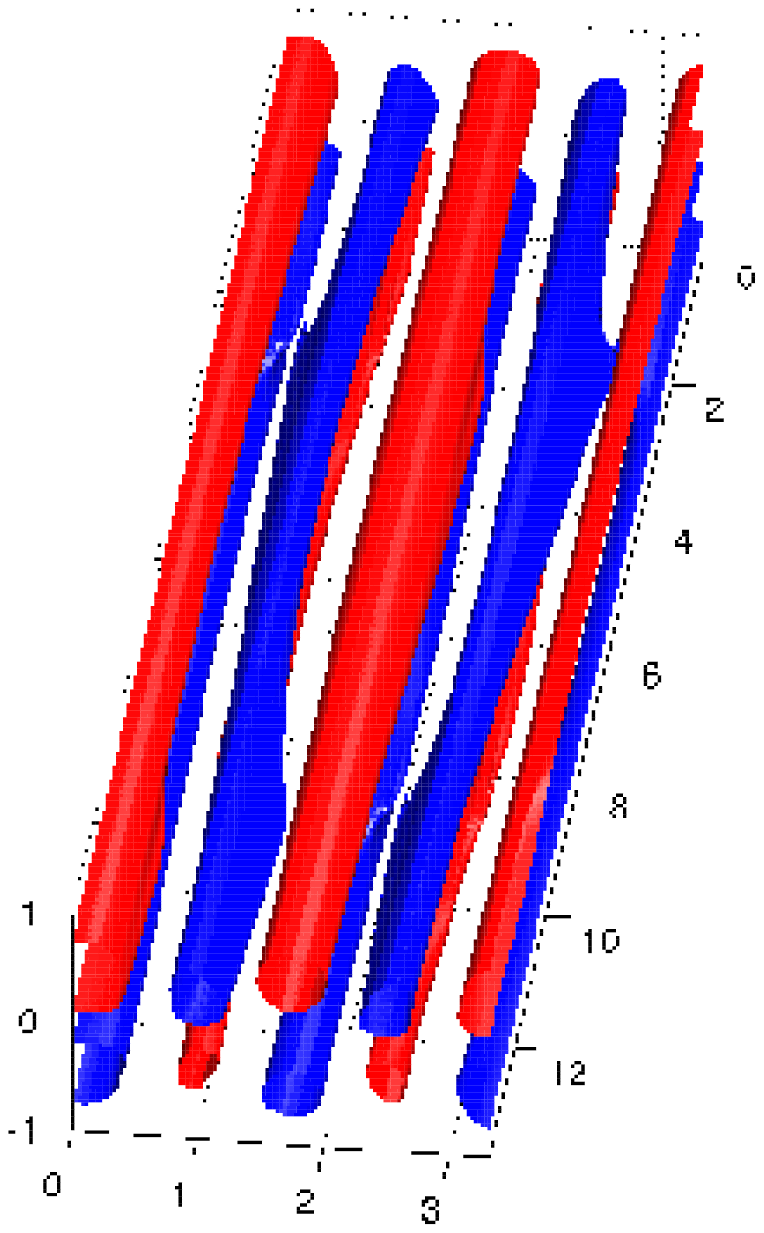}
  \includegraphics[width=0.32\textwidth]{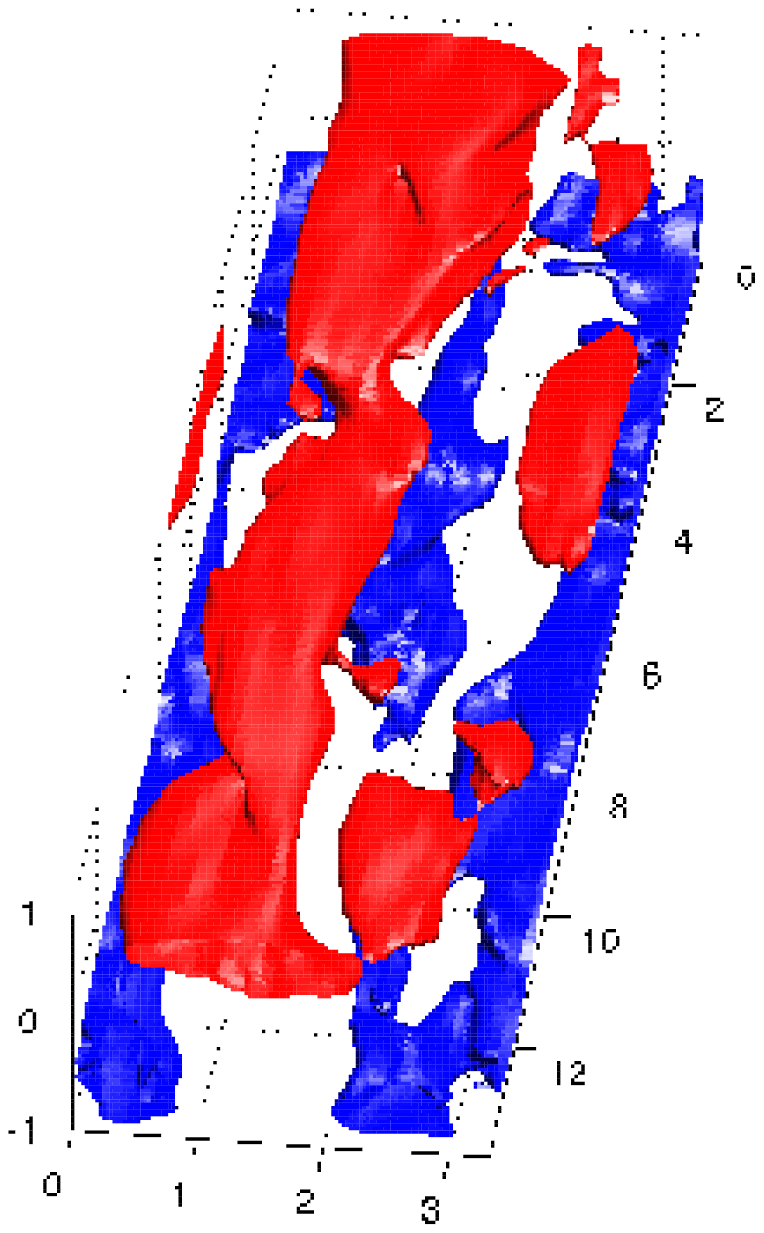}} 
  \caption{Iso surfaces of streamwise velocity $u$, at 60\% of maximum and minimum values, for the QLOP at
    $E_{0}= 2.2 \times 10^{-6}$ (left), the minimal seed (centre) and
    a turbulent seed above the edge at $E_{0}= 2.3 \times 10^{-6}$
    (right), at times 0, 150 ,250, 350.}
\label{iso_BF}
\end{figure}

Finally it is worth examining the 3D structure of the time-evolving
QLOP, the minimal seed and the turbulent seed just above the edge. In
figure \ref{iso_BF} we plot iso-contours of the streamwise velocity
for times 0, 150, 250 and 350 (in unit of $h/U$). From the plots it is
clear that the minimal seed is initially quite localized but quickly
`unpacks' itself into a series of a streamwise streaks. This unpacking
process appears to be achieved by the well-known Orr mechanism
followed by the lift-up mechanism. The minimal seed flow then remains
in this configuration, whereas the streaks destabilise and there is
transition to turbulence for the higher energy turbulent seed `above'
the edge.

To summarise, in this geometry and at this $Re$, it appears that a
well-converged NLOP does not exist prior to our algorithm uncovering
turbulence-triggering initial conditions at $E_0=E_{fail}$. Our
algorithm only fails to converge if there are turbulent seeds present
so $E_{fail} \geq E_c$ and we find no evidence for inequality
consistent with PWK11's first conjecture. The minimal seed (as is
apparent in figures \ref{contours_BF} and \ref{iso_BF}) is
qualitatively different from the QLOP, and so the optimals
do not converge to the minimal seed as $E_{0} \rightarrow E_{c}^-$,
a clear counterexample to PWK11's second conjecture.

%
% Fig 6
%

\section{M11 geometry} \label{comp}

\begin{figure}
  \centerline{\includegraphics[width=0.49\textwidth]{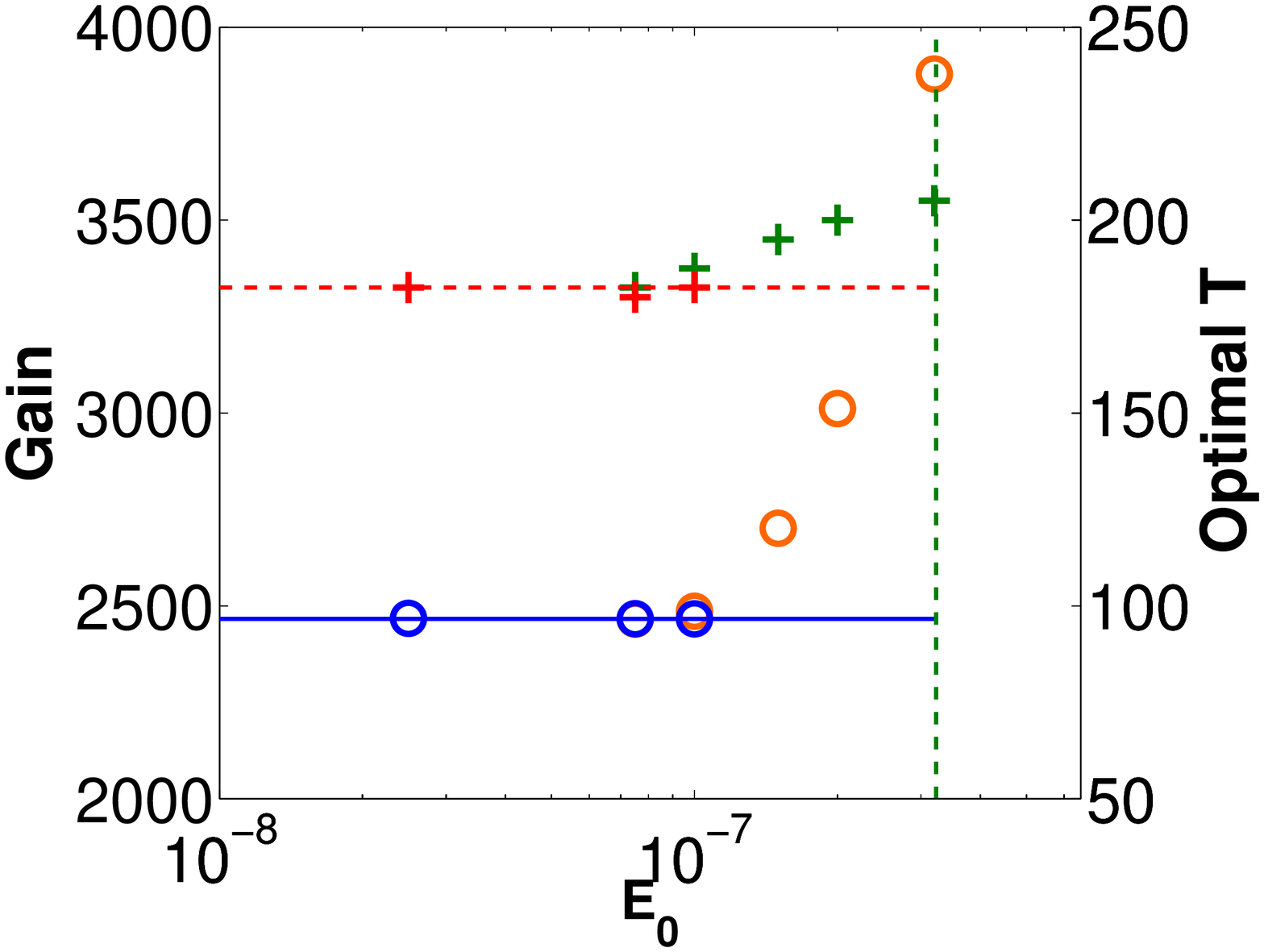}
  \includegraphics[width=0.49\textwidth]{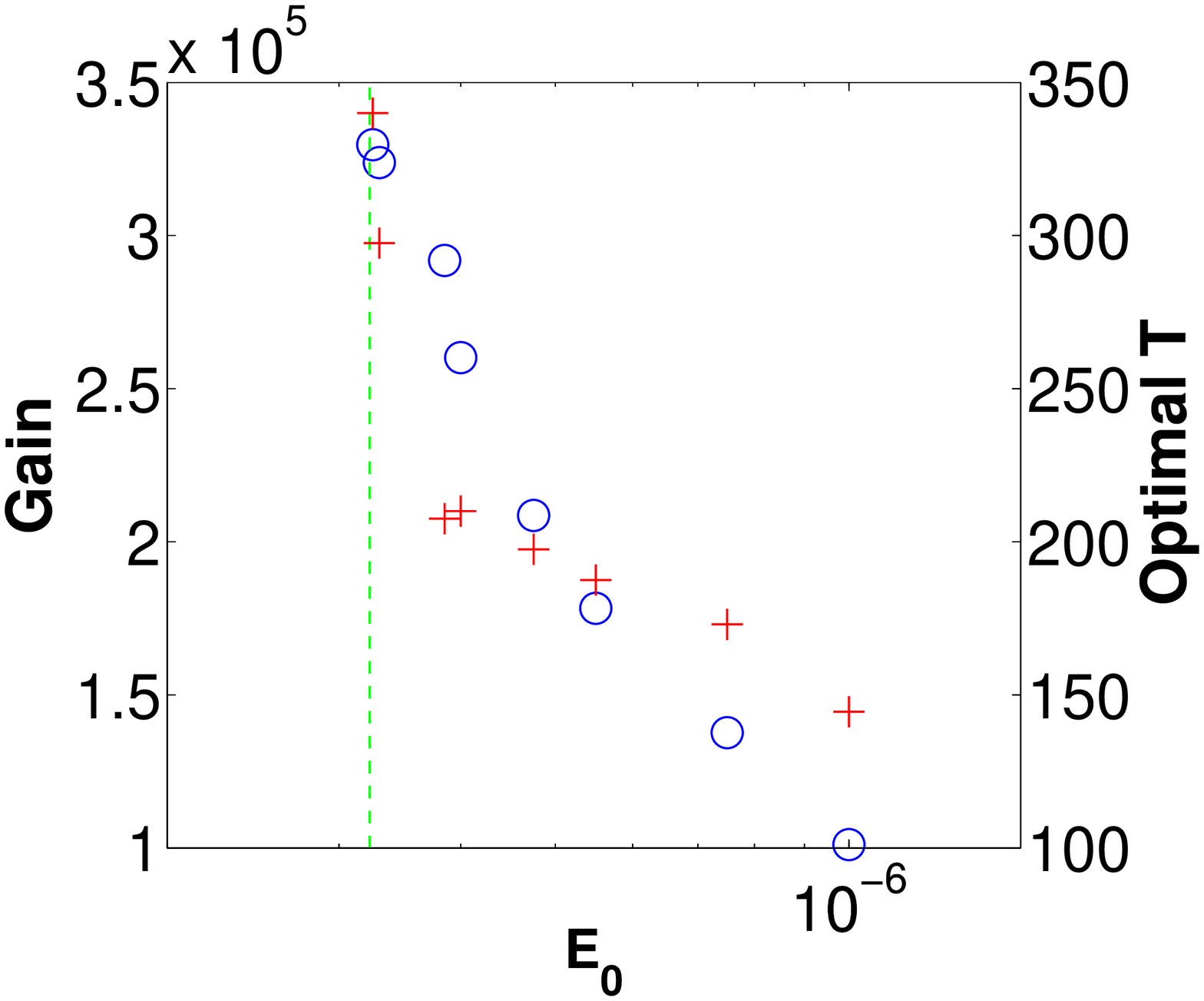}}% Images in 100% size
  \caption{(\textit{a}) Gain, $E(T)/E(0)$, of QLOP (blue circles)
    and NLOP (orange circles) and associated optimal time of QLOP (red
    crosses) and NLOP (green crosses), $T$, against $E_{0}$. $E_{c}$ is
    marked by vertical green dashed line. LOP gain horizontal blue
    solid line, LOP optimal time horizontal red dashed
    line. (\textit{b}) Gain, $E(T)/E(0)$, (blue circles) and
    associated optimal time (red crosses), $T$, against
    $E_{0}$. $E_{c}$ is marked by vertical green dashed
    line.}
\label{GT_DH}
\end{figure}

In this section we examine a second, larger geometry of dimensions $4
\pi \times 2 \times 2 \pi $ (essentially twice as wide as that in
BF92) at a higher Reynolds number $\Rey=1500$. We demonstrate in this
geometry that now a NLOP exists at energies below $E_{c}$ and
investigate whether in the limit $E_0 \rightarrow E_c^-$ it converges
to the minimal seed.  Choosing the geometry and Reynolds number used
by M11 has the added benefit that we can compare our results to those
obtained using an entirely different functional. M11 optimized the
total dissipation over a long time interval rather than the energy
gain achieved at a specific target time.  We find a critical energy
value $E_c=3.3 \times 10^{-7}$, plotted in figures \ref{GT_DH} (a) and
(b), which agrees well with M11, who find $3 \times 10^{-7}< E_c< 4
\times 10^{-7}$ (see their figure 1). Note $\epsilon_0$ in M11 is $E_0$ here as $||\quad||_E$ in their
  equation (1) is strictly a kinetic norm with a $\half$ included
  (Monokrousos, personal communication). Our calculated time for transition at $E_{0}=4.0 \times 10^{-7}$ is
approximately $200$ not too dissimilar from the time of $150$ in
M11. This suggests that the particular choice of optimizing functional
is not important for the calculation of a minimal seed (or more
accurately to lose convergence), provided the functional attains
heightened values for turbulent flows (as discussed in PWK11).

%
% fig 7  M11
%

%
% fig 8
%
\begin{figure}
  \centerline{\includegraphics[width=0.49\textwidth]{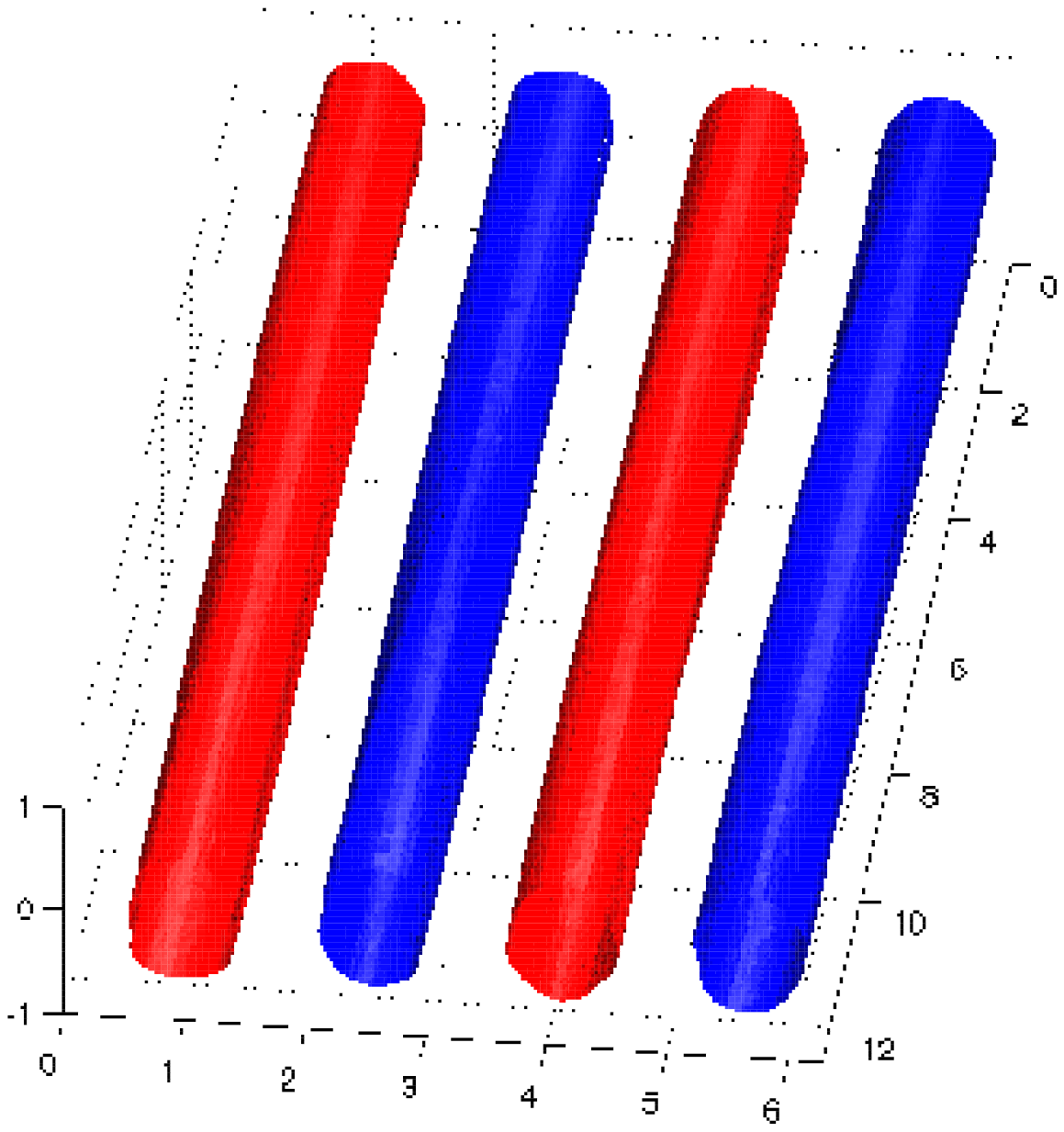}
  \includegraphics[width=0.49\textwidth]{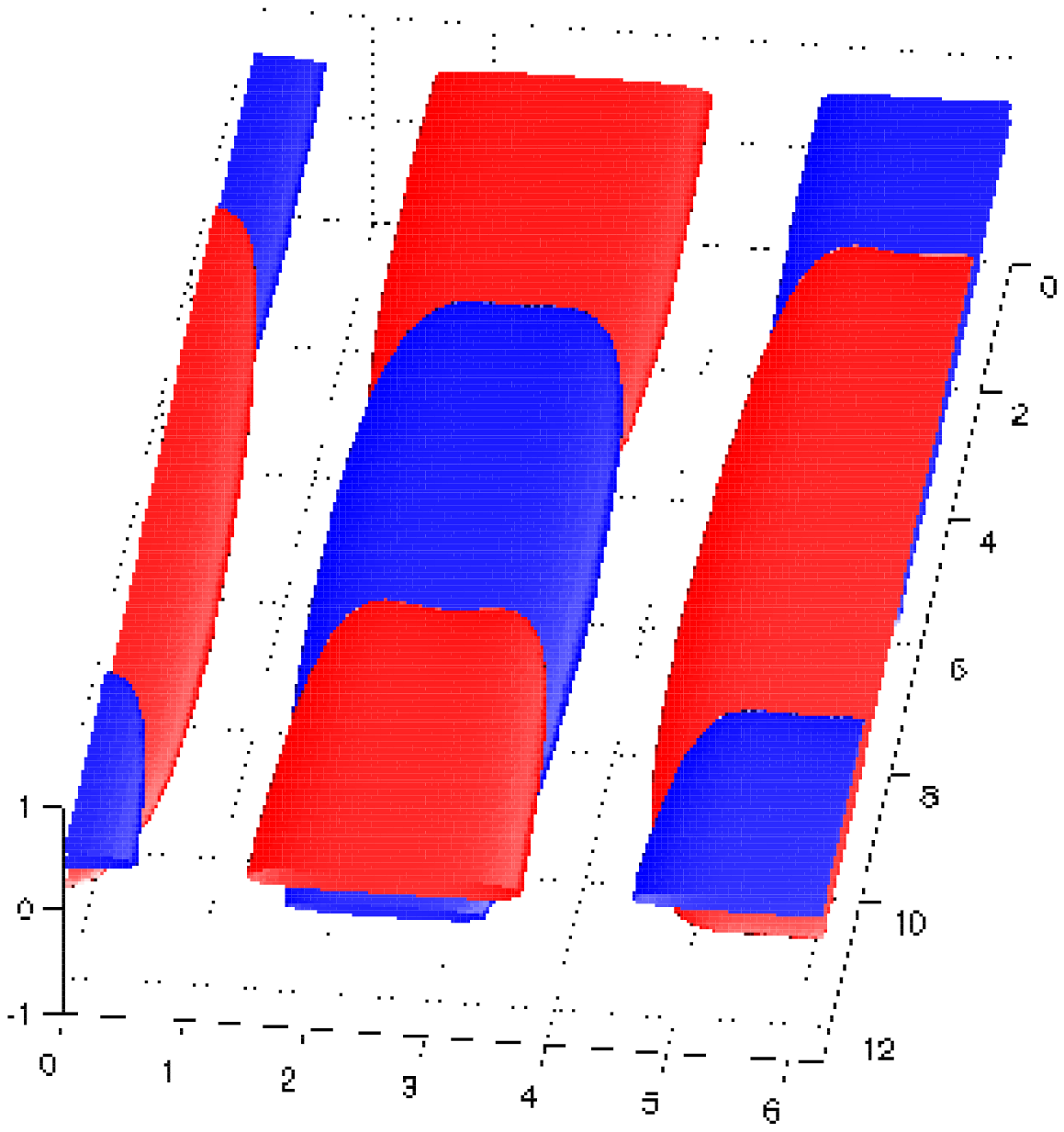}} 
 \caption{Iso surfaces of streamwise velocity $u$, at 60\% of maximum and minimum values, for
   (\textit{a}) QLOP at $E_{0}= 5.0 \times 10^{-8}$ and (\textit{b})
   NLOP at $E_{0}= 3.2 \times 10^{-7}$. It is clear that the NLOP is
   distinct from the QLOP.}
\label{iso_Q/NLOP}
\end{figure}

\begin{figure}
%\centerline{\includegraphics[width=0.32\textwidth]{New_figures/DH/Iter_v_G_G_DH_2.5e-8.eps}
%            \includegraphics[width=0.32\textwidth]{New_figures/DH/Iter_v_G_G_DH_1.0e-7.eps}
\centerline{\includegraphics[width=0.5\textwidth]{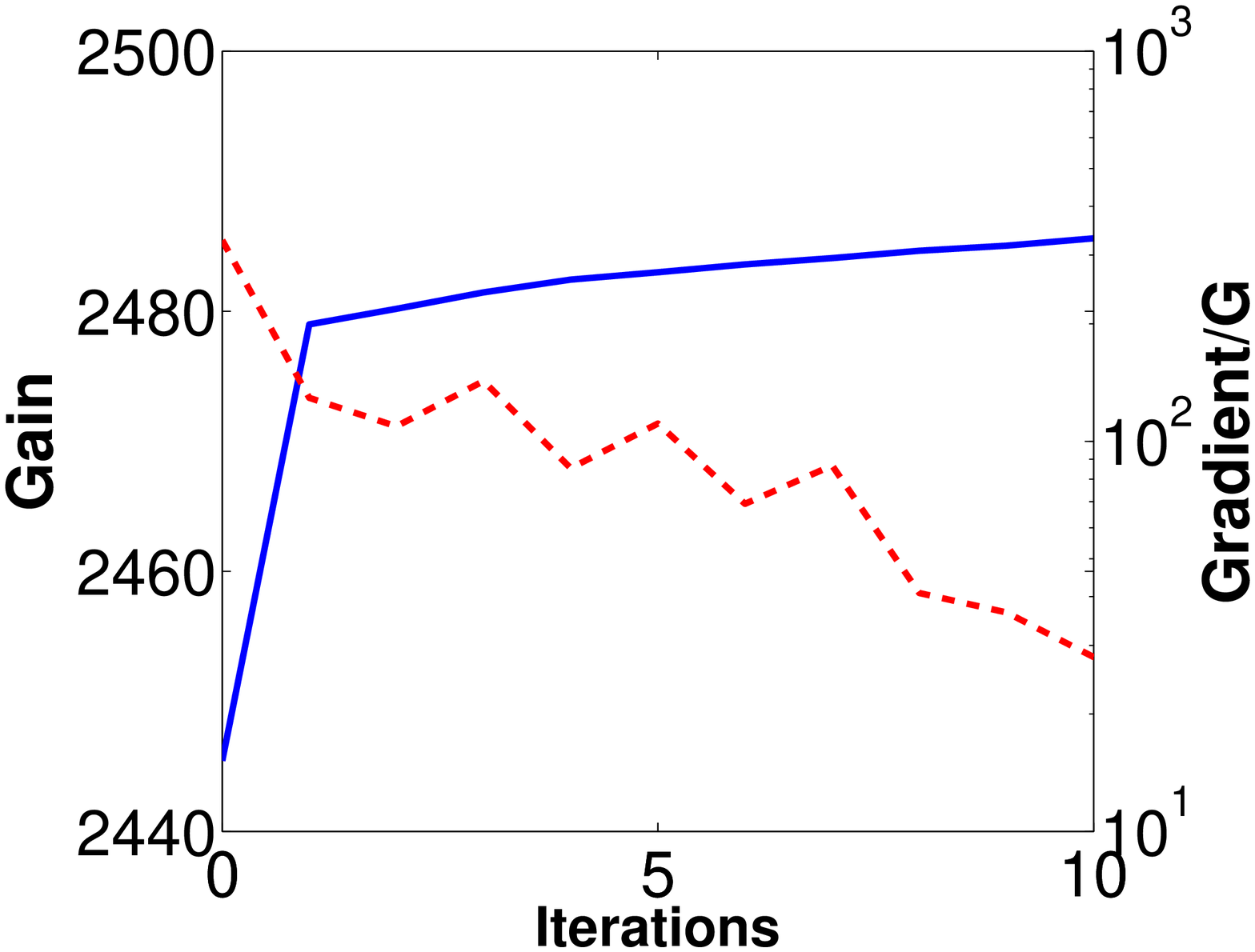}
\includegraphics[width=0.5\textwidth]{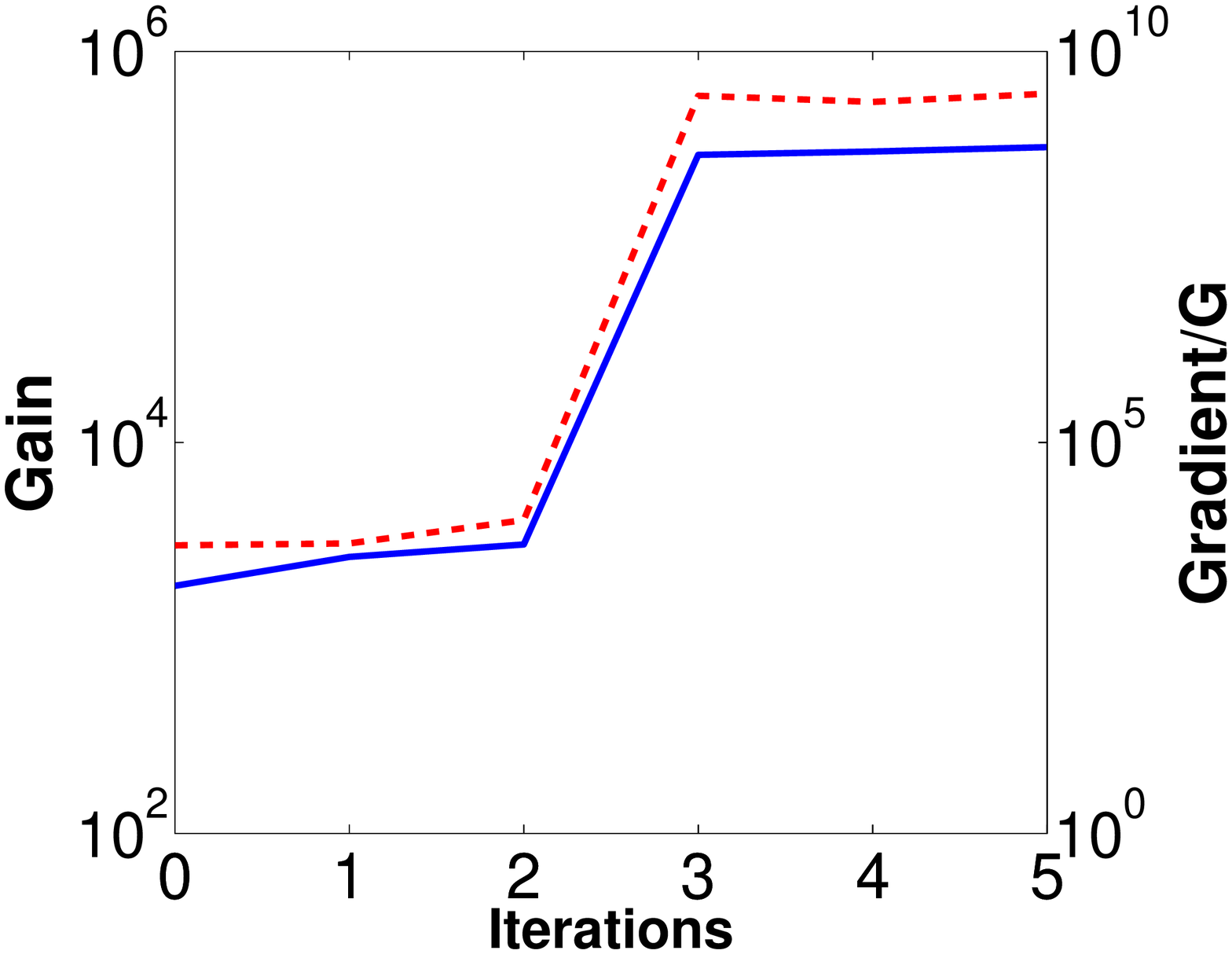}}% Images in 100% size
  \caption{G (blue solid line) and $\langle \delta \mathcal{L} /
    \delta \mathbf{u}_{0}, \delta \mathcal{L} / \delta \mathbf{u}_{0}
    \rangle^{\frac{1}{2}}/G$ (red dashed line) plotted against iteration for (left)
    $E_{0}=1.0 \times 10^{-7}$ and (right) $E_{0}=3.3 \times
    10^{-7}$. The NLOP (left) appears to be converging well, whereas
    the calculation which throws up turbulent seeds does not.}
\label{Iter_DH}
\end{figure}

Figure \ref{GT_DH}(\textit{a}) indicates that three energy regimes
exist in this geometry rather than the two in BF92. As before, below a
certain initial energy value, a QLOP is selected and above a critical
energy $E_{c}$ initial conditions significantly different from the
QLOP trigger turbulence. Between these two energy regions, however,
there now exists a range of initial energies where our algorithm
generates an initial condition different from the QLOP (in the sense
that it appears to have a qualitatively different spatial structure) -
see figure \ref{iso_Q/NLOP}. Using the nomenclature described in the
introduction, we call this qualitatively different optimal
perturbation a NLOP after \citet{pringle2010a} and PWK11. Figure
\ref{Iter_DH} contrasts the convergence for the NLOP with the
non-convergence in the turbulent seed region $E_0>E_c$. Only
five points are plotted in figure \ref{Iter_DH} as for additional
iterations (however small we made our step size in the direction of the
gradient) it was not possible to find a new $\mathbf{u}_{0}$ with a
gain that improved on the previous iteration.

%
% fig 9
%

As a consequence of the kinetic energy gains of the NLOP and QLOP
becoming very similar around $E_{0}=1.0 \times 10^{-7}$, the
cross-over between the NLOP and QLOP is hard to pinpoint. If the NLOP
is used to initialise the algorithm for energies slightly above
$E_{c}$, the algorithm is found to converge to an initial condition
very similar to the NLOP. In fact, the algorithm started with random
noise will still sometimes converge to the NLOP for values of $E_{0}$
twice as large as $E_{c}$. As a consequence, approaching $E_c$ from
above proved a better strategy. Random noise was used at $E_{0} \sim
2.5 \times 10^{-6} \approx 7.5E_c$ to find {\em a} turbulent seed and
then this was used sequentially to initiate the algorithm as $E_0$ was
gradually decreased. This experience clearly emphasizes the main
hazard of nonlinear optimisation: it is easy to get stuck near local
maxima. Although not a cure, an obvious strategy to reduce this
possibility is to look for robustness of result over a suite of initial
conditions.

\begin{figure}
  \centerline{\includegraphics[width=0.32\textwidth]{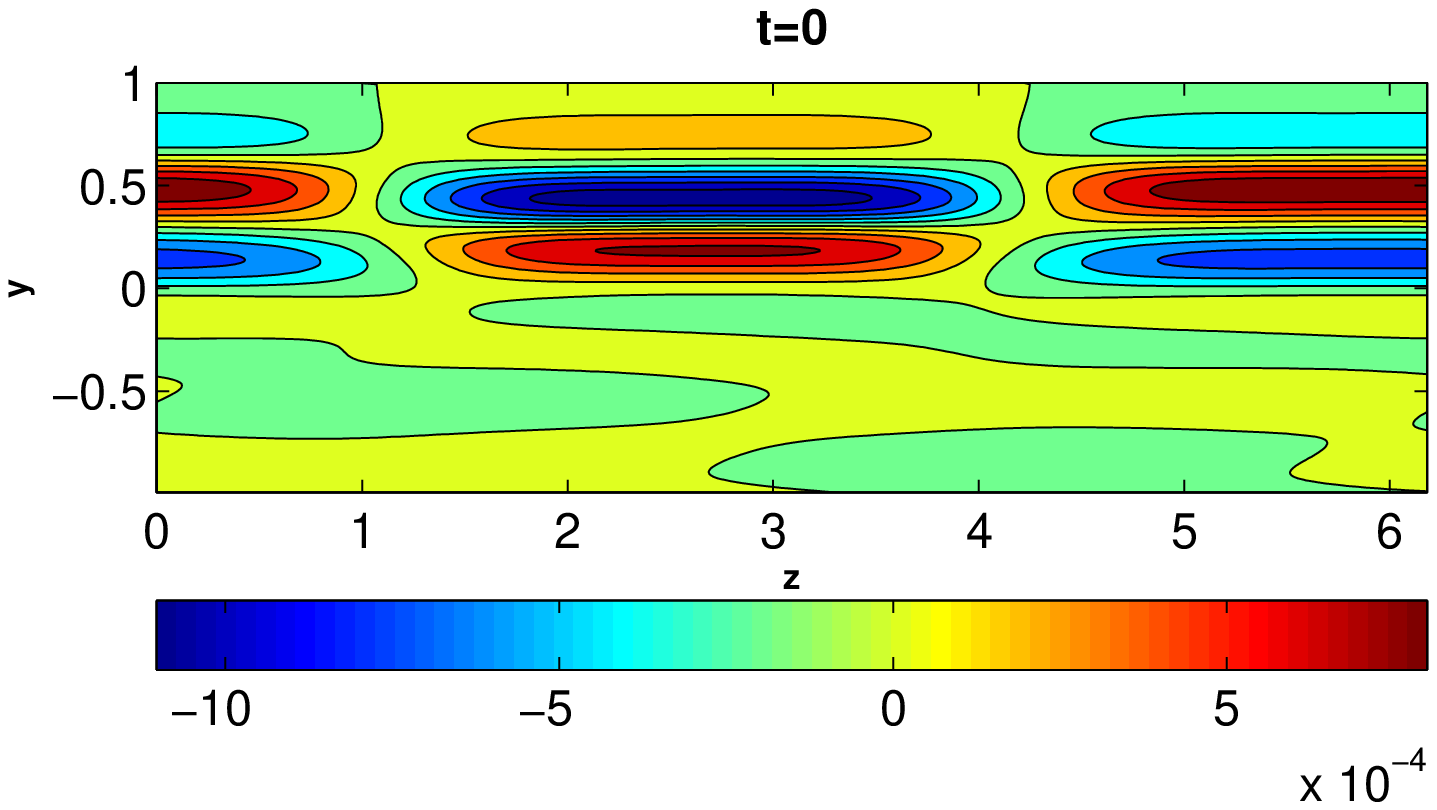}
    \includegraphics[width=0.32\textwidth]{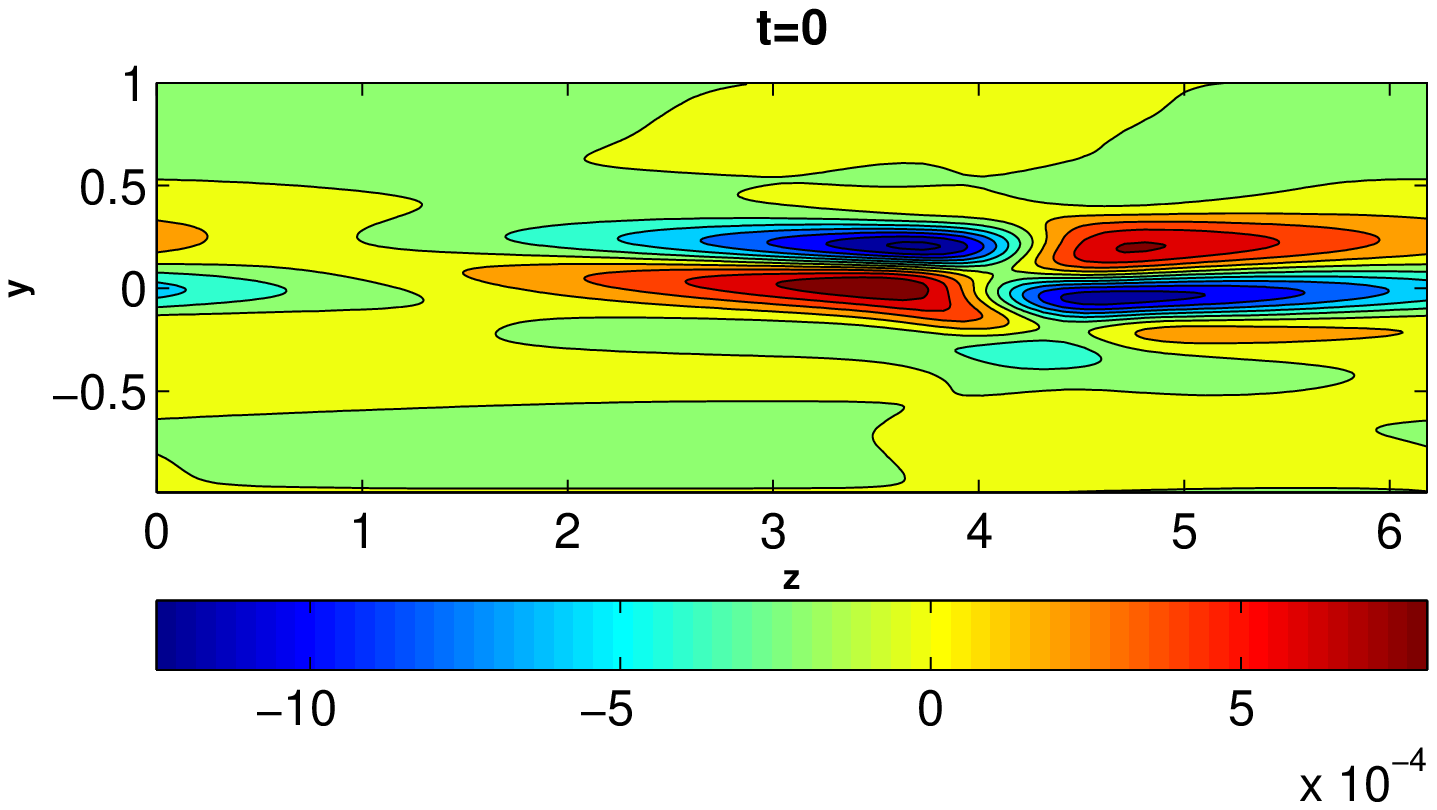}
    \includegraphics[width=0.32\textwidth]{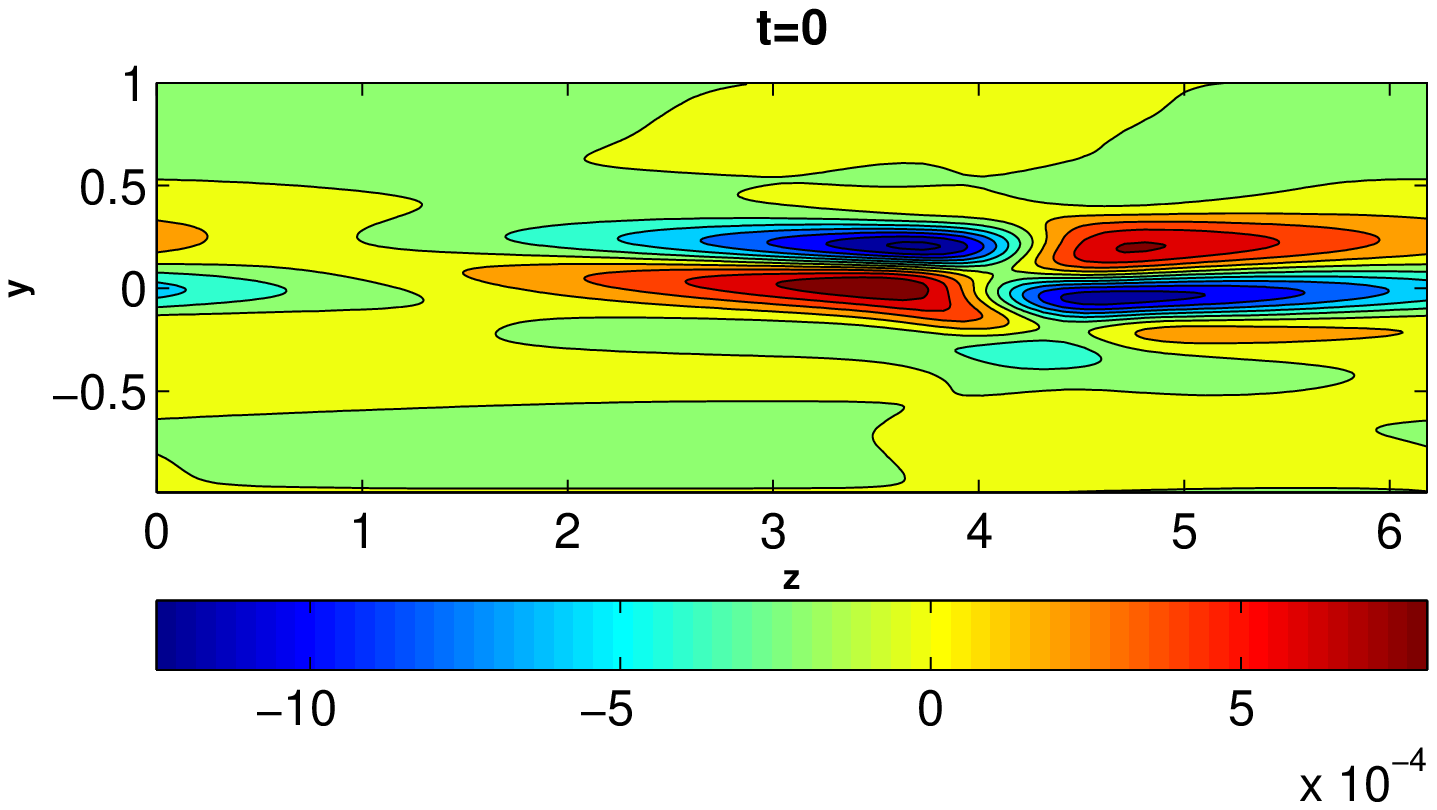}}
  \centerline{\includegraphics[width=0.32\textwidth]{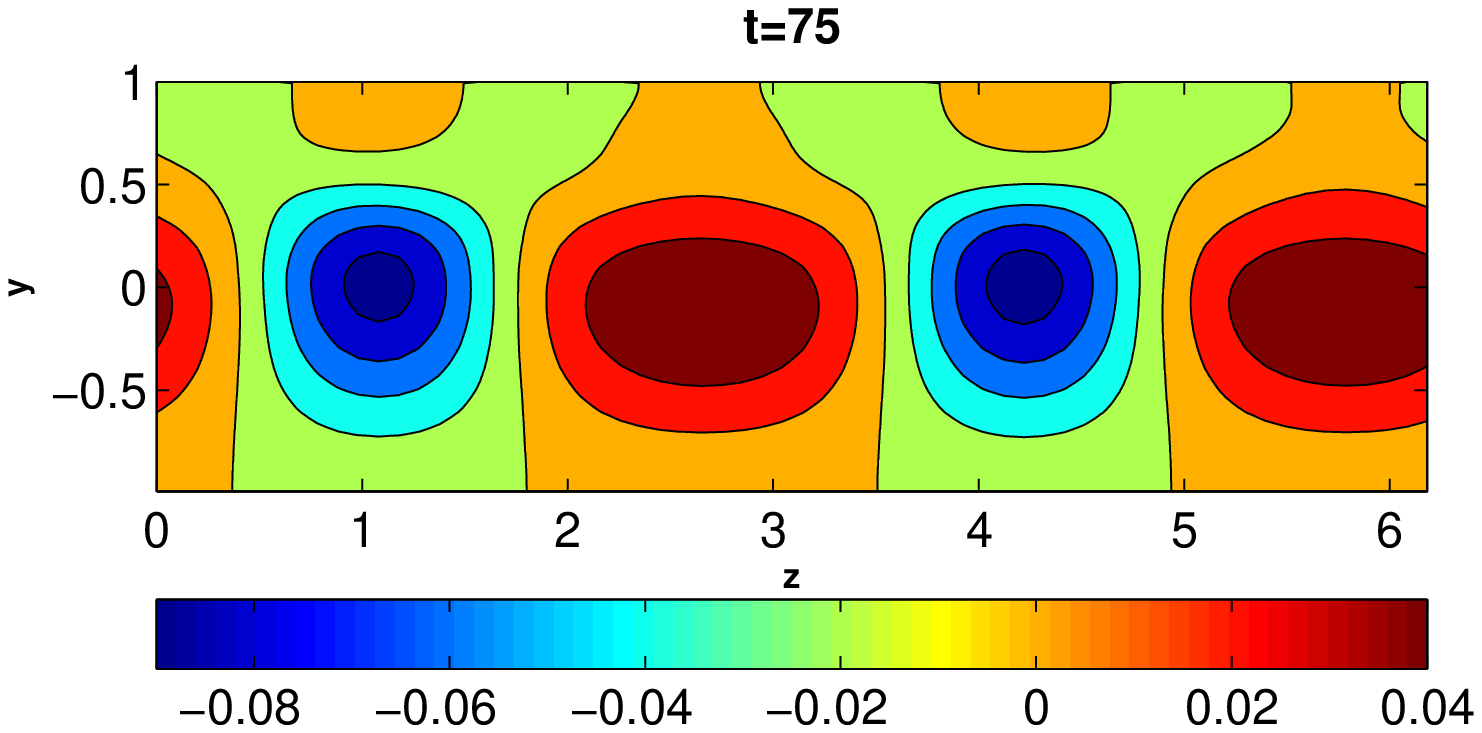}
    \includegraphics[width=0.32\textwidth]{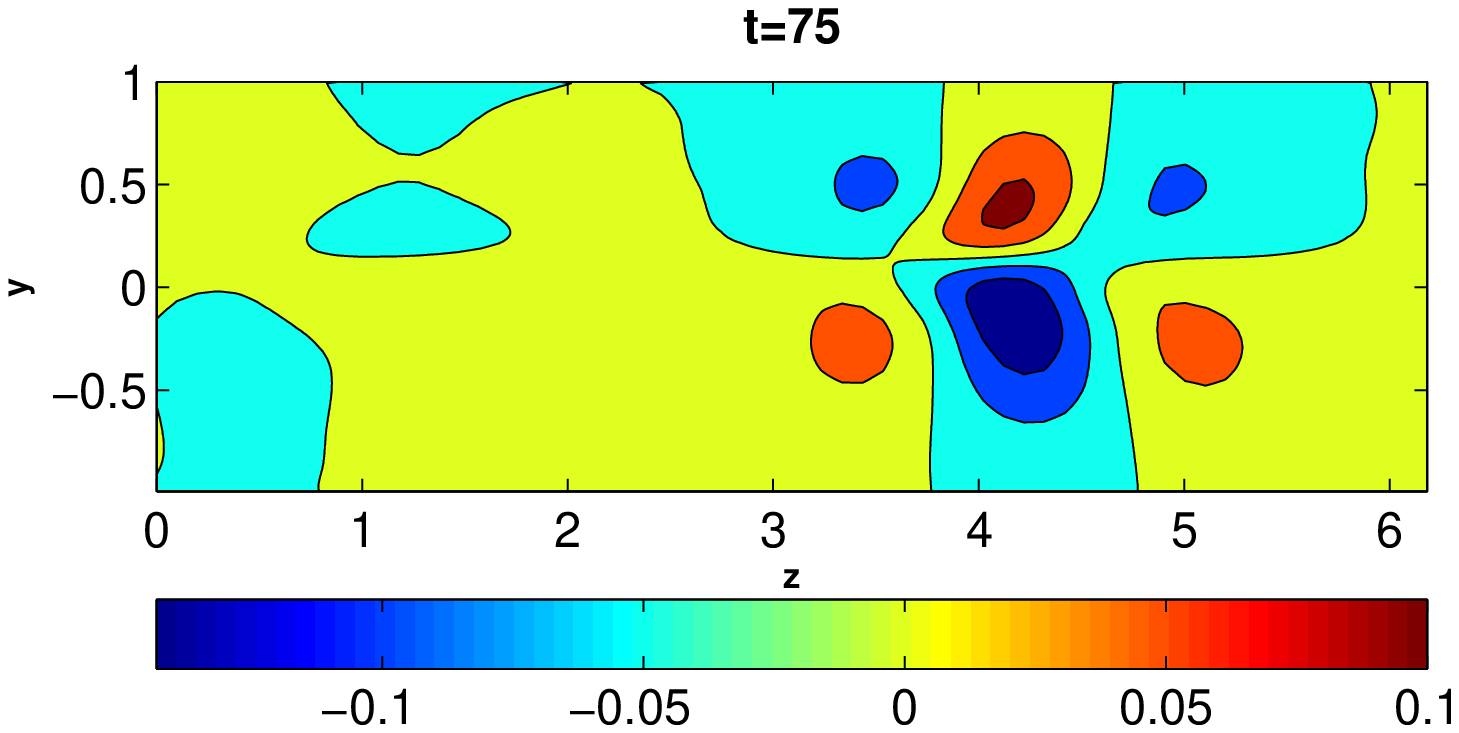}
    \includegraphics[width=0.32\textwidth]{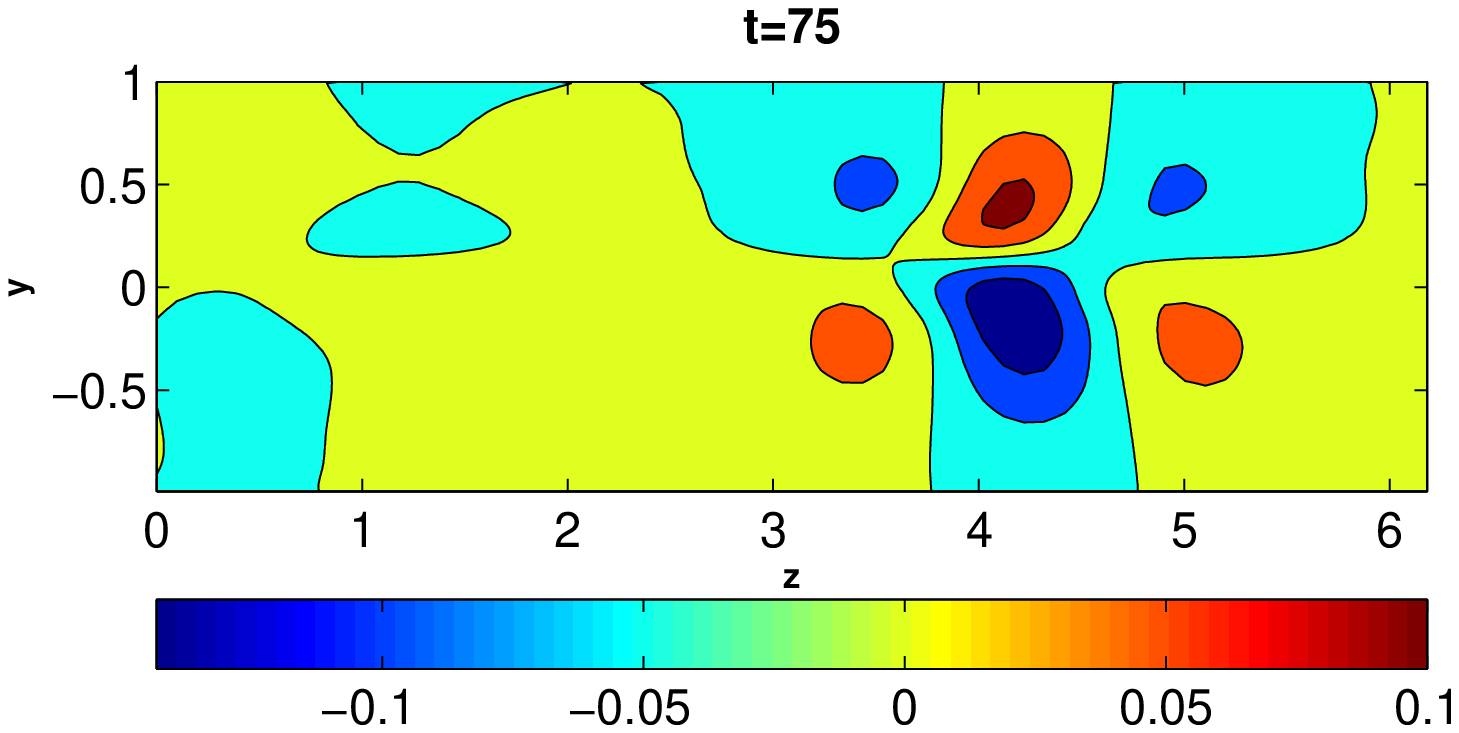}}
  \centerline{\includegraphics[width=0.32\textwidth]{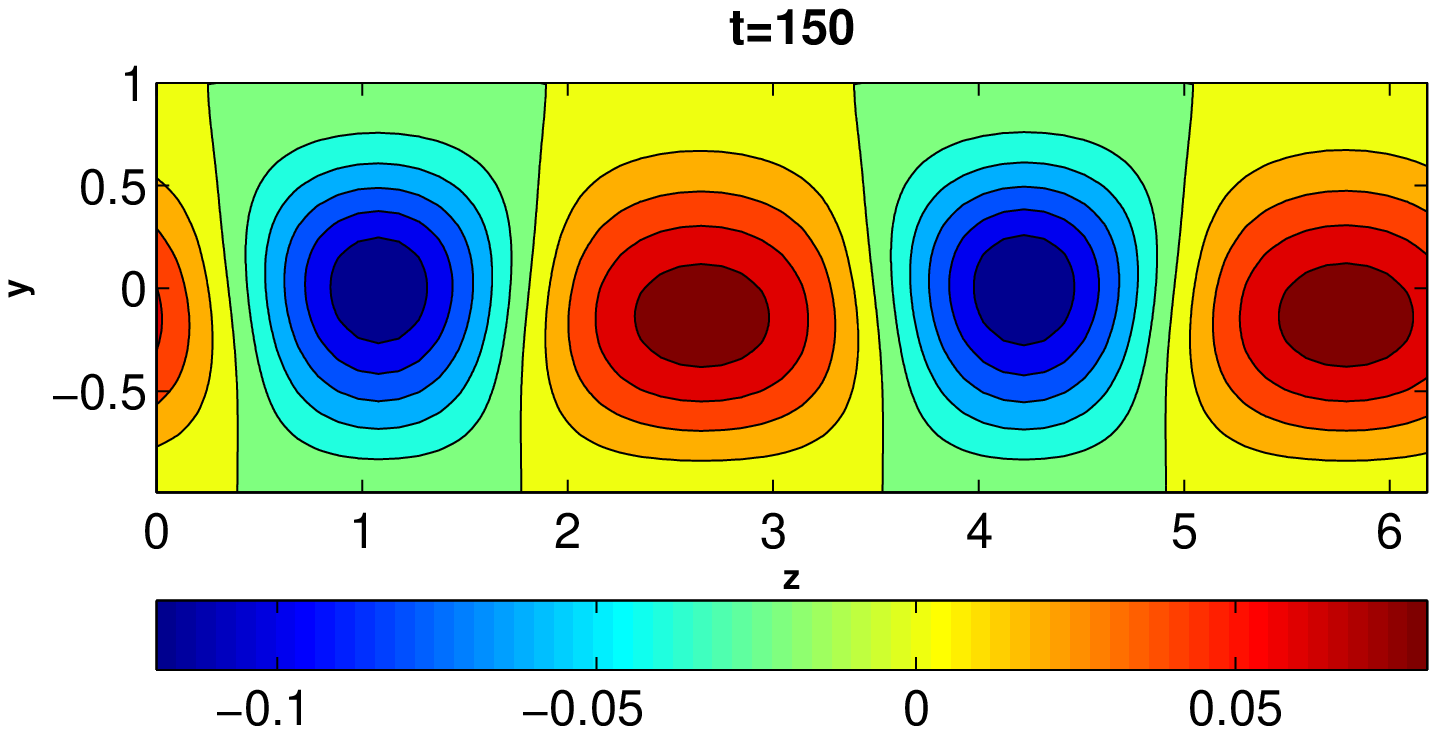}
    \includegraphics[width=0.32\textwidth]{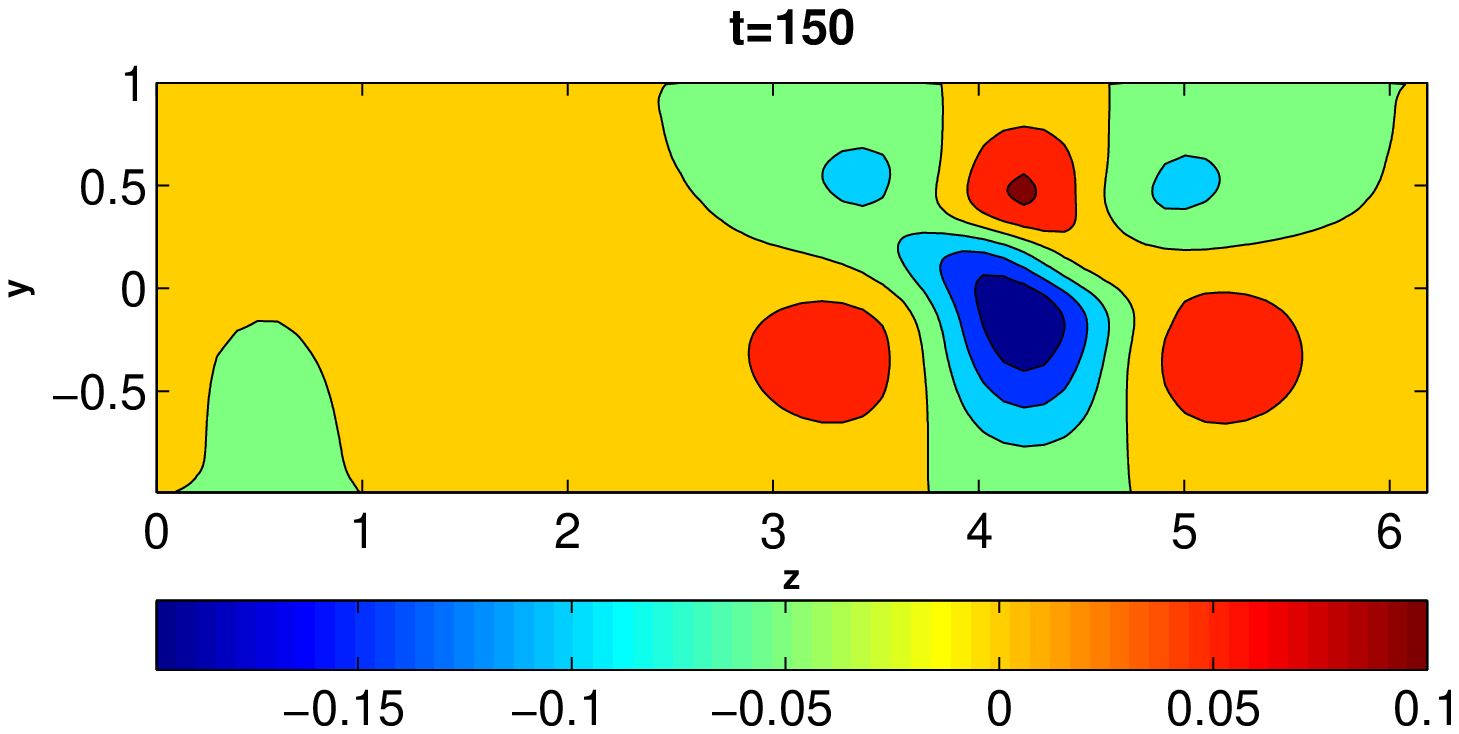}
    \includegraphics[width=0.32\textwidth]{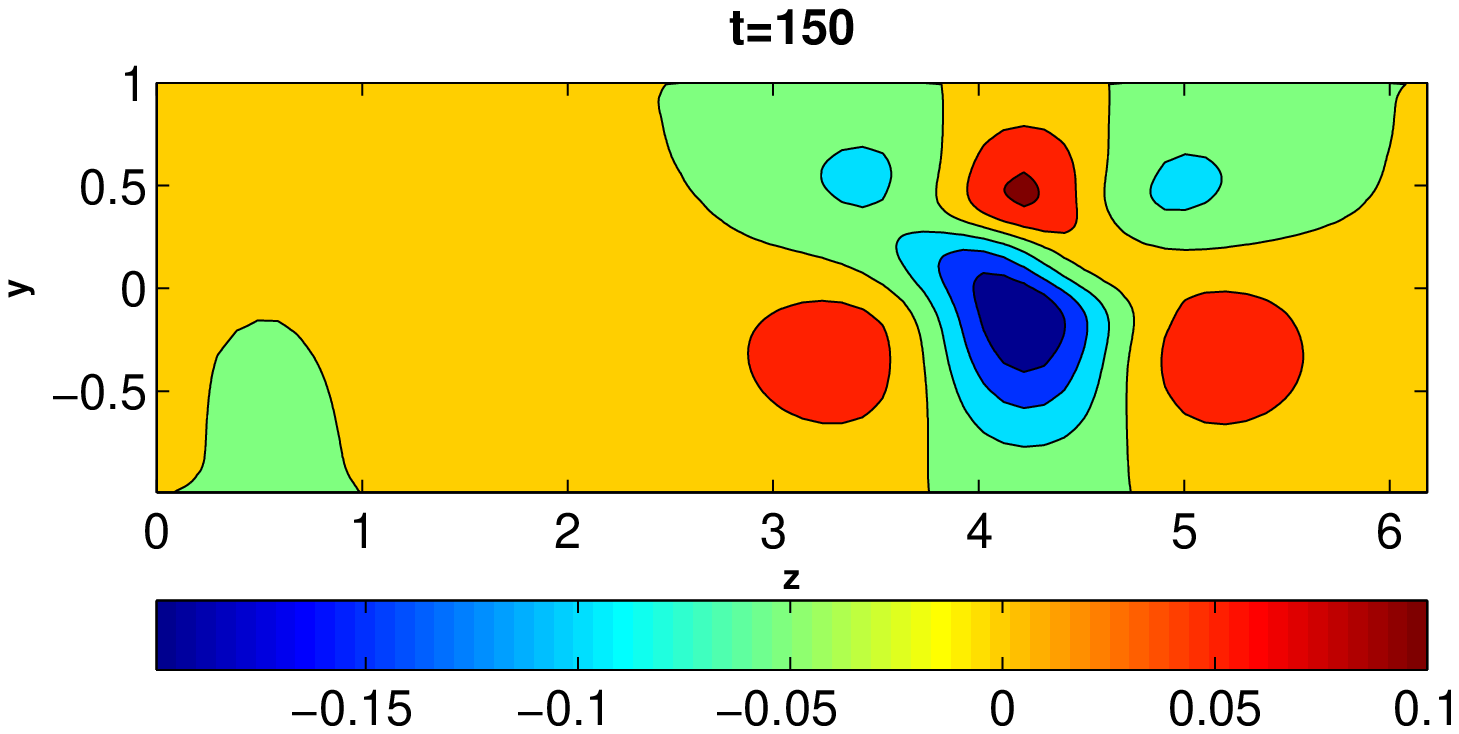}}
  \centerline{\includegraphics[width=0.32\textwidth]{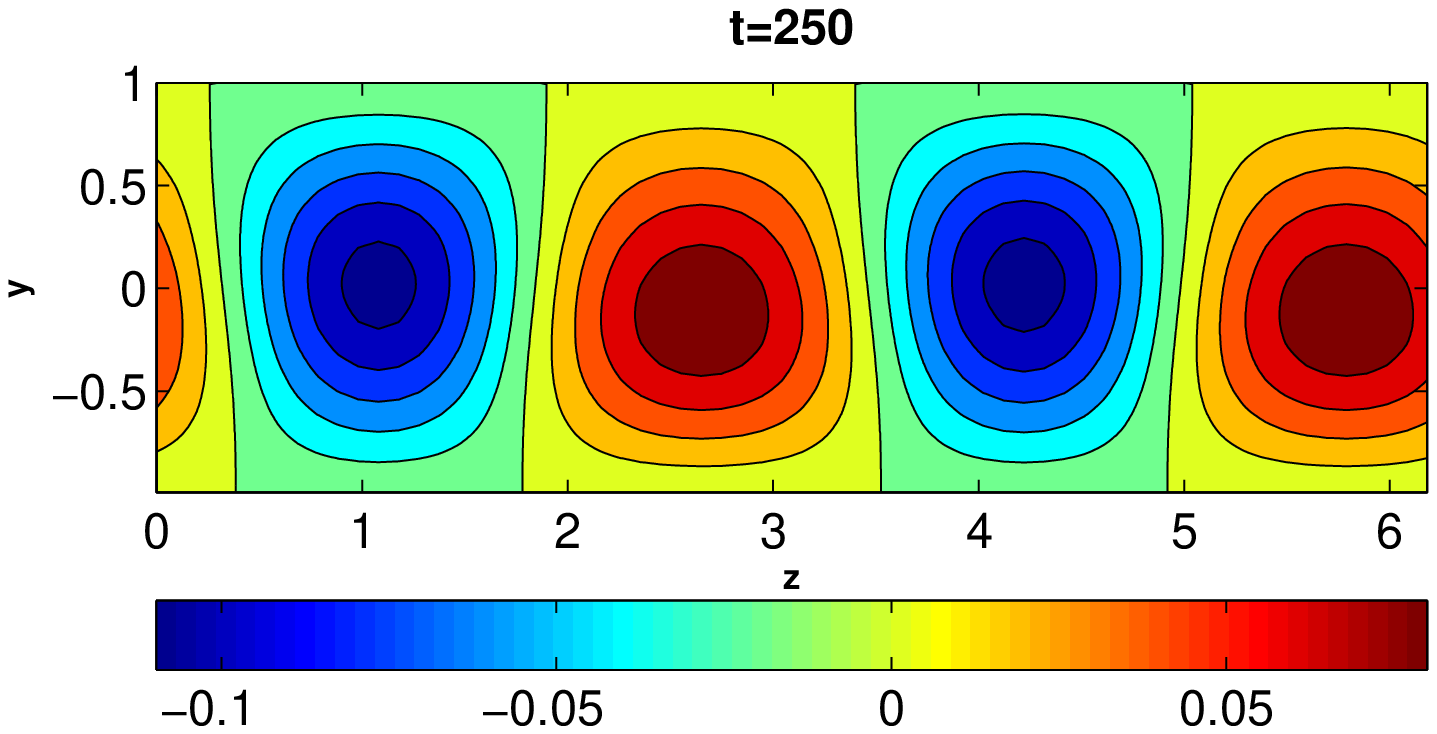}
    \includegraphics[width=0.32\textwidth]{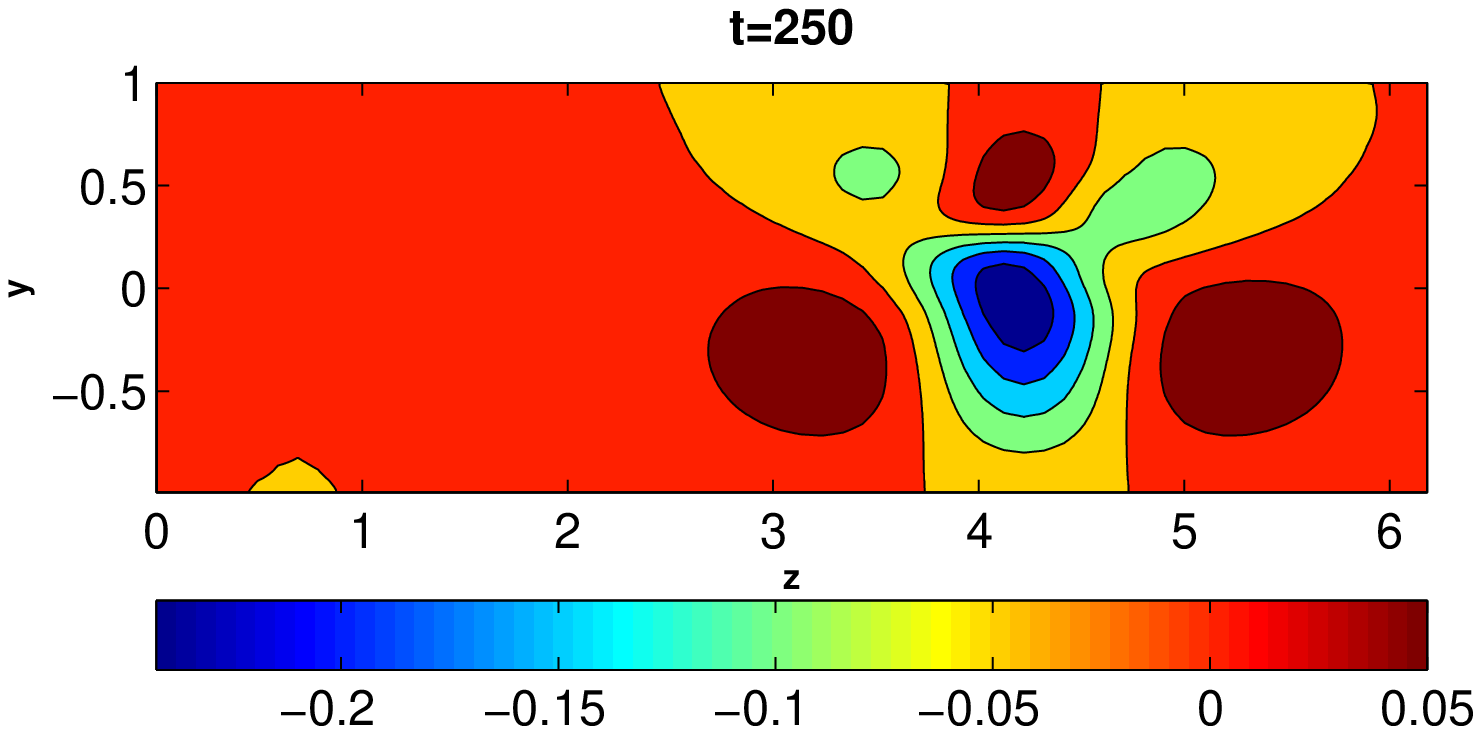}
    \includegraphics[width=0.32\textwidth]{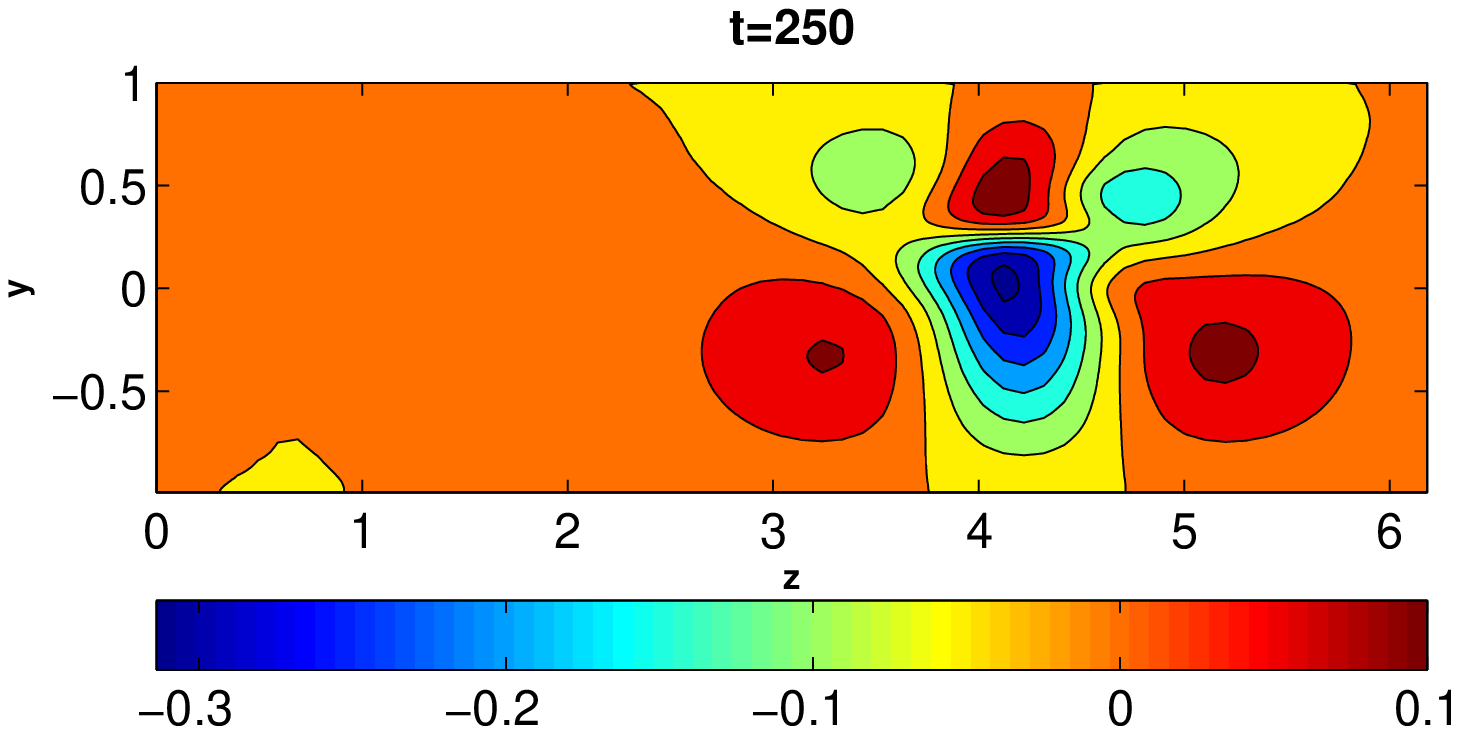}}
  \centerline{\includegraphics[width=0.32\textwidth]{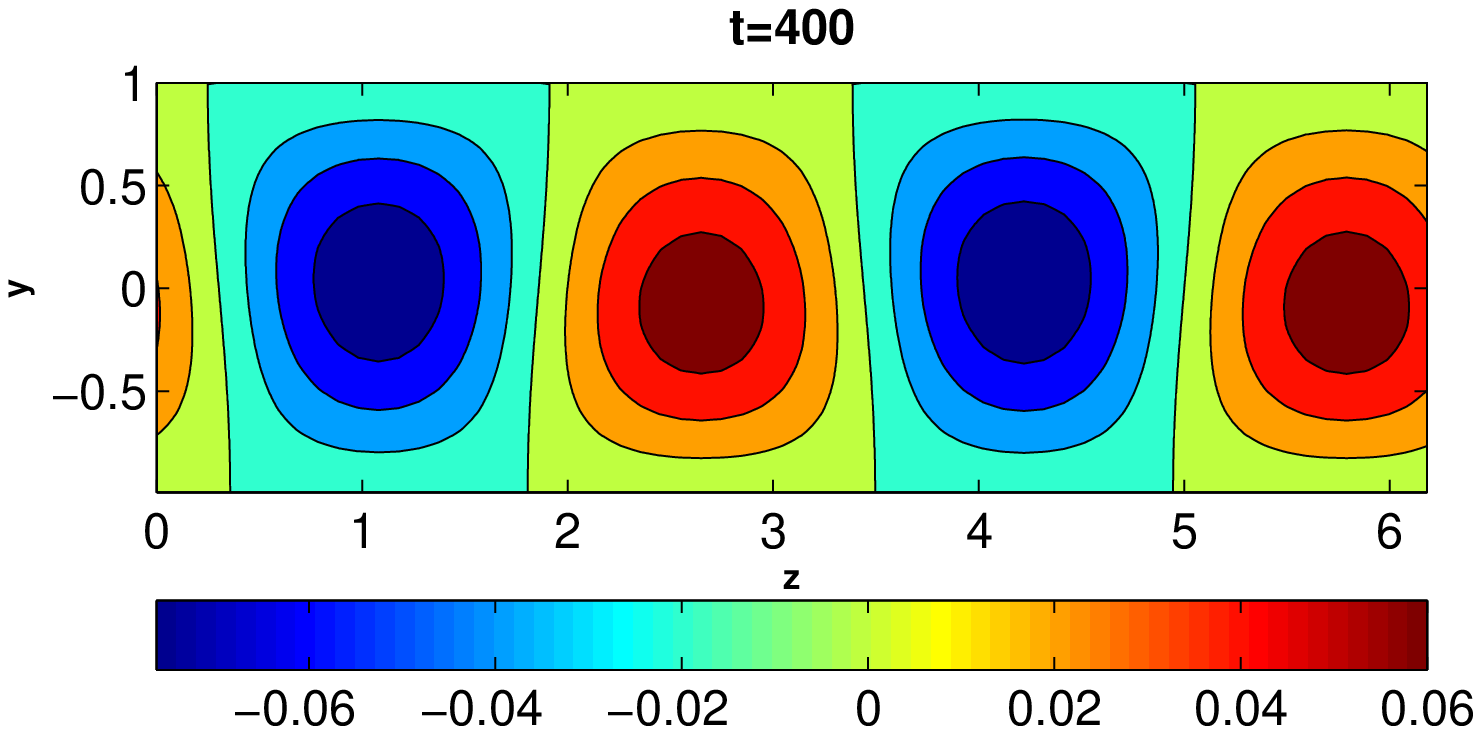}
    \includegraphics[width=0.32\textwidth]{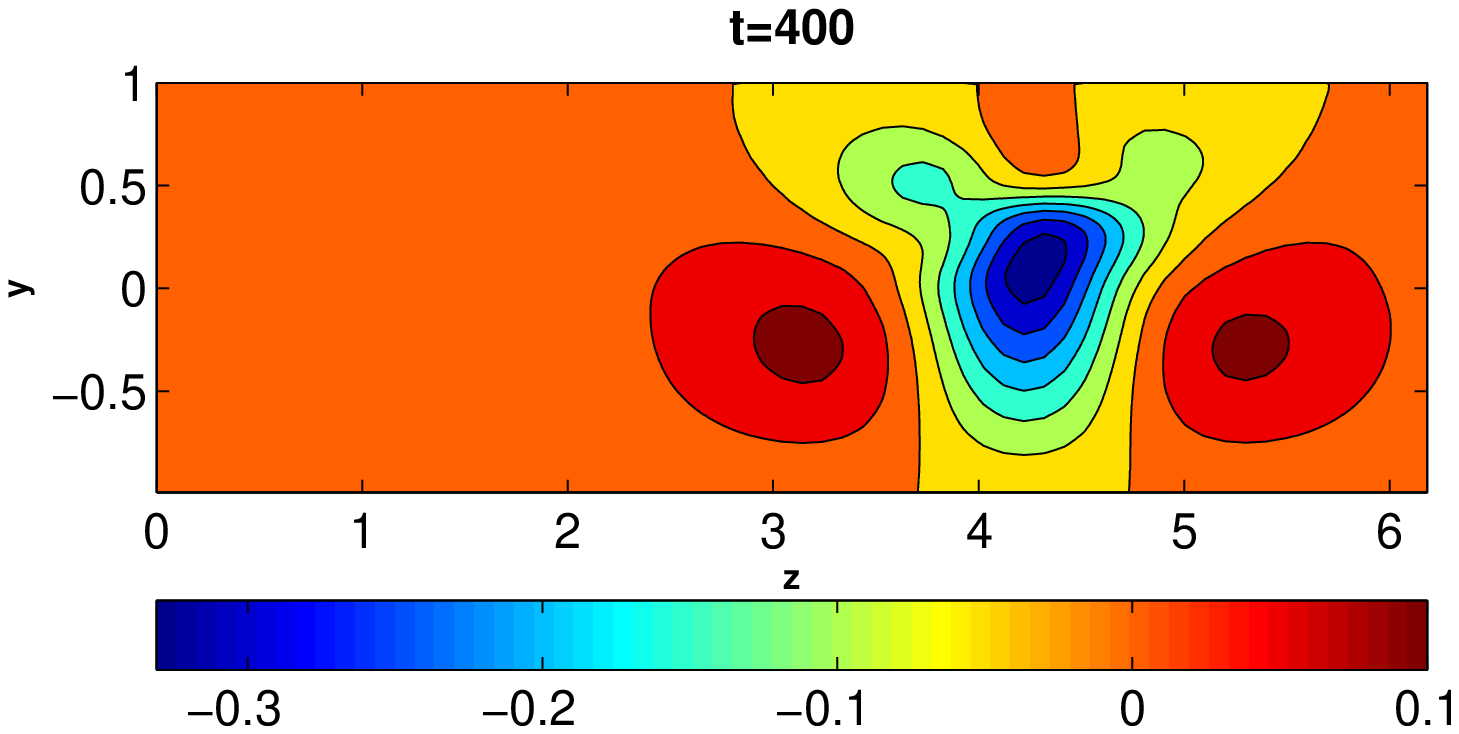}
    \includegraphics[width=0.32\textwidth]{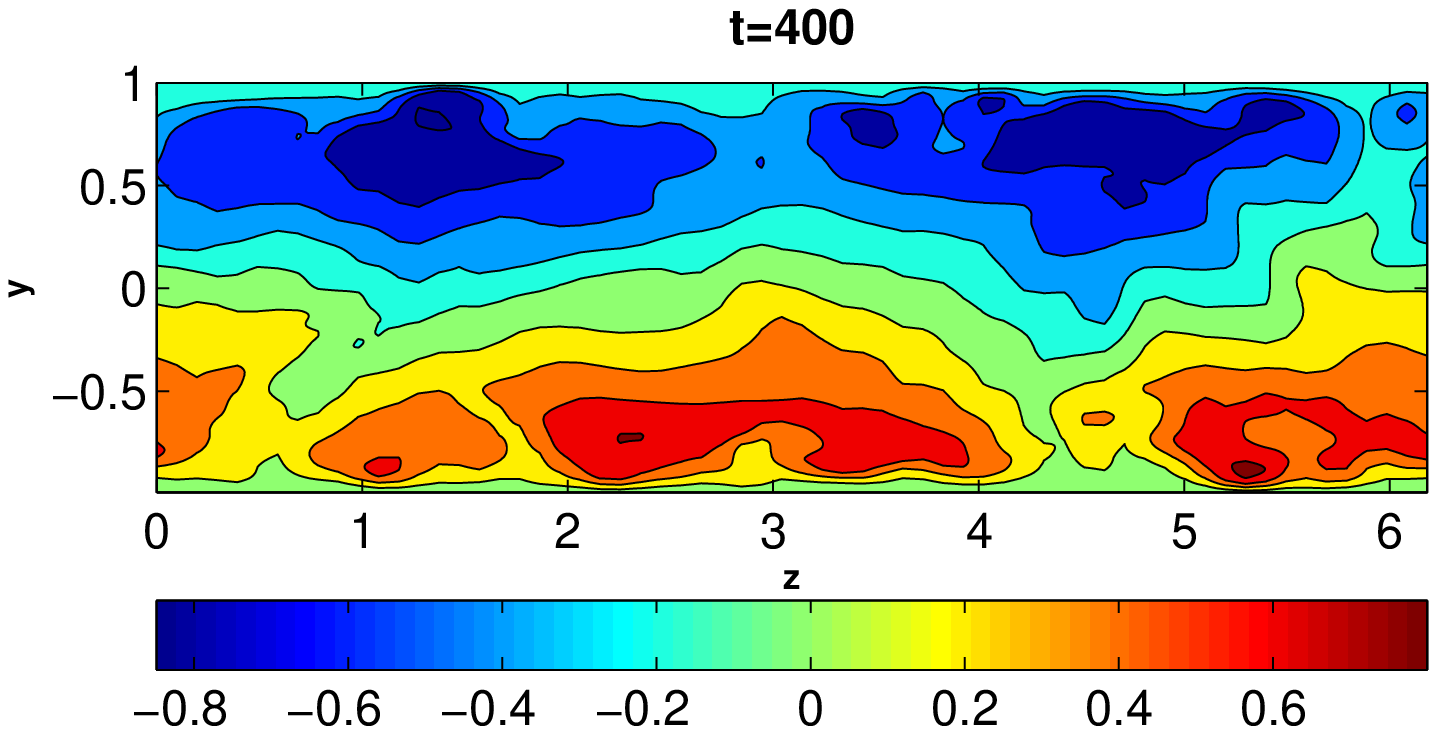}}
\caption{Contours of streamwise velocity $u$ for NLOP
  $E_{0}= 3.2 \times 10^{-7}$ (left), approximated minimal seed and
  turbulent seed at $E_{0}= 3.3 \times 10^{-7}$ (right), at times 0,
  75, 150, 250 and 400. At intermediate times, the minimal seed flow
  on the edge and that initiated from the turbulent seed remain
  relatively similar. Contour levels are: going down the left column                    
  (min,spacing,max)=$(-10,2,8)\times 10^{-4}$, 
                     $(-0.08,0.02,0.04)$,
                     $(-0.1,0.02,0.08)$,
                     $(-0.1,0.02,0.08)$,
                 and $(-0.06,0.02,0.06)$; 
going down the centre (min,spacing,max)=$(-10,2,8)\times 10^{-4}$, 
                                         $(-0.1,0.05,0.1)$,
                                         $(-0.15,0.05,0.1)$,
                                         $(-0.2,0.05,0.05)$
                                   and    $(-0.3,0.05,0.1)$;
going down the right column (min,spacing,max)=$(-10,2,8)\times10^{-4}$, 
                                              $(-0.1,0.05,0.1)$,
                                              $(-0.15,0.05,0.1)$,
                                              $(-0.3,0.05,0.1)$ 
                                          and $(-0.8,0.2,0.8)$.}
\label{contours_DH}
\end{figure}

\begin{figure}
  \centerline{\includegraphics[width=0.32\textwidth]{New_figures/DH_4plots/iso_NLOP1.eps}
    \includegraphics[width=0.32\textwidth]{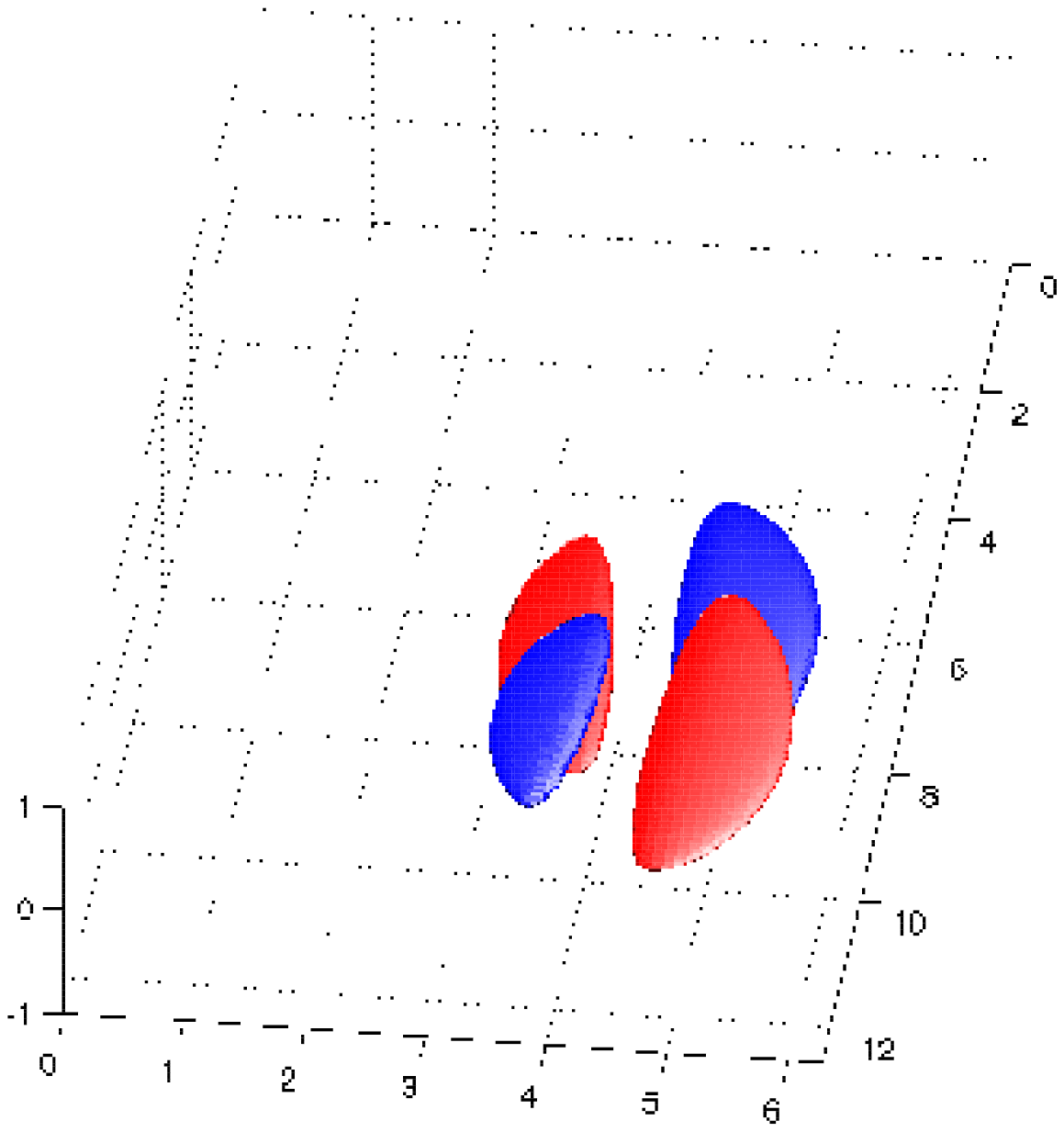}
    \includegraphics[width=0.32\textwidth]{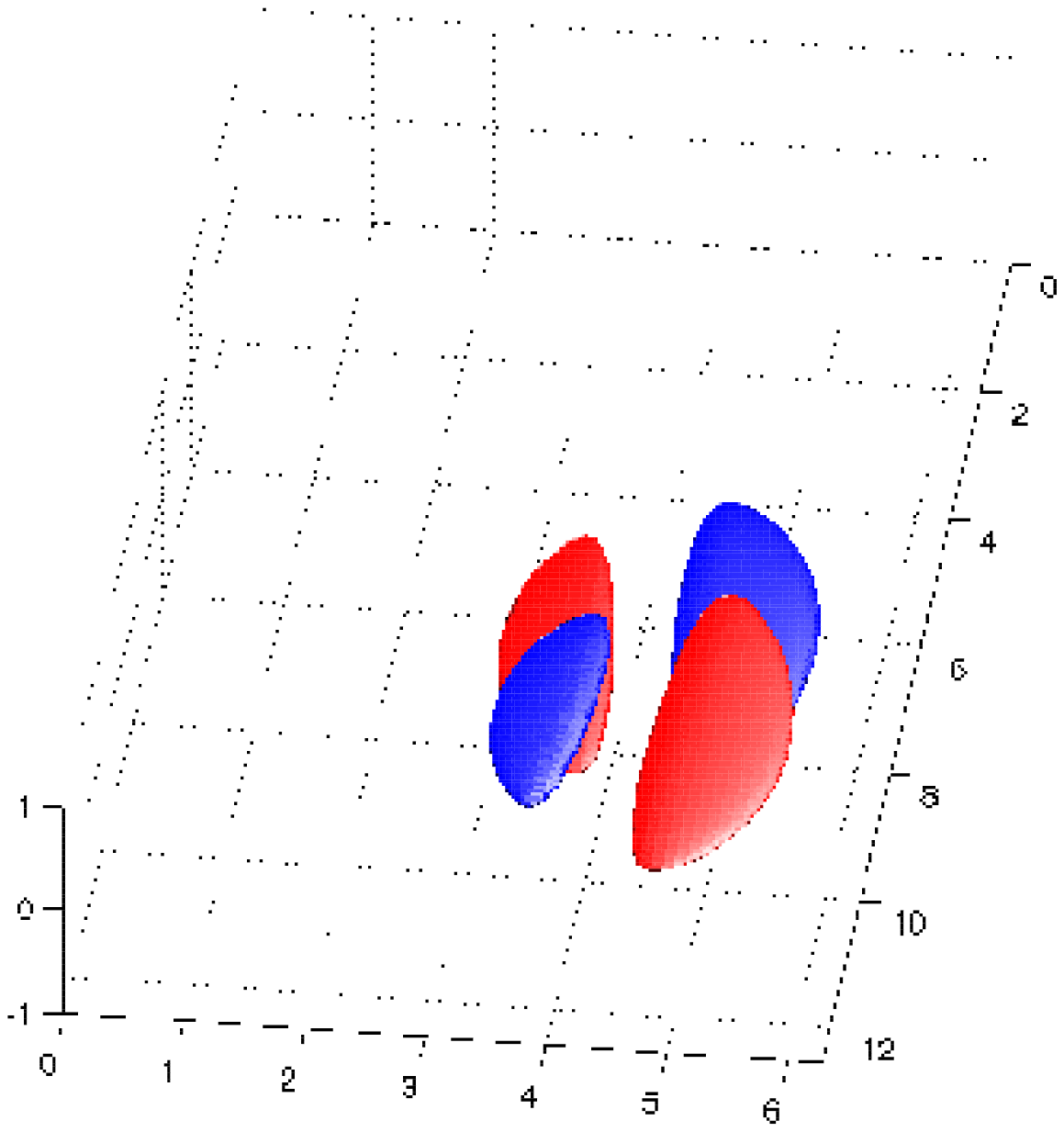}}
  \centerline{\includegraphics[width=0.32\textwidth]{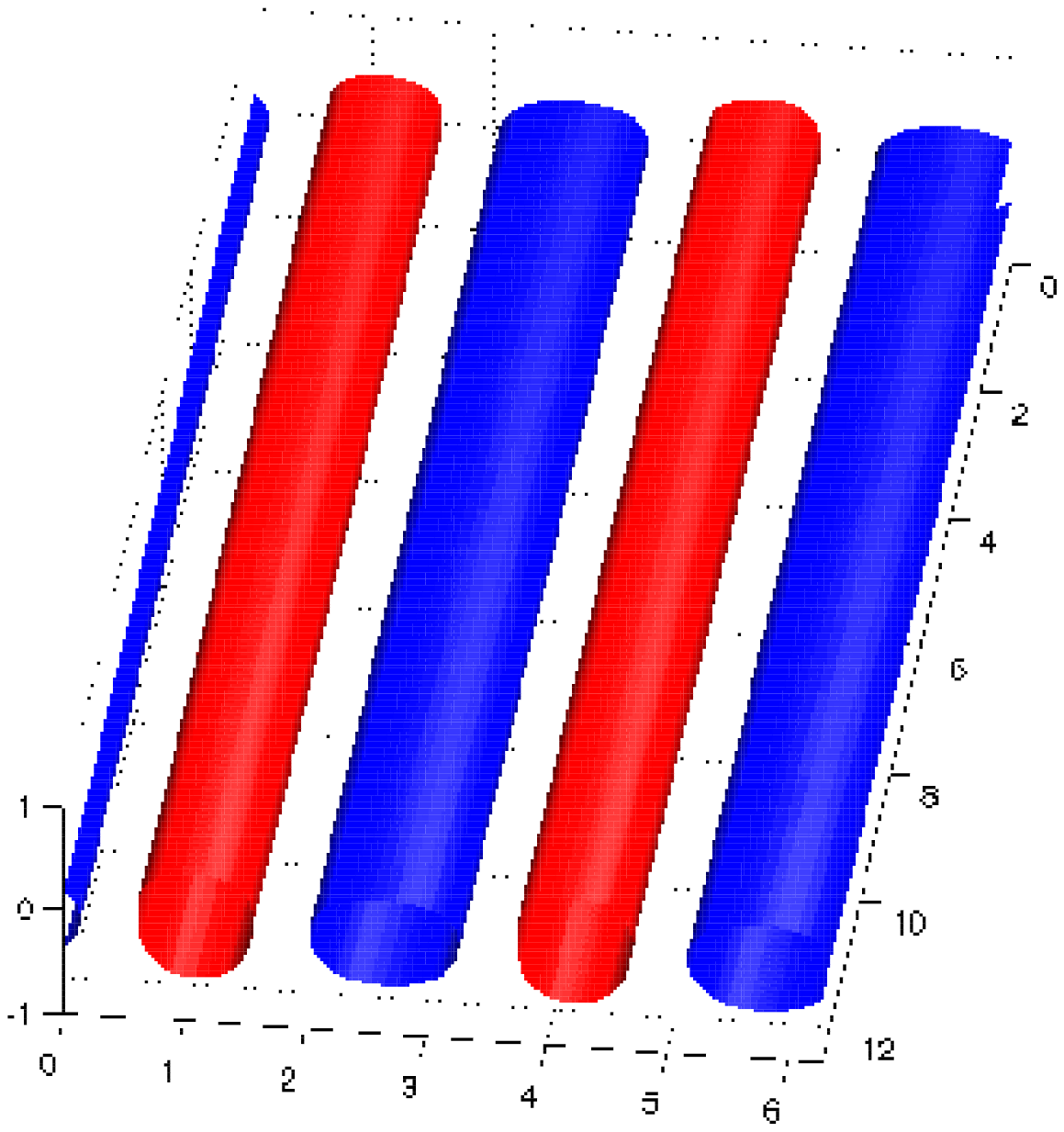}
    \includegraphics[width=0.32\textwidth]{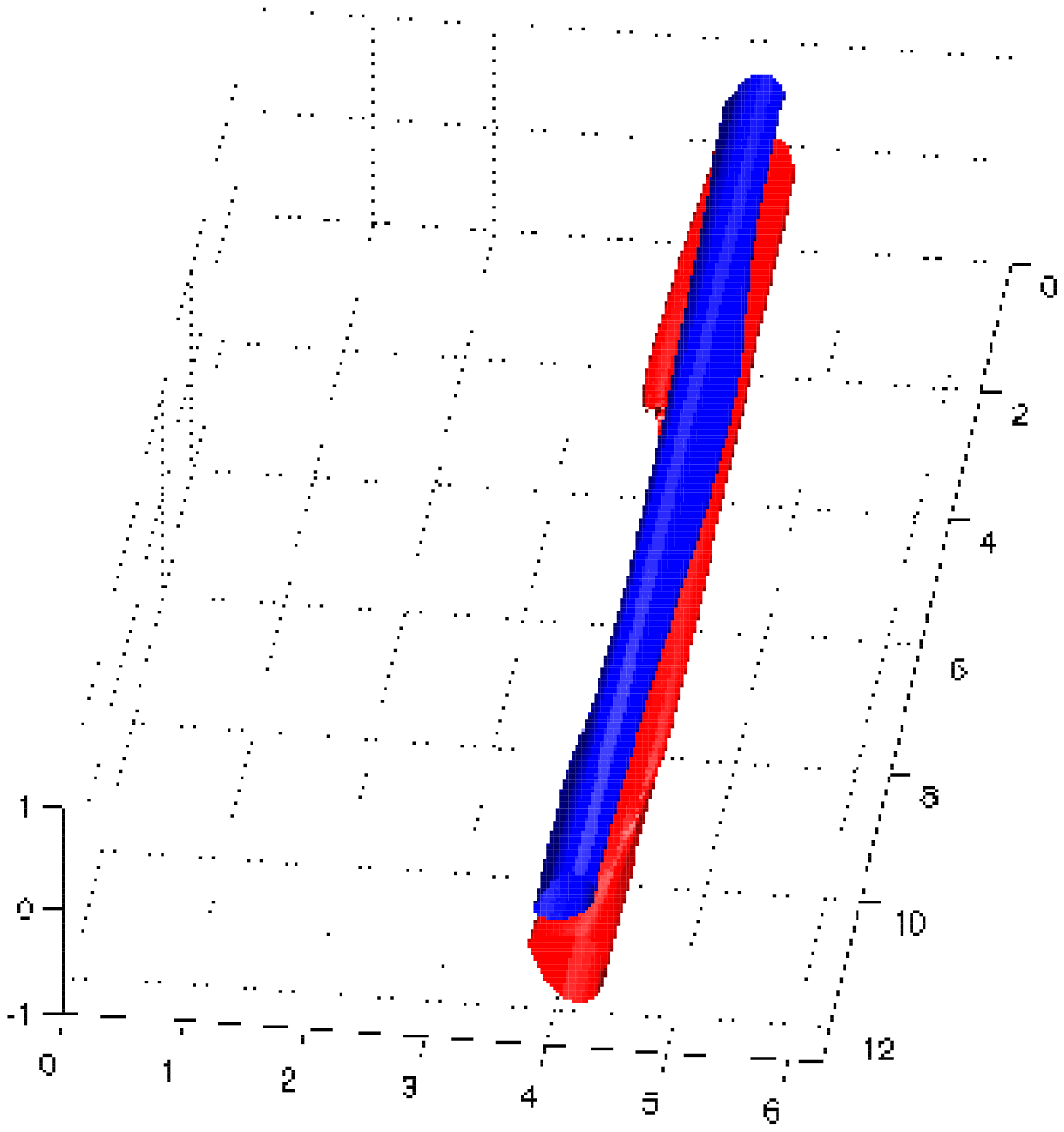}
    \includegraphics[width=0.32\textwidth]{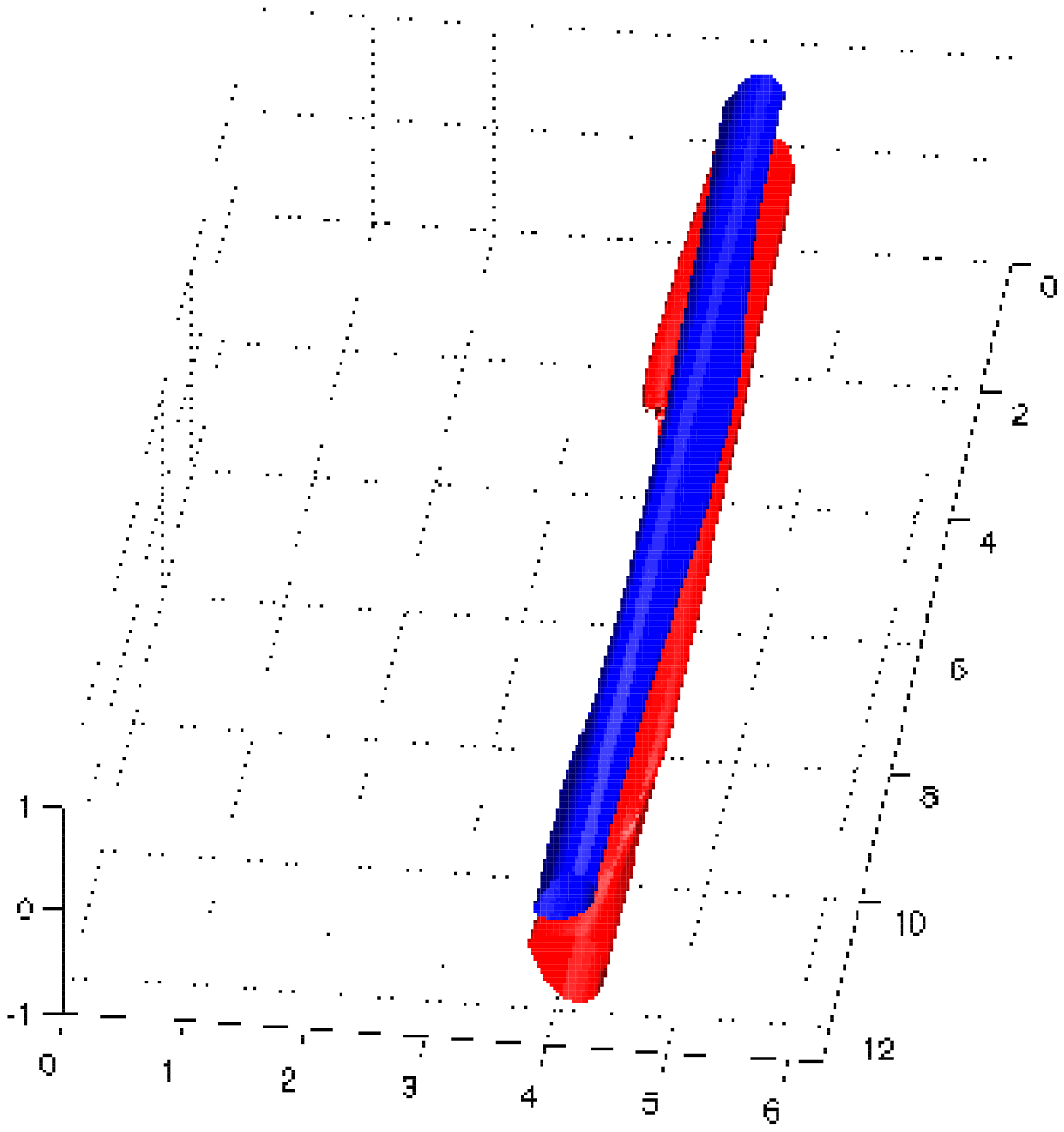}}
  \centerline{\includegraphics[width=0.32\textwidth]{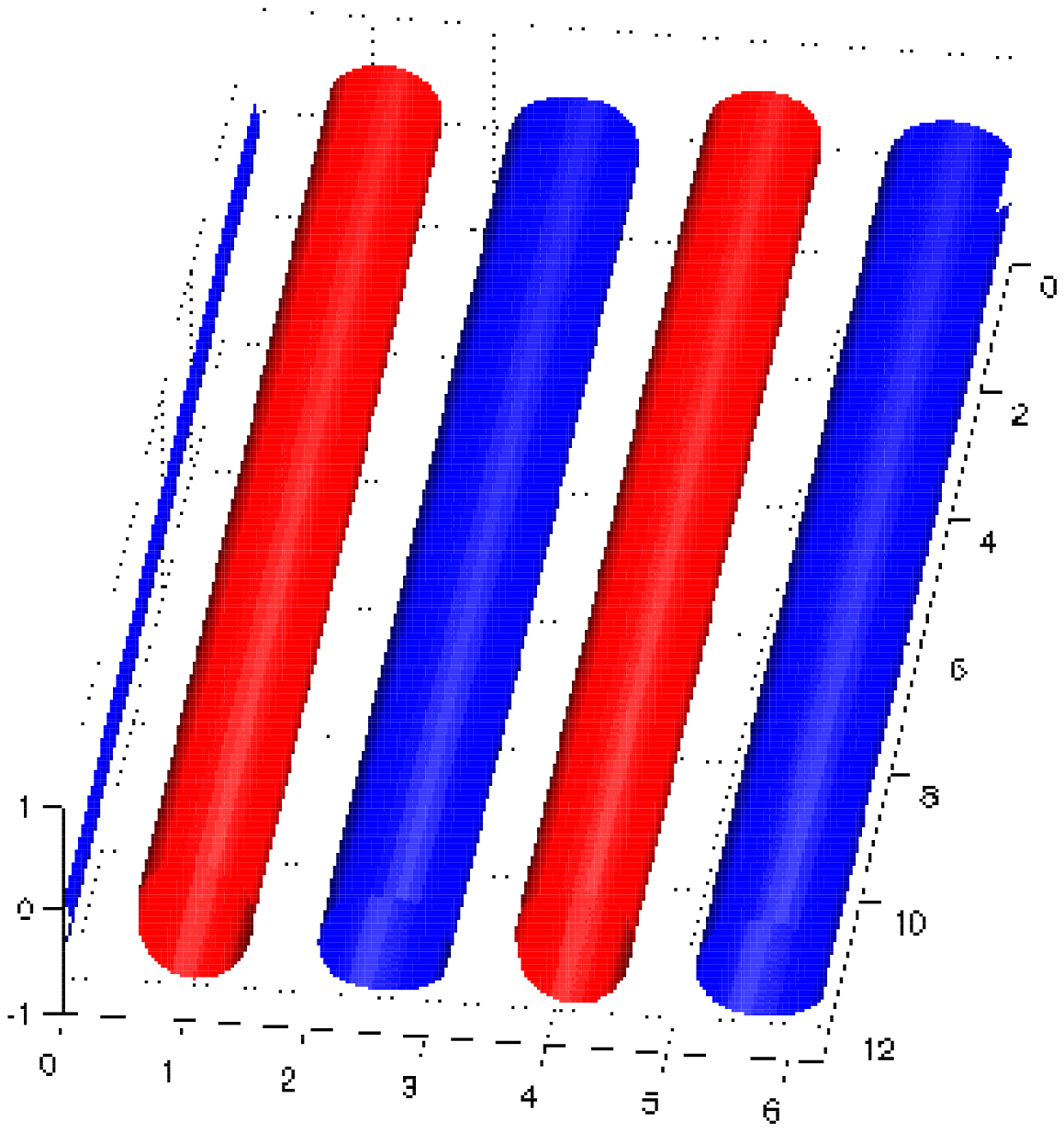}
    \includegraphics[width=0.32\textwidth]{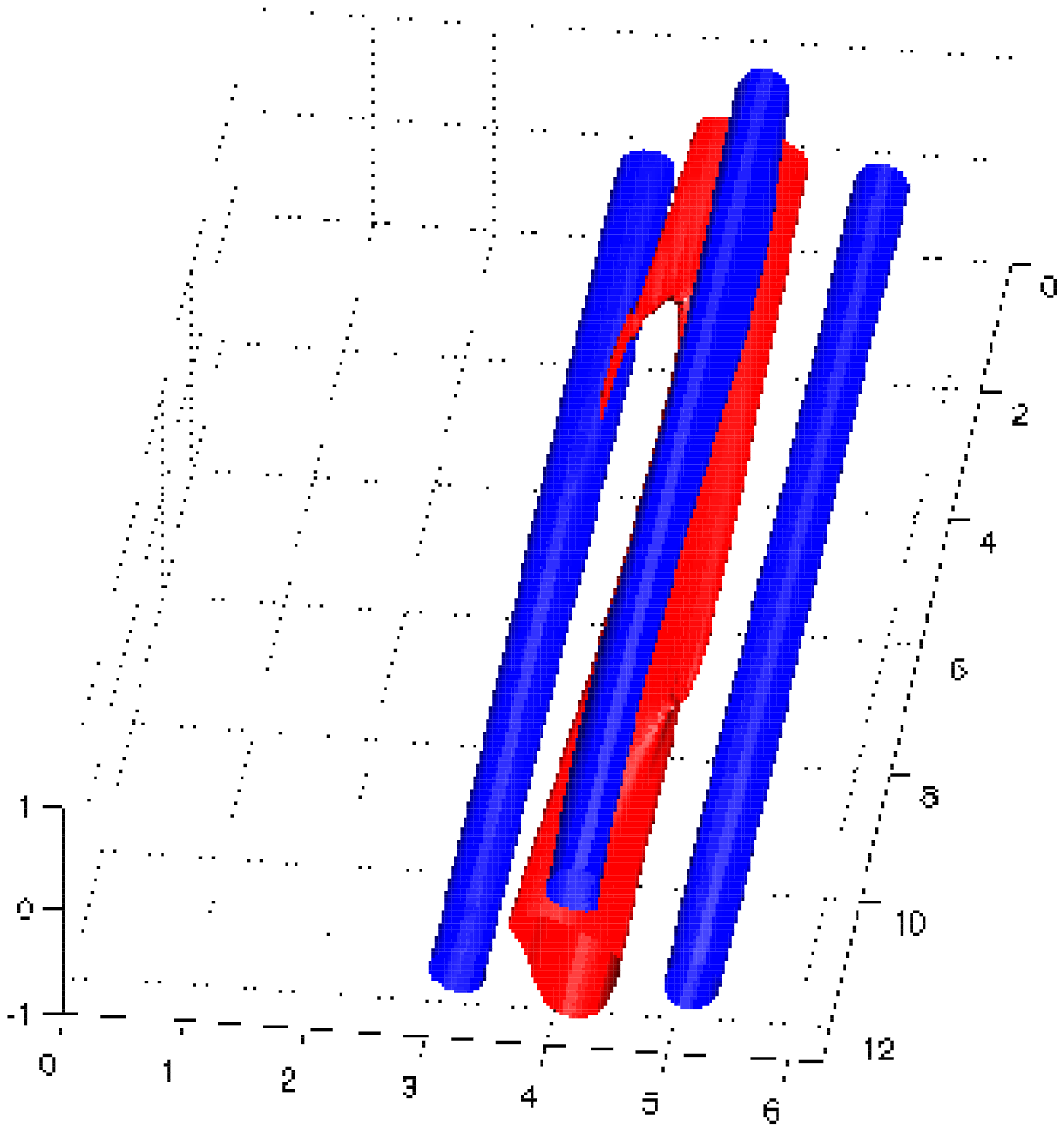}
    \includegraphics[width=0.32\textwidth]{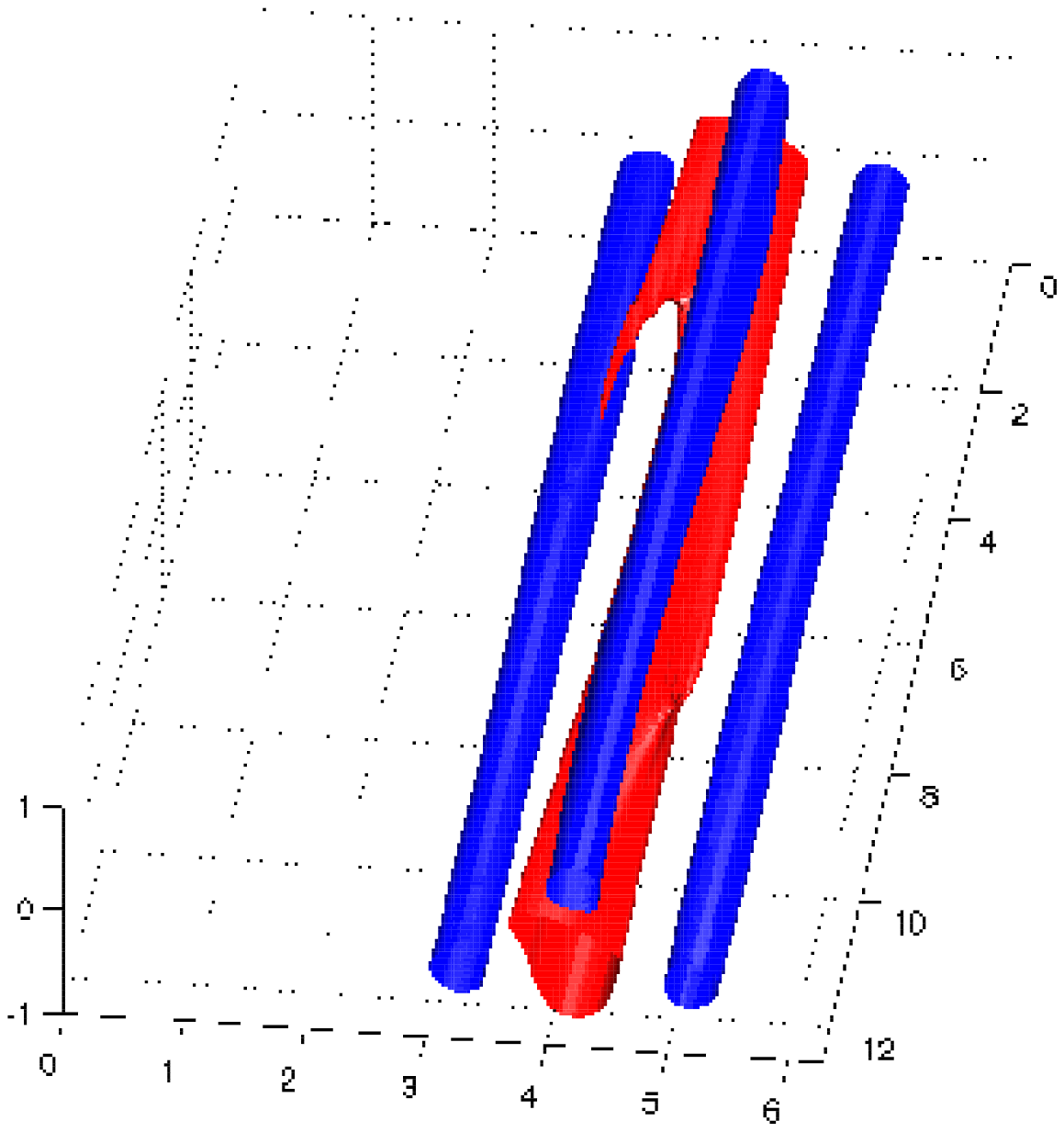}}
  \centerline{\includegraphics[width=0.32\textwidth]{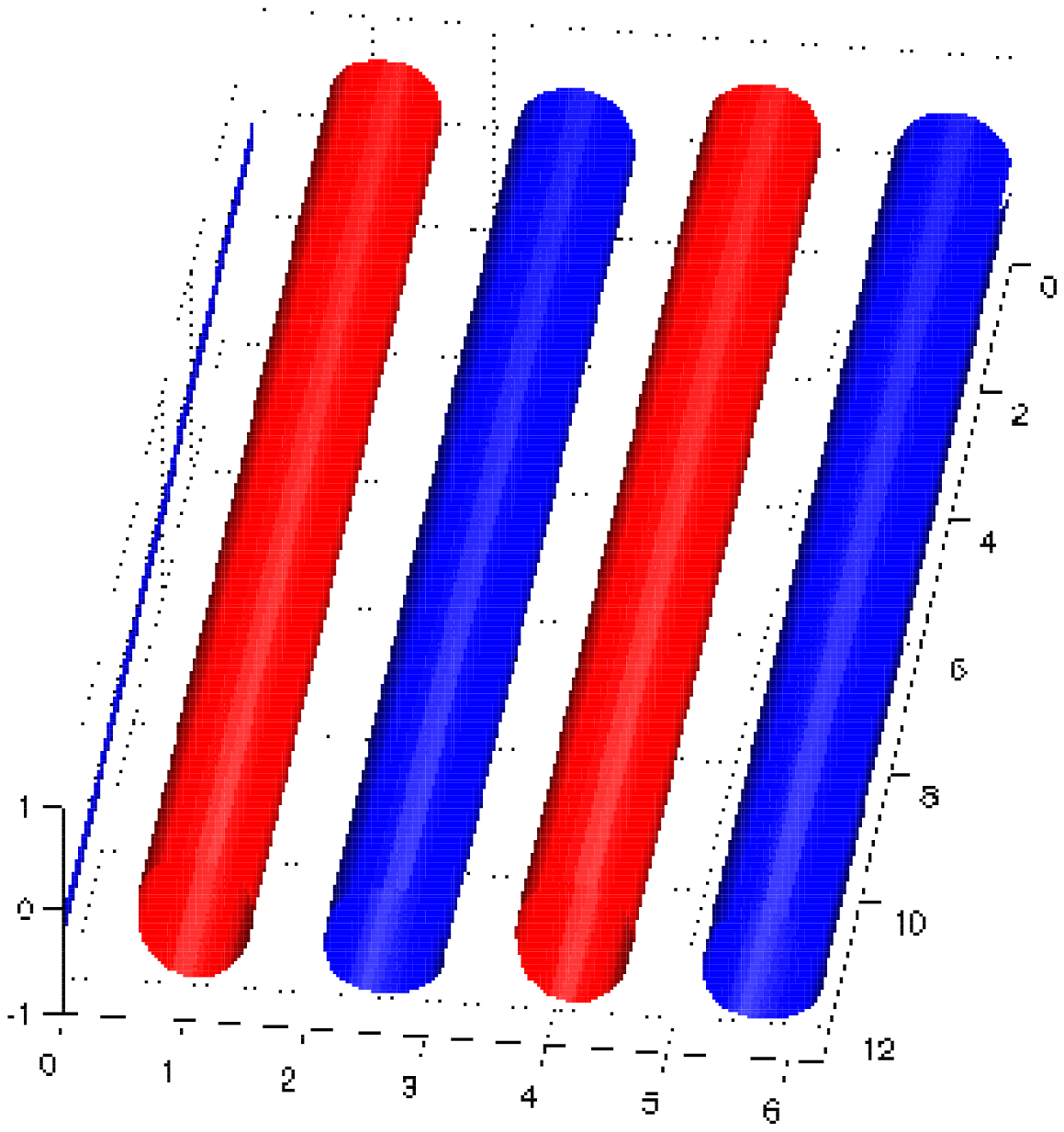}
    \includegraphics[width=0.32\textwidth]{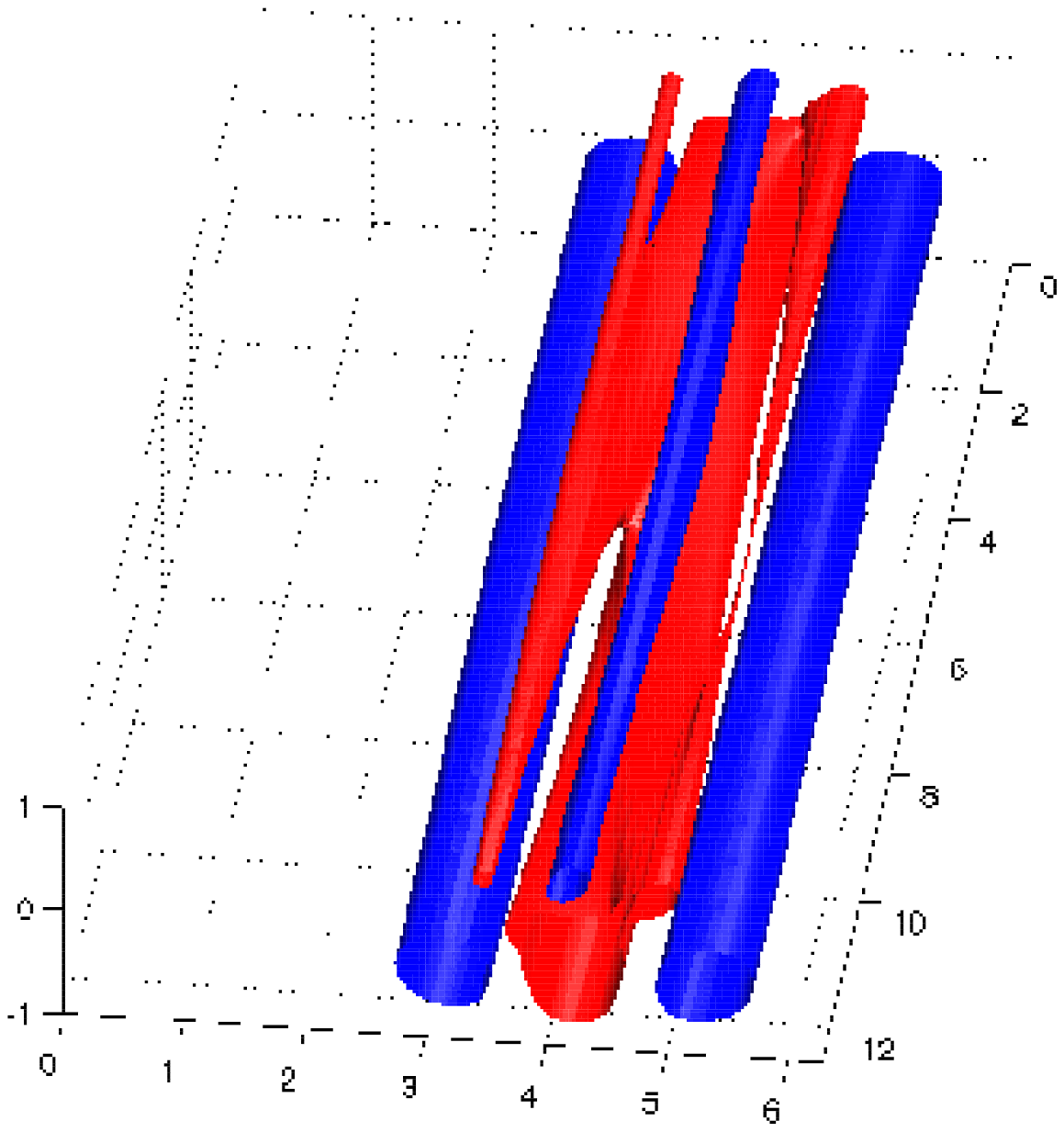}
    \includegraphics[width=0.32\textwidth]{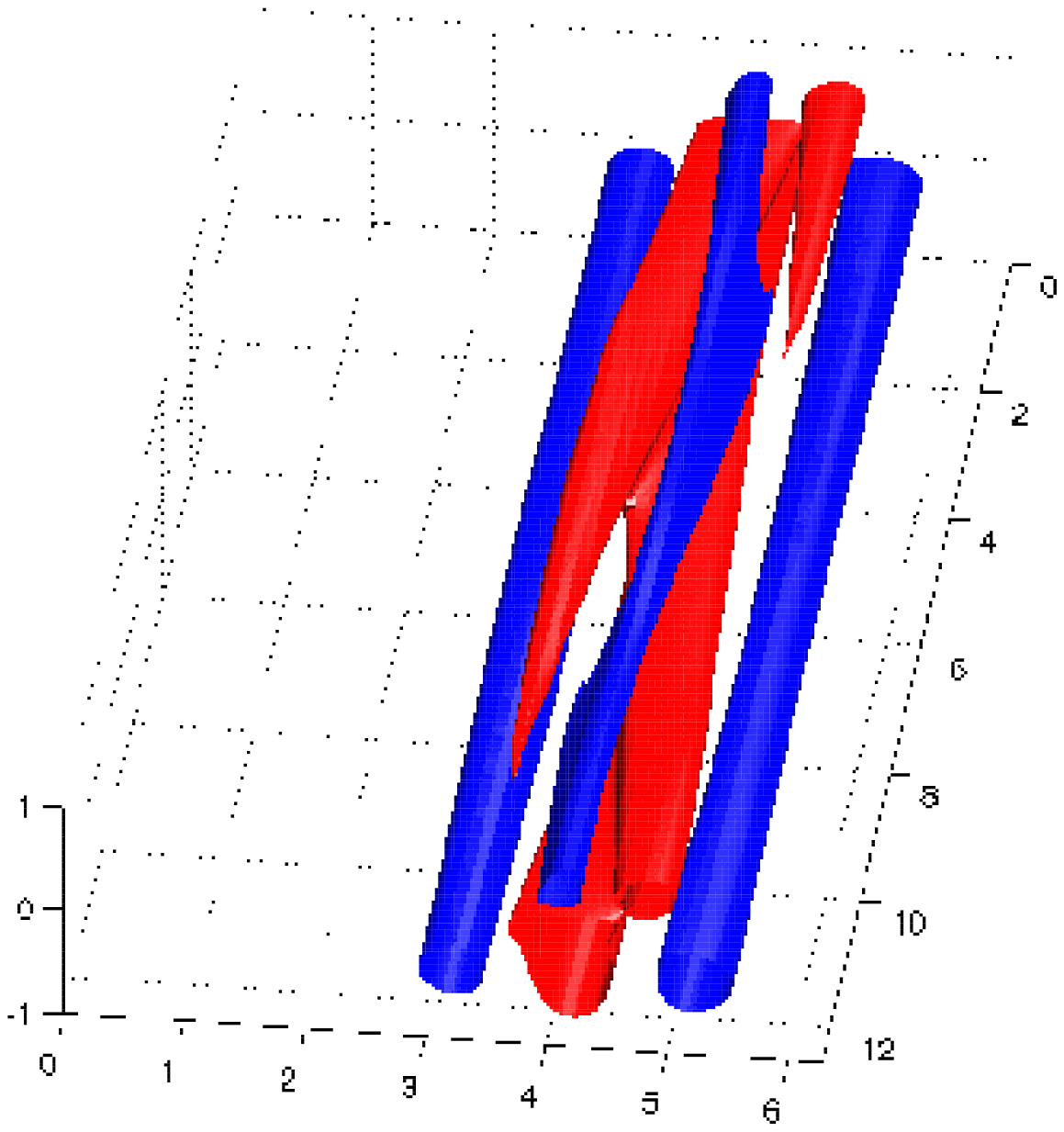}}
  \centerline{\includegraphics[width=0.32\textwidth]{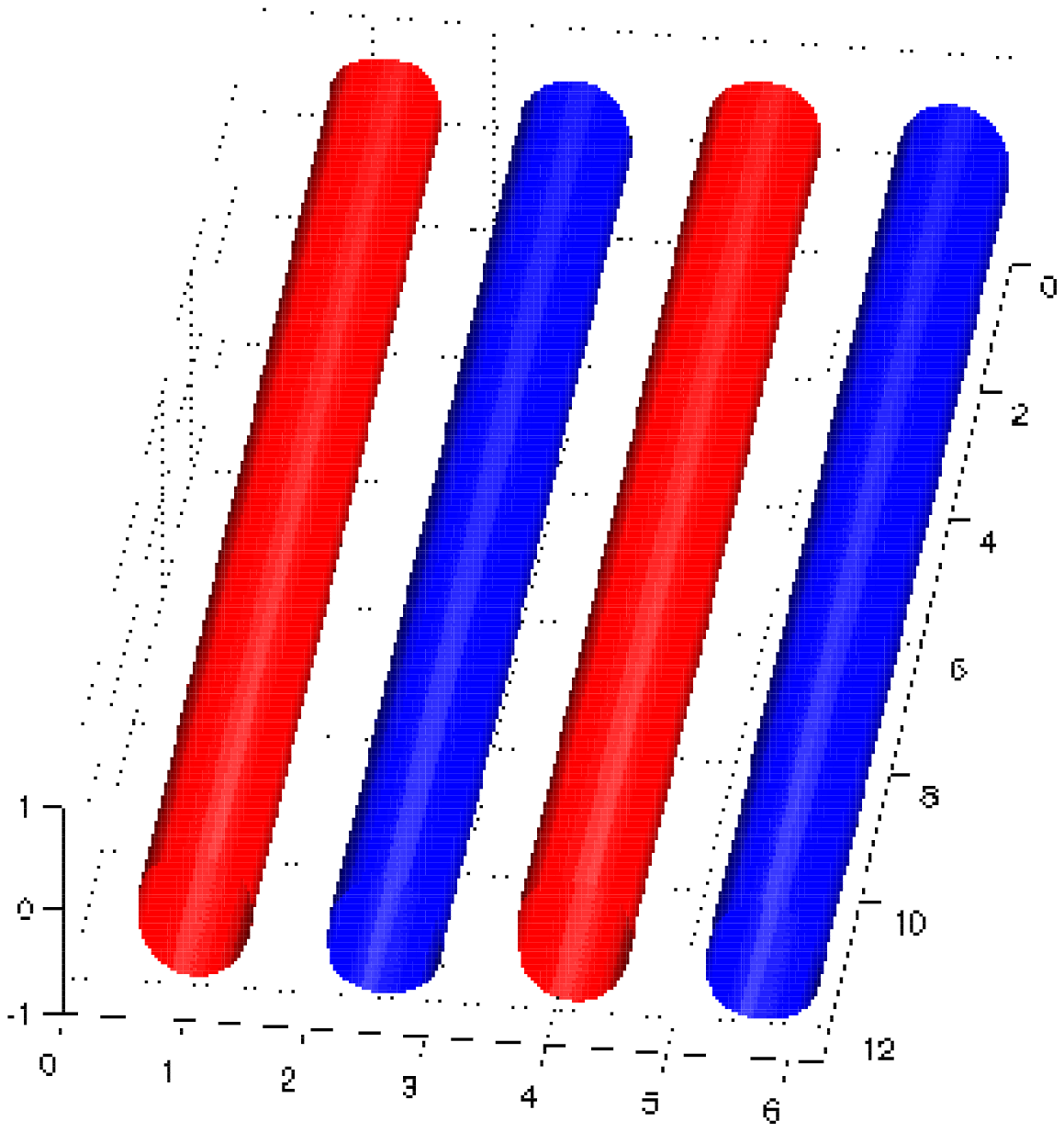}
    \includegraphics[width=0.32\textwidth]{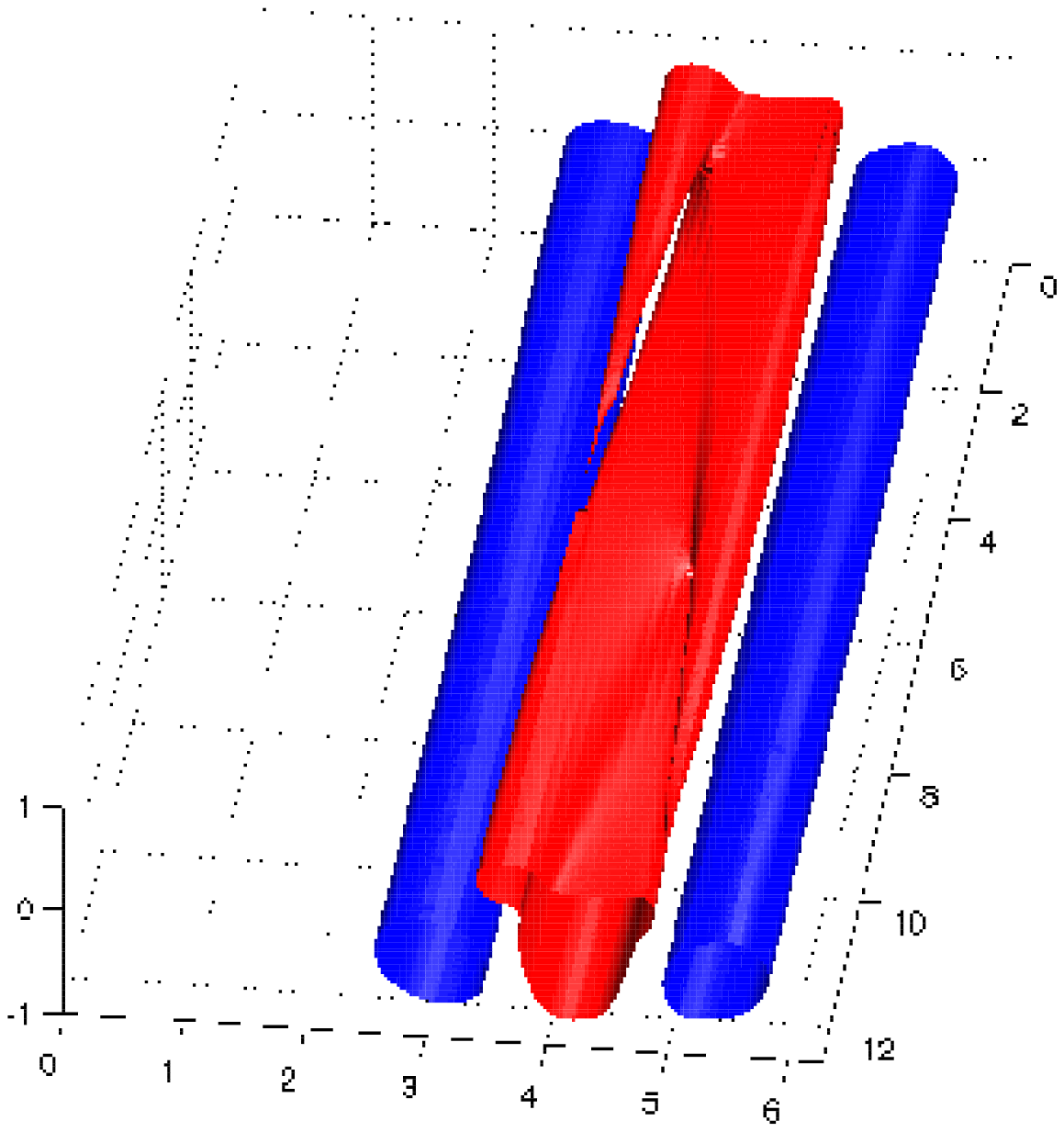}
    \includegraphics[width=0.32\textwidth]{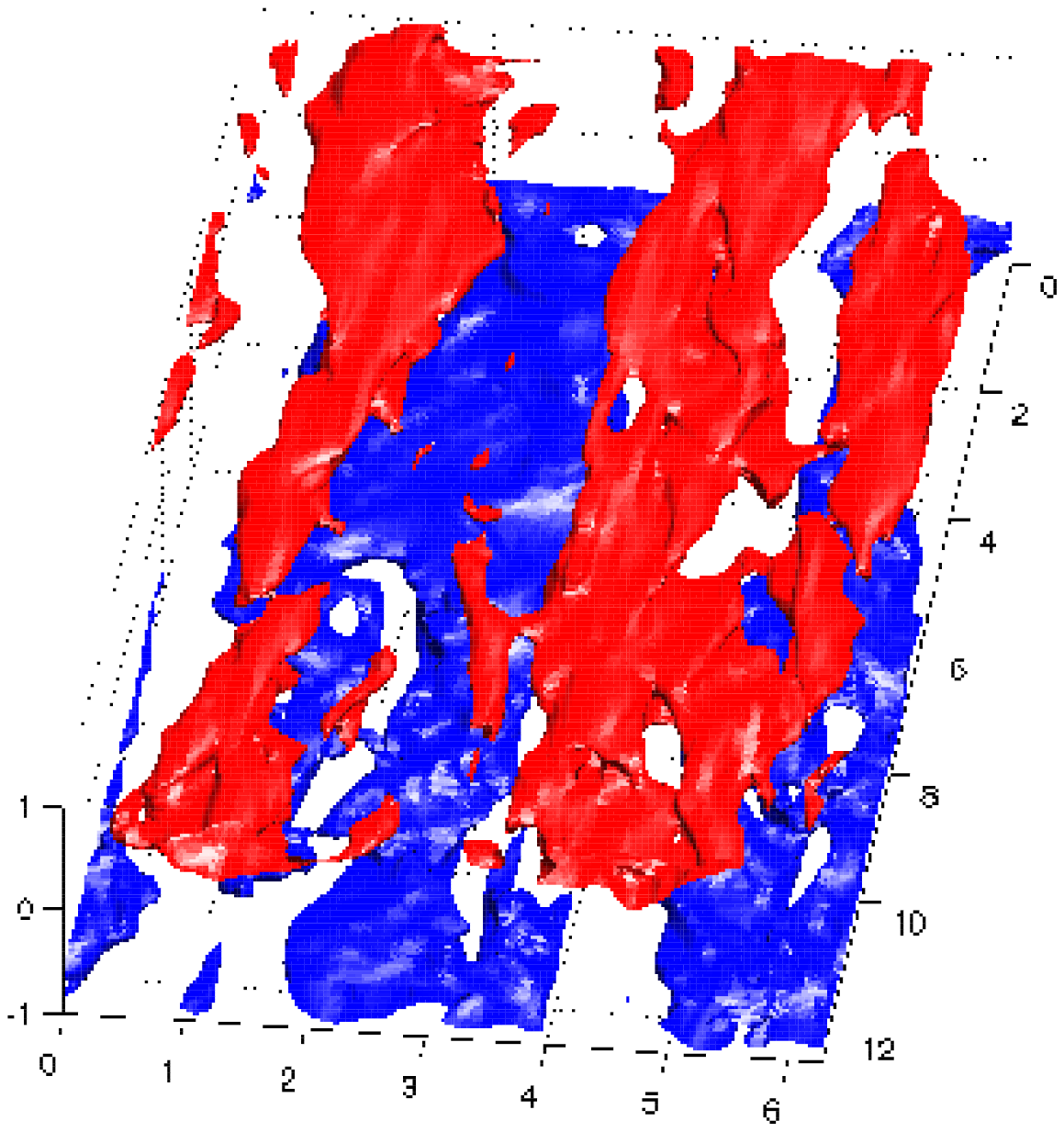}}
 \caption{Iso surfaces of streamwise velocity $u$, at 60\% of maximum and minimum value, for NLOP
   $E_{0}= 3.2 \times 10^{-7}$ (left), approximated minimal seed
   (centre) and turbulent seed at $E_{0}= 3.3 \times 10^{-7}$ (right),
   at times 0, 75, 150, 250 and 400. The minimal and turbulent seeds
   are initially localized but quickly unpack into streamwise streaks,
   which are stable for the minimal seed but unstable for the
   turbulent seed ultimately leading to breakdown.}
\label{iso_DH}
\end{figure}

Figure \ref{GT_DH} also indicates that the turbulent seeds show the same
trend, as in the BF92 geometry, with regards to the optimal target
time, namely that it increases drastically as $E_0 \rightarrow
E_{c}^+$. We conclude that the turbulent seeds remain near the edge
for an even greater period of time than the turbulent seeds in the BF92
geometry. This may be because they are closer to the edge and/or that
the edge is less repelling. As before, we have traced the edge up to
$t=400\, h/U$ using a slightly rescaled turbulent seed to find a
similar plot to figure \ref{3plots_BF} (not shown). A comparison of
cross sectional streamwise velocities for the NLOP at $E_0=3.2 \times
10^{-7}$, the minimal seed and the turbulent seed at
$3.3 \times 10^{-7}$ in figure \ref{contours_DH} again emphasizes
their different temporal evolutions despite being so close
energetically.  Also interestingly, the NLOP is localised in
the cross-stream direction and is not dissimilar from the minimal seed
although they are clearly not the same. This is made obvious by
comparing  their streamwise structure: see the top row of figure
\ref{iso_DH}.

%\begin{figure}
%  \centerline{\includegraphics[width=1.0\textwidth]{New_figures/DH/3Es_v_T_DH}}% Images in 100% size
%  \caption{Energy against time for edge state. Upper bound of edge
%    state blue, lower bound red. Every 100 time units, the edge state
%    is rescaled to produce new upper and lower bounds. The
%    continuation of the old upper bound after it leaves the edge is an
%    orange dashed line, continuation of lower bound after it leaves
%    the edge green dashed line. Therefore minimal seed for $E_{0} =3.3
%    \times 10^{-7}$ is the first blue line which continues into orange
%    dashed line and r-minimal seed is first red/ green dashed
%    line. While all three initially evolve in the same manner the
%    minimal seed flow eventually transitions to turbulence, while the
%    edge state remains approximately the same, and the r-minimal seed
%    begins to decay.}
%\label{3plots_DH}
%\end{figure}

%
% Fig 10
%

%
% figure 11
%

Figure \ref{iso_DH} also shows how the NLOP unpacks into a series of
streamwise streaks. An examination of the early time suggests that it
is a combination of the Orr and lift-up mechanisms (as discussed in
PWK11) that is responsible for the localized flow unpacking into
streamwise streaks. The minimal and turbulent seeds also unpack in the
streamwise and cross-stream direction producing streamwise streaks
which are still spanwise localised. If there is sufficient energy in
these streaks they are unstable (the turbulent seed) otherwise not
(the minimal seed). By comparing the isosurface of the QLOP at time
zero (figure \ref{iso_Q/NLOP}) and the isosurfaces depicting the time
evolution of the NLOP it is clear that the NLOP evolves into a
structure which is very similar to the QLOP. This suggests that while
at early time there exists a distinct NLOP structure, which is able to
extract enhanced gain from the base flow by `unpacking', it later
exploits the same `lift-up' mechanism as the QLOP at intermediate
times.

It is significant that the minimal and turbulent seeds are
spanwise-localised (at least until the turbulence is reached) in this
$2\pi$ wide geometry and not in the $1.05\pi$ wide geometry of
BF92. Of course the higher $Re$ must be a contributory factor but the
spanwise dimension does seem important. \cite{pringle2010a} originally
found an azimuthally-localised (and radially-localised) NLOP in a
short pipe where one could talk about a `spanwise' (azimuthal)
lengthscale of $2 \pi$ (radii or half-channel heights). The emergence
of a NLOP in the wider geometry is also noteworthy. The LOP, and by
definition QLOP, are global periodic states which are largely
insensitive to the geometry, whereas the gathering evidence is that
the NLOP is an attempt by the fluid to localise in order to maximise
the energy gain for given {\em global} kinetic energy. As a result,
the general trend should be for the energy cross-over from QLOP to
NLOP to {\em de}crease with increasing domain size. Clearly this
cross-over is above $E_c$ for the BF92 geometry at $Re=1000$ and below
$E_c$ for the M11 geometry at $Re=1500$.

The evolution shown in the right column of figure \ref{iso_DH} looks
very similar to that shown in figure 4 of M11 (albeit at a slightly
different initial energy) indicating that we have generated an
approximation to the same (presumably unique) minimal seed. This is
further supported by the aforementioned correspondence in their and
our estimates for $E_c$. Beyond validating each others results (which
is important for nonlinear optimisation problems), this points to an
insensitivity in the choice of the functional to be maximised for
finding the minimal seed. There is one proviso, of course, that the
functional must be selected so that it detects turbulent flows by
assuming large values as discussed in PWK11.

\section{Discussion} \label{conc}

In this paper, we have sought the disturbance to plane Couette flow of
a given finite kinetic energy $E_0$ which will experience the largest
subsequent energy gain $G=E(T)/E_0$ where the time of maximum gain $T$
is an {\em output} of our variational formulation. Two geometry-$Re$
situations have been considered: $(L_x \times L_y \times L_z,\,\Rey)
=(4.08\pi \times 2 \times 1.05\pi, \,1000)$ as it was used for the
original calculations of linear optimals in \cite{butler1992} and $(4 \pi
\times 2 \times 2\pi,\,1500)$ for which analogous calculations have
recently been performed optimising the total dissipation over a
specified period in \cite{monokrousos2011}. 
Our results can be summarised
as follows.\\
\begin{enumerate}
\item A nonlinear optimal (NLOP) distinct from the `nonlinearised'
  linear optimal (QLOP) exists only in the wider geometry at the $Re$
  considered. \\
\item In both situations, there exists an energy $E_{fail}$ beyond which
  the variational algorithm no longer converges due to the existence
  of turbulence-triggering initial conditions. This means $E_{fail}
  \geq E_c$, the energy above which turbulence can be
  triggered. PWK11's first conjecture is that $E_{fail}=E_c$ if the
  energy hypersurface is sufficiently sampled and we find nothing to
  contradict this.\\
\item The QLOP or NLOP are not found to converge to the minimal seed
  (the disturbance of lowest energy which can trigger turbulence) as
  $E_0 \rightarrow E_c^-$ contradicting PWK11's second conjecture.\\
\item The failure of our variational algorithm to optimise energy gain
  appears to give the same estimate for $E_c$ and the minimal seed as
  optimising the total dissipation over a long time period
  \cite{monokrousos2011}.  This confirms the `robustness of failure'
  of the variational approach discussed by PWK11 providing the
  functional to be optimised assumes large values for turbulent flows.\\
\end{enumerate}

 The underlying motivation for all these types of optimising
 calculations is the hope that the `optimal' perturbation at $E_0<E_c$
 for an appropriately selected functional bears some relation to
 disturbances of lowest energy which actually trigger turbulence. By
 optimising the energy gain, PWK11 found this was at least
 approximately the case for their pipe flow set-up. Here though, this
 is clearly not true in either PCF situation studied. Ironically,
 however, the variational procedure {\em does} identify these
 turbulence-triggering disturbances and the critical energy for
 transition $E_c$ but only indirectly by failing to
 converge. Importantly, we have also found evidence that this
 revealing failure to converge is insensitive to the exact functional
 selected providing the functional takes on heightened values for
 turbulent flows as argued in PWK11. Taken together, the variational
 approach adopted here seems to offer a fairly robust new theoretical
 tool to examine the {\em nonlinear} stability of fluid flows. 

Many applications suggest themselves, but here we just note one -
assessing the stabilizing or destabilizing influence of applied flow
perturbations or controls. Normally, this would be attempted either by
investigating the linearised operator around the base (laminar) flow
or by carrying our exhaustive numerical simulations. The current work
suggests a third way where the movement (in phase space) of the
laminar-turbulent boundary towards (destabilisation) or away
(stabilisation) from the base flow is investigated. We hope soon to report
on some calculations along these lines.

%
%Acknowledgements
%
\vspace{1cm}
\noindent
{\it Acknowledgements}: We would like to thank A. Monokrousos and
D. Henningson for helping us make contact with their work (M11). SMER
would like to thank S. Dalziel and A. Holyoake for their invaluable
help with computational issues and G. Chandler for insightful
discusions. He is supported by a doctoral training award from EPSRC,
and would like to acknowledge the Entente cordiale scholarship program
for support during the production of this manuscript.  This work was
performed using the Darwin Supercomputer of the University of
Cambridge High Performance Computing Service
(http://www.hpc.cam.ac.uk/), provided by Dell Inc. using Strategic
Research Infrastructure Funding from the Higher Education Funding
Council for England.  The research activity of CPC is supported by
EPSRC Research Grant EP/H050310/1 ``AIM (Advanced Instability Methods)
for industry".  CPC would also like to acknowledge the generous
hospitality of the Hydrodynamics Laboratory (LadHyX) \'Ecole
Polytechnique/CNRS during the production of this manuscript.

\bibliography{Paper}
\bibliographystyle{jfm}

\end{document}